\documentclass[12pt]{article}

\usepackage{a4,graphicx,amssymb,slashed}
\usepackage[small,bf]{caption}

\usepackage{hhline}

\newcommand{\note}[1]{}                    

\newcommand{\mnote}[1]{}                   

\def\lsim{\mathrel{\rlap{\lower4pt\hbox{\hskip1pt$\sim$}}
    \raise1pt\hbox{$<$}}}                
\def\gsim{\mathrel{\rlap{\lower4pt\hbox{\hskip1pt$\sim$}}
    \raise1pt\hbox{$>$}}}                

\newcommand{\ts}{\textstyle}

\newcommand{\tophat}{$\phantom{I^{I^{I^I}}}\hspace{-6.5mm} $}

\newcommand{\q}{\hspace*{4mm}}

\newcommand{\ind}[1]{\rm\scriptscriptstyle #1}

\begin{document}

\title{
\vspace{-3.25cm}
\flushright{\small ADP-19-19/T1099} \\
\vspace{-0.35cm}
{\small DESY 19-149} \\
\vspace{-0.35cm}
{\small Liverpool LTH 1211} \\
\vspace{-0.35cm}
{\small December 31, 2019} \\
\vspace{0.5cm}
\centering{\Large \bf Patterns of flavour symmetry breaking in hadron
           matrix elements involving u, d and s quarks}}

\author{\large
        J.~M. Bickerton$^a$,
        R. Horsley$^b$, Y. Nakamura$^c$, \\
        H. Perlt$^d$, D. Pleiter$^e$,
        P.~E.~L. Rakow$^f$, G. Schierholz$^g$, \\
        H. St\"uben$^h$, R.~D. Young$^a$ and J.~M. Zanotti$^a$ \\[1em]
        \small -- QCDSF-UKQCD-CSSM Collaboration -- \\[1em]
        \footnotesize $^a$ CSSM, Department of Physics,
               University of Adelaide, \\[-0.5em]
        \footnotesize Adelaide SA 5005, Australia \\[0.25em]
        \footnotesize $^b$ School of Physics and Astronomy,
               University of Edinburgh, \\[-0.5em]
        \footnotesize Edinburgh EH9 3FD, UK \\[0.25em]
        \footnotesize $^c$ RIKEN Center for Computational Science, \\[-0.5em]
        \footnotesize Kobe, Hyogo 650-0047, Japan \\[0.25em]
        \footnotesize $^d$ Institut f\"ur Theoretische Physik,
               Universit\"at Leipzig, \\[-0.5em]
        \footnotesize 04103 Leipzig, Germany \\[0.25em]
        \footnotesize $^e$ J\"ulich Supercomputer Centre,
               Forschungszentrum J\"ulich, \\[-0.5em]
        \footnotesize 52425 J\"ulich, Germany \\[-0.5em]
        \footnotesize Institut f\"ur Theoretische Physik,
               Universit\"at Regensburg, \\[-0.5em]
        \footnotesize 93040 Regensburg, Germany \\[0.25em]
        \footnotesize $^f$ Theoretical Physics Division,
               Department of Mathematical Sciences, \\[-0.5em]
        \footnotesize University of Liverpool,
               Liverpool L69 3BX, UK \\[0.25em]
        \footnotesize $^g$ Deutsches Elektronen-Synchrotron DESY, \\[-0.5em]
        \footnotesize 22603 Hamburg, Germany \\[0.25em]
        \footnotesize $^h$ Universit\"at Hamburg,
               Regionales Rechenzentrum, \\[-0.5em]
        \footnotesize 20146 Hamburg, Germany}

\date{}

\maketitle


\clearpage

\begin{abstract}
By considering a flavour expansion about the $SU(3)$-flavour symmetric 
point, we investigate how flavour-blindness constrains octet baryon 
matrix elements after $SU(3)$ is broken by the mass difference between 
quarks. Similarly to hadron masses we find the expansions to be constrained 
along a mass trajectory where the singlet quark mass is held constant,
which provides invaluable insight into the mechanism of flavour symmetry 
breaking and proves beneficial for extrapolations to the physical point.
Expansions are given up to third order in the expansion parameters.
Considering higher orders would give no further constraints on the
expansion parameters. The relation of
the expansion coefficients to the quark-line-connected and 
quark-line-disconnected terms in the three-point correlation functions
is also given. As we consider Wilson clover-like fermions, the addition
of improvement coefficients is also discussed and shown to be included
in the formalism developed here. As an example of the method
we investigate this numerically via a lattice calculation of
the flavour-conserving matrix elements of the vector first class
form factors.

\end{abstract}


\clearpage


\tableofcontents 

\clearpage


\section{Introduction} 


Understanding the pattern of flavour symmetry breaking and mixing, and
the origin of CP violation, remains one of the outstanding problems in
particle physics. The big questions to be answered are
(i) What determines the observed pattern of quark and lepton
mass matrices and (ii) Are there other sources of flavour symmetry
breaking? In \cite{bietenholz10a,bietenholz11a}
we have outlined a programme to systematically investigate the pattern
of flavour symmetry breaking. The program has been successfully applied
to meson and baryon masses involving up, down and strange quarks.
In this article we will extend the investigation to include matrix elements.

The QCD interaction is flavour-blind. Neglecting
electromagnetic and weak interactions, the only difference
between flavours comes from the quark mass matrix. We have our best 
theoretical understanding when all three quark flavours
have the same masses, because we can use the full power of flavour
$SU(3)$. The strategy is to keep the average bare quark mass
$\bar{m} = (m_u+m_d+m_s)/3$ constant and expand the matrix elements
about the flavour symmetric point $m_u = m_d = m_s$. Thus all the
quark mass dependence will be expressed as polynomials in
$\delta m_q = m_q - \bar{m}$, $q = u$, $d$, $s$.
It should be mentioned that this is a completely different 
approach for studying the manifestations of low-energy QCD than chiral 
perturbation theory. It is a complementary method and based on group 
theory rather than effective field theory.

The programme has been successfully applied
to meson and baryon masses in \cite{bietenholz10a,bietenholz11a}
including an extension to incorporate QED effects 
\cite{Horsley:2015eaa,Horsley:2015vla,Horsley:2019wha}.
Besides constraining the quark mass dependence of hadron masses, 
which helps in extrapolations to the physical point, 
it provides valuable information on the physics of flavour symmetry 
breaking. For example, the order of the polynomial can be associated 
with the order of $1/N_c$ corrections, \cite{Jenkins:2009wv}.
Furthermore, similar to the analysis of Gell-Mann and Okubo 
\cite{gell-mann62a,okubo62a}, the order of the polynomial classifies
the order of $SU(3)$ breaking, \cite{bietenholz10a,bietenholz11a}.
As opposed to the conventional method of keeping the
strange quark mass fixed, our method has the further advantage that flavour
singlet quantities which are difficult to compute can now be disentangled
in the extrapolation, and are largely constant on the $\bar{m}$ constant line.

\begin{figure}[htb]
   \begin{tabular}{cc}
      \includegraphics[width=6.50cm]{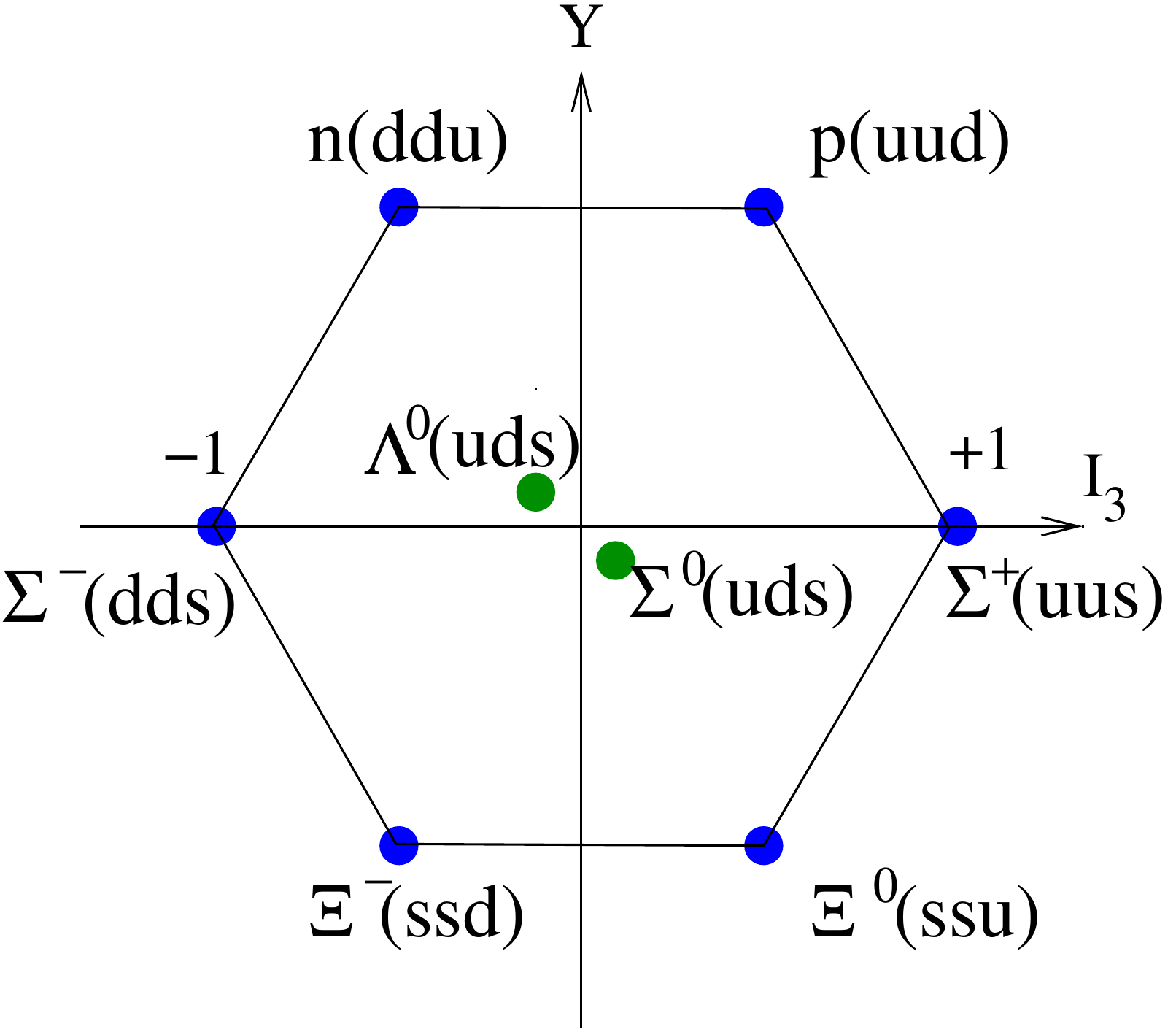}
      \hspace{1.0cm}
      \includegraphics[width=6.50cm]{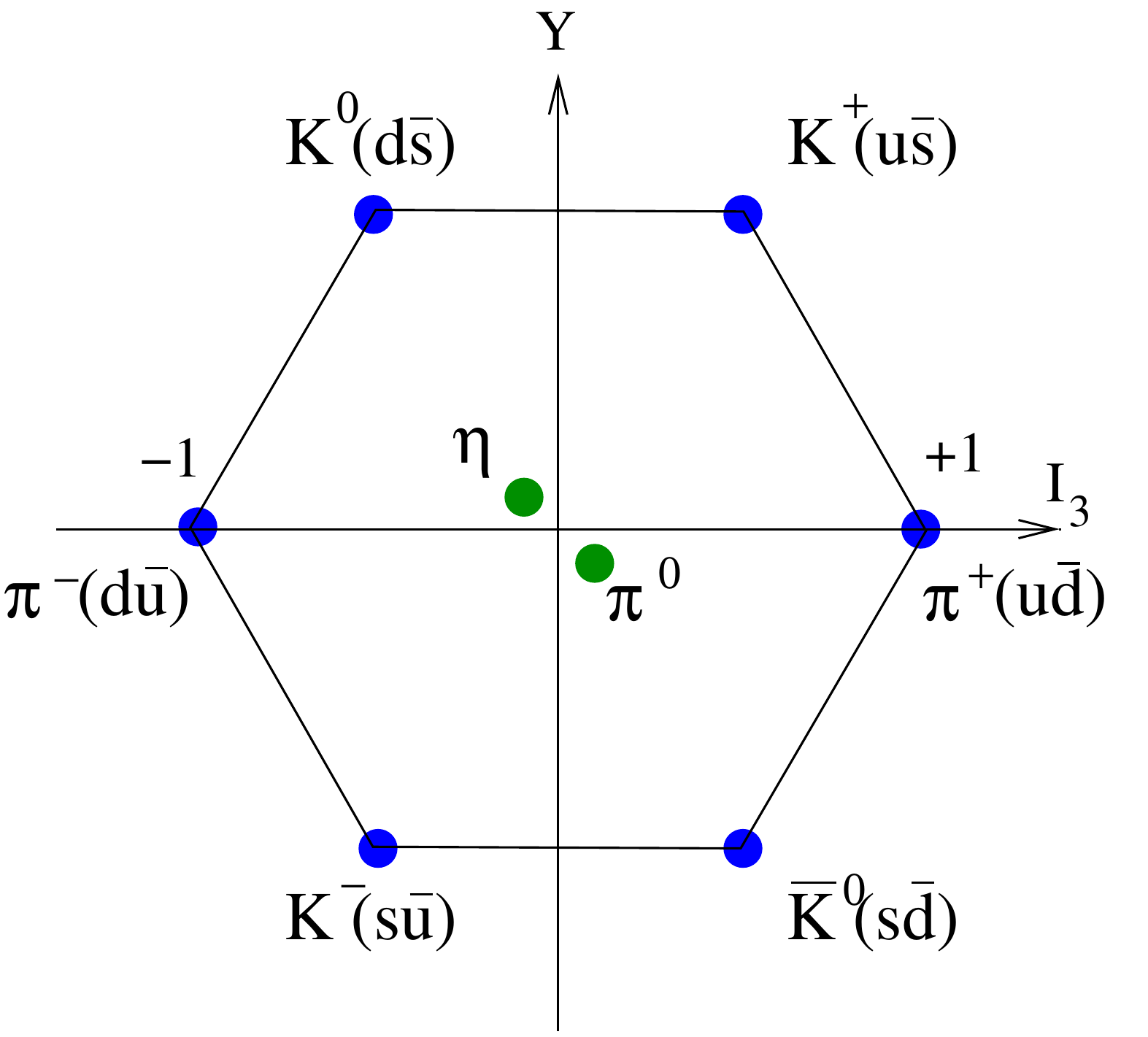}
   \end{tabular}
   \caption{Left panel: The baryon octet. Right panel: The meson octet.}
   \label{meson_baryon_octet}
\end{figure}

In this article we shall concentrate on matrix elements for the
baryon octet as sketched in the $Y$ -- $I_3$ plane in the left
hand panel of Fig.~\ref{meson_baryon_octet}.
It is easy to translate the results to octet mesons
sketched in the right hand panel of Fig.~\ref{meson_baryon_octet}.
Furthermore we restrict ourselves to the case of $n_f = 2+1$, i.e.\
the case of degenerate $u$ and $d$ quark masses, $m_u = m_d \equiv m_l$.
(Initial results were given in \cite{Cooke:2012xv}.)
However our method is also applicable to isospin breaking effects
arising from non-degenerate $u$ and $d$ quark masses. We postpone this analysis
to a separate paper, including electromagnetic effects, \cite{QCDSF}.
The formalism is general. In our application we consider
for definiteness just local currents, but covering all possible 
Dirac gamma matrix structure%
\footnote{It can also easily be extended to currents including covariant
derivatives.}.

While of intrinsic interest in itself, an obvious application
of this formalism is the determination of semileptonic decay form
factors and the associated CKM matrix element, $|V_{us}|$. In general
disentangling quark mass and momentum dependencies is helpful
for determining generalised form factors of baryons, as described
for example in the forthcoming Electron Ion Collider (EIC) programme,
\cite{Accardi:2012qut}.

The structure of this article is as follows.
In section~\ref{gen_currents}, we discuss all
possible currents (which we call `generalised currents' here) and
also their splitting into `first' class and `second' class currents.
Then in sections~\ref{qm_expansions}, \ref{method_ME}, \ref{mass_amp}
we discuss the group theory. In section~\ref{qm_expansions} we define our
expansion parameter, $\delta m_l$ and the general structure of
our expansions. Also discussed there (and at the beginning of
section~\ref{singop}) are simple cases
which have previously been determined. In particular the
singlet case will be used later in this article. The next
section, section~\ref{method_ME} gives our sign conventions
(commonly employed in chiral perturbation theory). As we have
mass degenerate $u$ and $d$ quarks then there is an $SU(2)$ isospin
symmetry. We then use the Wigner-Eckart theorem to give the reduced
matrix elements, contrasting the difference here to the usual conventions.
Then in section~\ref{mass_amp}, after discussing the group theory
classification of $SU(3)$ tensors, we determine those relevant to our 
study (with complete tables being given in Appendix~\ref{non-zero_tensor}),
and then in section~\ref{LO_coeff_tables} give the LO expansions.
Higher-order terms are given in section~\ref{higher_order}.
These sections giving the expansion coefficients form the heart of
this report. This is followed by section~\ref{am_sym_pt} where 
we briefly restrict ourselves to a discussion of the amplitudes at the
symmetric point. 

Continuing with the main thread, in section~\ref{FanSection} linear 
combinations of the matrix elements are constructed for the various baryons, 
leading to functions that all have the same value at the
$SU(3)$ flavour symmetric point. Four different `fan' plots are
constructed, two detailed in section~\ref{FanSection} and a further
two given in Appendix~\ref{doub-sing_rep}.

Lattice QCD determinations of matrix elements involve the
computation of $3$-point correlation functions, which fall
into two classes -- quark-line connected diagrams and
quark line disconnected diagrams. In section~\ref{con+disc},
we discuss the implications of this splitting for the
$SU(3)$ symmetry flavour breaking expansions at LO. In particular
for the connected terms, there are further constraints on the
expansion coefficients. In section~\ref{matrix_els} this is
applied to the baryon-diagonal matrix elements (and as
a special case to the electromagnetic current). The quark-line
connected expansions are given there with the general expressions
described in Appendix~\ref{LO_flav_diag_ME} while the
quark-line disconnected expansions are given in
Appendix~\ref{LO_flav_diag_ME_dis}.

In section~\ref{renorm+improv} we discuss improvement coefficients 
for the currents, see e.g.\ \cite{Bhattacharya:2005rb},
and show that they lead to (small) modifications of 
the $SU(3)$ flavour symmetric breaking expansion coefficients. 
Using the vector current as an example, we show how we can determine 
two improvement coefficients (and the renormalisation constant).
Section~\ref{lat_comp} briefly describes how matrix elements
(i.e.\ form factors) are computed from the ratios of $3$-point
to $2$-point correlation functions.
In section~\ref{lattice_details}, we describe our $n_f = 2+1$ flavour
Wilson clover action used and provide some numerical details.
In section~\ref{results}, specialising to the vector current again
we give some flavour singlet `$X$'-plots, showing their constancy 
for the $F_1$ and $F_2$ form factors. This is followed by some fan plots
revealing $SU(3)$-breaking effects. The momentum transfer 
($Q^2$) dependence of the expansion coefficients is also investigated.
The numerical values of two improvement coefficients are also determined.
Finally in section~\ref{conclusions} we give our conclusions.

 
\section{Baryon matrix elements and generalised currents}
\label{gen_currents}


We take here `generalised currents' to be
\begin{eqnarray}
   J^{\ind F \,({\cal M})}
     = \overline{q} F \gamma^{\ind{({\cal M})}} q
       \equiv \sum_{f_1,f_2= 1}^3 F_{f_1f_2} \,
              \overline{q}_{f_1} \gamma^{\ind{({\cal M})}} q_{f_2} \,,
\label{gen_current}     
\end{eqnarray}
where $q$ is a flavour vector, $q = (u,d,s)^T$, $F$ is a flavour
matrix and $\gamma^{\ind{({\cal M})}}$ is some Dirac gamma matrix.
In particular we have $\gamma^{\ind{({\cal M})}} = \gamma^{\ind{({\cal M})}\mu}$,
$\gamma^{\ind{({\cal M})}\mu}\gamma^{\ind{({\cal M})}}_{\ind{5}}$,
$I$, $i\gamma^{\ind{({\cal M})}}_{\ind{5}}$ and $\sigma^{\ind{({\cal M})}^{\mu\nu}}$
for the vector $V^{\ind{({\cal M})}\mu}$, axial $A^{\ind{({\cal M})}\mu}$,
scalar $S^{\ind{({\cal M})}}$, pseudoscalar $P^{\ind{({\cal M})}}$
and tensor $T^{\ind{({\cal M})}\mu\nu}$ generalised currents respectively.
The further generalisation to operators including covariant derivatives
is straightforward. With our gamma matrix conventions, we obviously have
\begin{eqnarray}
   J^{\ind{F({\cal M})}\dagger}
     = \bar{q} F^{\ind{T}} \gamma^{\ind{({\cal M})}} q \,,
\label{hermitian_gen_current}     
\end{eqnarray}
and so are Hermitian if the flavour matrix, $F$, is symmetric 
and anti-Hermitian if $F$ is antisymmetric.

We use Minkowski space%
\footnote{The conventions used include
          $\eta^{\mu\nu} = \mbox{diag}(1,-1,-1,-1)$,
          $\gamma^{\ind{({\cal M}})\mu\,\dagger}
             = \gamma^{\ind{({\cal M})}0}\gamma^{\ind{({\cal M})}\mu}
                                                 \gamma^{\ind{({\cal M})}0}$,
          $\gamma_{\ind{5}}^{\ind{({\cal M})}}
             = i \gamma^{\ind{({\cal M})}0}\gamma^{\ind{({\cal M})}1}
               \gamma^{\ind{({\cal M})}2}\gamma^{\ind{({\cal M})}3}$
          and $\sigma^{\ind{({\cal M})}\mu\nu} 
          = i/2 [ \gamma^{\ind{({\cal M})}\mu}, \gamma^{\ind{({\cal M})}\nu}]$.},
and to emphasise this we use the superscript: $^{\ind{({\cal M})}}$.
The expansion described later will be valid whether we are working
in Minkowski or Euclidean space (when we will drop the superscript). 
We wish to compute matrix elements for $B \to B^\prime$
\begin{eqnarray}
   A(B \to B^\prime)
   = \langle B^\prime, \vec{p}^{\,\prime}, \vec{s}^{\,\prime} | J^{\ind{F({\cal M})}}(q)
                                    | B, \vec{p}, \vec{s}^{\,} \rangle 
   \equiv A_{\ind{\bar{B}^\prime FB}} \,,
\label{A_def}   
\end{eqnarray}
where $B$ and $B^\prime$ belong to the baryon octet, the members of
which are shown in Fig.~\ref{meson_baryon_octet}
(the quark content of each baryon is also depicted there).
This can thus include scattering processes for example $Be \to Be$
or semi-leptonic (or $\beta$-decays) $B \to B^\prime e\bar{\nu}_e$ from a
parent baryon, $B$, to a daughter baryon $B^\prime$. For semi-leptonic
decays in the standard model, neutral currents are flavour diagonal,
and hence there is an absence of flavour-changing neutral currents (FCNCs),
i.e.\ $s \to d$ transitions. In addition $\Delta S = \Delta Q$ 
violating modes are not seen. From Fig.~\ref{meson_baryon_octet} 
we see that this means that transitions from right to the left 
in the picture are suppressed or absent.
For example twelve allowed non-hyperon and hyperon $\beta$-decays,
are listed in Table $1$ of \cite{cabibbo03b}. Of course the present
formalism does not incorporate these constraints, but this can motivate
our choice of independent matrix elements, which are transitions from
the left to the right in Fig.~\ref{meson_baryon_octet}.

Momentum transfer $p^{\ind{({\cal M})}} - p^{\ind{({\cal M})} \prime}$
is more natural to take for semi-leptonic decays, as this is the momentum
carried by the lepton and neutrino. However for scattering processes
$p^{\ind{({\cal M})} \prime} - p^{\ind{({\cal M})}}$ is more natural. We wish to
adopt a unified notation here, so we define the momentum transfer as
\begin{eqnarray}
   q^{\ind{({\cal M})}} = p^{\ind{({\cal M})} \prime} - p^{\ind{({\cal M})}}
              = \left( E_{B^\prime}(\vec{p}^{\,\prime}) - E_B(\vec{p}),\, 
                       \vec{p}^{\,\prime} - \vec{p}^{\,} \right) \,.
\label{qMink}
\end{eqnarray}
The decompositions of the matrix elements in eq.~(\ref{A_def})
are standard, and we write 
\begin{eqnarray}
   \langle B^\prime, \vec{p}^{\,\prime}, \vec{s}^{\,\prime} | J^{\ind{F({\cal M})}}(q)
                                    | B, \vec{p}, \vec{s} \rangle
   = \bar{u}_{B^\prime}(\vec{p}^{\,\prime}, \vec{s}^{\,\prime})
                {\cal J}^{({\cal M})}(q) 
         u_B(\vec{p}, \vec{s}) \,,
\label{current_FF}                
\end{eqnarray}
with for ${\cal J}^{\ind{({\cal M})}}$
\begin{eqnarray}
   {\cal V}^{\ind{({\cal M})}\mu}
      &=& \gamma^{\ind{({\cal M})}\mu} F_1
           + i\sigma^{\ind{({\cal M})}\mu\nu}q^{\ind{({\cal M})}}_\nu{F_2
                                                \over M_B+M_{B^\prime}} 
           + q^{\ind{({\cal M})}\mu} {F_3 \over M_B + M_{B^\prime}} \,,
                                                           \nonumber  \\
   {\cal A}^{\ind{({\cal M})}\mu} 
      &=& \left( \gamma^{\ind{({\cal M})}\mu} G_1
                 + i\sigma^{\ind{({\cal M})}{\mu\nu}}
                     q^{\ind{({\cal M})}}_\nu{G_2 \over M_B+M_{B^\prime}} 
                 + q^{\ind{({\cal M})}\mu} {G_3 \over M_B + M_{B^\prime}}
          \right) \gamma^{\ind{({\cal M})}}_{\ind{5}} \,,
                                                           \nonumber  \\
   {\cal S}^{\ind{({\cal M})}}
      &=& g_S \,, 
                                                           \nonumber  \\
   {\cal P}^{\ind{({ \cal M})}}
      &=& i\gamma^{\ind{({\cal M})}}_{\ind{5}} g_P \,,
                                                     \label{FF_def}   \\
   {\cal T}^{\ind{({\cal M})\mu\nu}}
      &=& \sigma^{\ind{({\cal M})}\mu\nu} h_1
          + i(q^{\ind{({\cal M})}\mu}\gamma^{\ind{({\cal M})}\nu} -
             q^{\ind{({\cal M})}\nu}\gamma^{\ind{({\cal M})}\mu})
                  {h_2 \over M_B + M_{B^\prime}}
                                                           \nonumber  \\
      & & + i(q^{\ind{({\cal M})}\mu}P^{\ind{({\cal M})}\nu} -
             q^{\ind{({\cal M})}\nu}P^{\ind{({\cal M})}\mu})
                  {h_3 \over (M_B + M_{B^\prime})^2}
                                                           \nonumber  \\
      & & + i(\gamma^{\ind{({\cal M})}\mu}
                 \slashed{q}^{\ind{({\cal M})}}\gamma^{\ind{({\cal M})}\nu} -
                 \gamma^{\ind{({\cal M})}\nu}
                 \slashed{q}^{\ind{({\cal M})}}\gamma^{\ind{({\cal M})}\mu})
                   {h_4 \over M_B + M_{B^\prime}} \,,
                                                           \nonumber
\end{eqnarray}
where $P^{\ind ({\cal M})} = p^{\ind({\cal M})} + p^{\ind ({\cal M})\,\prime}$.
$F_i \equiv F_i^{\ind{\bar{B}^\prime FB}}$, $G_i \equiv G_i^{\ind{\bar{B}^\prime FB}}$,
$g_S \equiv g_S^{\ind{\bar{B}^\prime FB}}$, $g_P \equiv g_P^{\ind{\bar{B}^\prime FB}}$
and $h_i \equiv h_i^{\ind{\bar{B}^\prime FB}}$ are the form factors and are
functions of $q^{\ind{({\cal M})}\, 2}$ and the masses of the baryons
(or alternatively the quark masses). Each combination in
eqs.~(\ref{current_FF}, \ref{FF_def}) represents a current times
a  form factor (i.e.\ the coefficient).
For example the first term for the vector current reads
$\bar{u}_{B^\prime}(\vec{p}^{\,\prime}, \vec{s}^{\,\prime})
\gamma^{\ind{({\cal M})}\mu} u_B(\vec{p}, \vec{s}) \times
F_1^{\ind{\bar{B}^\prime FB}}(q^{\ind{({\cal M})}\, 2})$.
The goal of this article is to establish ways in which these
form factors depend on the transition taking place and on the quark masses.

From eqs.~(\ref{hermitian_gen_current}, \ref{A_def}) we have
\begin{eqnarray}
   A_{\ind{\bar{B}F^{\ind{T}}B^\prime}}^* = A_{\ind{\bar{B}^\prime FB}} \,,
\label{A_herm}  
\end{eqnarray}  
and we now apply this to eq.~(\ref{current_FF}) with individual terms
defined by eq.~(\ref{FF_def}). Consider first the current pieces.
For example for the vector currents we find that the first and
second terms (i.e.\ currents) are unaltered,
$(\bar{u}_B\gamma^{\ind{({\cal M})}\mu} u_{B^\prime})^*
= \bar{u}_{B^\prime}\gamma^{\ind{({\cal M})}\mu} u_B$,
$(\bar{u}_{B^\prime} i\sigma^{\ind{({\cal M})}\mu\nu} (-q_\nu^{\ind{({\cal M})}})u_B)^*
  = \bar{u}_B i\sigma^{\ind{({\cal M})}\mu\nu} q_\nu^{\ind{({\cal M})}}u_{B^\prime}$
while the third current changes sign, 
$(\bar{u}_B (-q^{\ind{({\cal M})}\mu}) u_{B^\prime})^*
= - \bar{u}_{B^\prime} q^{\ind{({\cal M})}\mu} u_B$.
Strong interactions are invariant under $T$-parity and from this it can
be shown that the form factors can be chosen to be all real.
Hence from eq.~(\ref{A_herm}) we must have
\begin{eqnarray}
   F_1^{\ind{\bar{B}F^{\ind{T}}B^\prime}} = F_1^{\ind{\bar{B}^\prime FB}} \,, \qquad
   F_2^{\ind{\bar{B}F^{\ind{T}}B^\prime}} = F_2^{\ind{\bar{B}^\prime FB}} \,,
\label{V_first}   
\end{eqnarray}
but
\begin{eqnarray}
   F_3^{\ind{\bar{B}F^{\ind{T}}B^\prime}} = - F_3^{\ind{\bar{B}^\prime FB}} \,.
\label{V_second}  
\end{eqnarray}
$F_1$ and $F_2$ are called first class form factors while $F_3$ is 
called a second class form factor. This can be applied to all the 
further currents. These properties of the form factors thus give rise 
to the notation, \cite{Weinberg:1958ut}
\begin{eqnarray}
   \begin{array}{rl}
      \mbox{first class}  & F_1, F_2, G_1, G_3, g_S, g_P, h_1, h_2, h_3  \\
      \mbox{second class} & F_3, G_2, h_4  \\
   \end{array} \,,
\label{1+2class}   
\end{eqnarray}
(with the meaning given by eqs.~(\ref{V_first}, \ref{V_second})).
Note that when $B^\prime = B$, then the second class currents 
(i.e.\ form factors) vanish. This occurs, either for a scattering
process (i.e.\ a diagonal current in flavour space, so the matrix $F$
is symmetric and the current is Hermitian) or for semi-leptonic processes
at the quark mass symmetric point.

We now consider the flavour structures, i.e.\ the possible flavour
matrices in eq.~(\ref{gen_current}). In Table~\ref{ind8} we give the
possible octet states,
\begin{table}[!htb]  
\begin{center} 
   \begin{tabular} {c|ccc}
      Index & Baryon ($B$) & Meson ($F$) & Current ($J^{\ind{F}}$) \\
      \hline
      \rule{0pt}{1.0\normalbaselineskip} 
      1     & $n$ & $K^0$ & $ \bar d \gamma s $  \\
      2     & $p$ & $K^+$ & $ \bar u \gamma s $  \\  
      3     & $\Sigma^-$ &  $\pi^-$ & $ \bar d \gamma u $  \\  
      4     & $\Sigma^0$ & $\pi^0$ & $ \frac{1}{\sqrt{2}}
                         \left(\bar u \gamma u - \bar d \gamma d \right) $ \\   
      5     & $\Lambda^0$&  $\eta$  & $ \frac{1}{\sqrt{6}}
                                        \left(\bar u \gamma u +
                         \bar d \gamma d -2  \bar s \gamma s \right) $ \\   
      6     & $\Sigma^+$ & $\pi^+$ & $\bar u \gamma d $ \\  
      7     & $\Xi^-$ & $K^-$ & $\bar s \gamma u $ \\  
      8     & $\Xi^0$ & $\bar K^0$ & $\bar s \gamma d $  \\
      \hline
      \rule{0pt}{1.0\normalbaselineskip}
      0     &         & $\eta^\prime$ & ${1 \over \sqrt{3}}
                        \left(\bar u \gamma u + \bar d \gamma d +
                               \bar s \gamma s \right)$  \\
   \end{tabular} 
\caption{Our numbering and conventions for the generalised currents.
         For example, $B_3 = \Sigma^-$, $F_3 = \pi^-$,
         $J^{F_3} \equiv J^{\pi^-}$. We use the convention that current 
         (i.e.\ operator) numbered by $i$ has the same effect as absorbing
         a meson with the index $i$. $\gamma$ represents an arbitrary 
         Dirac matrix.}
\label{ind8}
\end{center} 
\end{table} 
$i = 1, \ldots, 8$ and in addition the singlet state, labelled
by $i=0$. As we are primarily concerned with the flavour structure of
bilinear operators, we use the corresponding meson name for the
flavour structure of the bilinear quark currents. So for example
the $i = 5$ current is given by the flavour matrix
$F_{\eta} = \mbox{diag}(1,1,-2)/\sqrt{6}$. 
We shall use the convention that the current $i$ has the same
effect as absorbing a meson with the same index.
In the operator expressions $q$ is the annihilation operator and
$\bar{q}$ the creation operator. As an example, we note that absorbing
a $\pi^+$ annihilates one $d$ quark and creates a $u$ quark. That is
\begin{eqnarray}
   J^{\pi^+}|0\rangle \propto |\pi^+\rangle \,,
\end{eqnarray}
while $\langle p| \bar{u}\gamma d|n\rangle = \langle p|J^{\pi^+}|n \rangle$
represents $p = \pi^+ n$.

As an example of this (current) notation the quark electromagnetic current
can be written by defining an appropriate flavour matrix $F$ or
alternatively as
\begin{eqnarray}
   J_{{\rm em}\,\mu}
     &=& {2\over 3} \bar{u}\gamma_\mu u
        -{1\over 3} \bar{d}\gamma_\mu d
        -{1\over 3} \bar{s}\gamma_\mu s
                                                        \nonumber  \\
     &\equiv& {1 \over \sqrt{2}} V^{\pi^0}_\mu
        +{1 \over \sqrt{6}} V^{\eta}_\mu  \,.
\label{em_current_def}        
\end{eqnarray}        
Furthermore the charged $W$s currents are a mixture of the
charged $\pi$ and $K$ currents, while the $Z$ current is diagonal and thus
a mixture of the $\pi^0$, $\eta$ and $\eta^\prime$ currents.
The $K^0$ current is a FCNC, so only contributes to 
beyond the standard model (BSM) or higher-order processes.

The previous discussion on first and second class currents can now be
reformulated in terms of these flavour matrices and isospin rotations%
\footnote{This discussion follows \cite{georgi84a}.}.
The diagonal currents, and hence diagonal matrix elements, discussed
here are given by $i = 4$, $5$ and $0$ with $F_{\pi^0}$, $F_{\eta}$
and $F_{\eta^\prime}$ respectively. As a result $F_3$, $G_2$, $g_P$, $h_2$
and $h_3$ all vanish for these currents. For the off-diagonal currents
consider the $SU(3)$-flavour symmetric point. As all the quark masses
have the same mass, and in particular the $u$ and $d$ quarks
then we first consider isospin, $I$, invariance. Isospin rotations
are $d$-$u$ rotations and relate off-diagonal currents to
diagonal currents. (For example $\langle p|J^{\pi^+}|n\rangle$
is related to $\langle p|J^{\pi^0}|p\rangle$, see section~\ref{su2rel}.)
Similarly for $U$-spin rotations $s$-$d$, and $V$-spin rotations $s$-$u$.
Hence we expect that for transitions within a given multiplet
(whether $I$, $U$ or $V$) at the $SU(3)$-flavour symmetric point
then again $F_3$, $G_2$, $g_P$, $h_2$ and $h_3$ all vanish.
Between isospin multiplets they need not vanish when
$SU(3)$ flavour symmetry is broken.
We later discuss this in more detail and our coefficient tables,
for example Table~\ref{coef_1_8}, reflect these results.

  
\section{Quark mass expansions}
\label{qm_expansions}


\subsection{Choice of quark masses} 
\label{choice_qm}


As mentioned already, we follow the strategy used in \cite{bietenholz11a}
of holding constant the average bare quark mass
\begin{eqnarray}
   \bar{m} = \frac{1}{3} (m_u + m_d + m_s) \,.
\end{eqnarray}
This greatly reduces the number of mass polynomials 
which can occur in Taylor expansions of physical quantities, 
and relates the quark-mass dependencies of hadron masses or
matrix elements within an $SU(3)$ multiplet. 
Since we expand about the symmetric point where all three quarks
have the same mass, it is useful to introduce the notation
\begin{equation} 
   \delta m_q \equiv m_q - \bar{m}\,, \qquad \qquad q = u, d, s \,,
\label{dml_def}  
\end{equation}
to describe the `distance' from the $SU(3)$ flavour symmetry point.
Note that it follows from the definition that we have the identity
\begin{equation} 
   \delta m_u + \delta m_d + \delta m_s = 0 \,,
\label{zerosum111} 
\end{equation}
so we can always eliminate one of the $\delta m_q$.
In this article we concentrate on the $n_f = 2+1$ case, i.e.\ we keep 
\begin{equation} 
   m_u = m_d \equiv m_l \,.
\label{mass_isospin}
\end{equation}
All our expansion coefficients are functions of $\bar{m}$.
The methods developed here can be generalised to the case of $n_f = 1+1+1$
non-degenerate quark-mass flavours.
For this case eq.~(\ref{zerosum111}) reduces to 
\begin{equation} 
   2\delta m_l + \delta m_s = 0  \,,
\label{zerosum21} 
\end{equation}
which we use to eliminate $\delta m_s$. Thus, all mass dependences will
be expressed as polynomials in the single variable $\delta m_l$. At the
physical point $m_l \ll \bar{m}$, so $\delta m_l$ is negative. However 
on the lattice in principle we are free to choose $\delta m_l$ positive,
and look at matrix elements on both sides of the symmetric point.


\subsection{Matrix elements}


In the following we want to use group theory in flavour
space to calculate the possible quark-mass dependence
of baryonic form factors. However for simplicity of notation
we shall continue to discuss matrix elements and amplitudes,
but it should be noted that for form factors the Lorentz/Dirac structure
has been factored out. So we shall consider the quark mass expansion for
\begin{eqnarray}
   \langle B_i | J^{F_j} | B_k \rangle \equiv A_{\bar{B}_i F_j B_k} \,.
\end{eqnarray}
The indices $i$ and $k$ will run from $1$ to $8$ for octet hadrons
(or $1$ to $10$ for decuplets). The currents/operators we are 
interested in are quark bilinears, so the index $j$ will run from
$1$ to $8$ for non-singlets, or $0$ for the singlet.
In the following the singlet will be considered separately.
When $i \ne k$ we get transition matrix elements; when $i=k$
within the same multiplet, we get operator expectation values.
This has already been indicated in Table~\ref{ind8}.

The allowed quark mass Taylor expansion for a hadronic 
matrix element must follow the schematic pattern 
\begin{eqnarray} 
   \langle B_i | J^{F_j} | B_k \rangle 
      &=& \sum (\mbox{singlet mass polynomial}) \times 
                           (\mbox{singlet tensor})_{ijk} 
                                                        \nonumber  \\
      & & + \sum (\mbox{octet mass polynomial}) \times 
                           (\mbox{octet tensor})_{ijk} 
                                                    \label{schema} \\
      & & + \sum (\mbox{27-plet mass polynomial}) \times 
                           (\mbox{27-plet tensor})_{ijk} 
                                                        \nonumber  \\
      & & + \ \cdots \,.
                                                        \nonumber
\end{eqnarray}
The mass polynomials have been determined and given in
Table~III of \cite{bietenholz11a}. The relevant part
of this table is given in Table~\ref{quadratic}
\begin{table}[!htb] 
   \begin{center} 
   \begin{tabular} {c|cccccc}
   Polynomial & \multicolumn{6}{c}{$SU(3)$}                      \\
   \hline 
   $1$ & $1$ &  &  &  &  &                                       \\
   \hline
   $\delta m_s$ &  & $8$ &  &  &  &                              \\
   $(\delta m_u - \delta m_d)$ &  & $8$ &  &  &  &               \\
   \hline
   $\delta m_u^2 + \delta m_d^2 +\delta m_s^2$ & $1$ &  &  &  & $27$ &  \\
   $3 \delta m_s^2 - (\delta m_u - \delta m_d)^2$ &  & $8$ &  &  & $27$ &  \\
   $\delta m_s (\delta m_d - \delta m_u)$ & & $8$    &  &  & $27$ &   \\
   \hline
   $\delta m_u \delta m_d \delta m_s $ & $1$ &  &  &  & $27$ & $64$  \\
   $\delta m_s (\delta m_u^2 + \delta m_d^2 +\delta m_s^2 )$
                         &  & $8$ &  &  & $27$ & $64$  \\
   $(\delta m_u - \delta m_d) (\delta m_u^2 
              + \delta m_d^2 +\delta m_s^2 )$ &  & $8$ &  &  & $27$ & $64$  \\
   $(\delta m_s - \delta m_u)(\delta m_s-\delta m_d)(\delta m_u-\delta m_d)$
                         &  &   & $10$ & $\overline{10}$ &  & $64$  \\
\end{tabular}
\caption{All the quark-mass polynomials up to $O(\delta m_q^3)$, classified by
         symmetry properties.}
\label{quadratic}
\end{center} 
\end{table} 
where we classify all the polynomials which could occur in a Taylor
expansion about the symmetric point, $\delta m_q = 0$, $q = u$, $d$, $s$
up to $O(\delta m_q^3)$. 
The tensors in eq.~(\ref{schema}) are $3$-dimensional arrays of 
integers and square-roots of integers; objects somewhat analogous to 
three-dimensional Gell-Mann matrices. 
We recover the standard results for unbroken $SU(3)$ by  only keeping
singlet tensors on the right-hand side of eq.~(\ref{schema}). 
Adding higher dimensional flavour tensors tells us the allowed mass
dependences of matrix elements. The dots in eq.~(\ref{schema})
represent terms that are cubic or higher in $\delta m_q$.
 
We now need to classify the three-index tensors according to their 
group transformations, using the same techniques we used for 
masses \cite{bietenholz11a}.
The new cases to look at will be $8 \otimes 8 \otimes 8$ and 
$10 \otimes 8 \otimes 10$  for octet and decuplet hadrons
respectively, $10 \otimes 8\otimes 8$ for transitions between 
octet and decuplet baryons, and $3 \otimes 8 \otimes 3$ 
for quark matrix elements, useful for considering 
renormalisation and improvement of quark bilinear operators. 
We shall only consider the octet (and singlet) baryon cases here.


\subsection{Simple cases I: Decay constants $f_\pi$ and $f_K$}


The vacuum is a singlet, so vacuum to meson, $M$, matrix elements
or decay constants $\langle 0 | J^{F_j} | M_k \rangle$, $j = 1, \ldots ,8$
are proportional to $1 \otimes 8 \otimes 8$ tensors, i.e.\ 
$8 \otimes 8$ matrices. So again the allowed mass dependence of
$f_\pi$ and $f_K$ is similar to the allowed dependence of
$M_\pi^2$ and $M_K^2$, as given in \cite{bietenholz11a}.
Results using this approach are given in \cite{Bornyakov:2016dzn}.
For example to LO we have
\begin{eqnarray}
   f_\pi &=& F_0 + 2G\delta m_l \,,
                                                          \nonumber \\
   f_K  &=& F_0 - G\delta m_l \,.
\end{eqnarray}
The same argument applies in principle to hyperon distribution
amplitudes $qqq$, and to baryon decays via $qqqe$
$4$-fermi grand unified theory (GUT) interactions, but in this work 
we shall only consider bilinear operators.


\section{Method for matrix elements}
\label{method_ME}


Recall from eq.~(\ref{A_def}) that we have used the notation for the
matrix element transition $B \to B^\prime$ of
\begin{eqnarray}
   A_{\bar{B}^\prime F B} = \langle B^\prime | J^F | B \rangle \,,
\end{eqnarray}
where $J^F$ is the appropriate operator from Table~\ref{ind8} and
$F$ denotes the flavour structure of the operator. But note that
as we are suppressing the Lorentz structure, this includes
first and second class form factors as given in eq.~(\ref{1+2class}).


\subsection{Sign conventions: Octet operators and octet hadrons}


In the case of a $n_f = 2+1$ simulation we only need to give the
amplitudes for one particle in each isospin multiplet,
and can then use isospin symmetry to calculate all other
amplitudes in (or between) the same multiplets. So, for example,
we can calculate the $\Sigma^-$ and $\Sigma^0$ matrix elements
if we are given all the $\Sigma^+$ matrix elements. Similarly,
given the $\Sigma^+ \to p$ transition amplitude, we can find
all the other $\Sigma \to N$ transition amplitudes.
All the symmetry factors will be listed in section~\ref{su2rel}.

In the next section we will calculate the allowed quark-mass
dependencies of the amplitudes between the baryons.
Within this set there are $7$ diagonal matrix elements, and $5$
transition amplitudes making $7+5 = 12$ in total. 
The $7$ diagonal elements are
\begin{eqnarray}
   A_{\bar{N}\eta N} \,\,,\, A_{\bar{\Sigma}\eta \Sigma} \,\,,\,
   A_{\bar{\Lambda}\eta \Lambda} \,\,,\, A_{\bar{\Xi}\eta \Xi} \,\,\,
   \mbox{and} \,\,\,
   A_{\bar{N}\pi N} \,,\, A_{\bar{\Sigma}\pi \Sigma} \,,\, A_{\bar{\Xi}\pi \Xi} \,,
\end{eqnarray}  
because there are four $I=0$ amplitudes,
one for each particle, but only three $I=1$ amplitudes, because isospin
symmetry rules out an $I=1$, $\Lambda^0 \leftrightarrow \Lambda^0$ amplitude.
There are only $5$ transition amplitudes
\begin{eqnarray}
   A_{\bar{\Sigma}\pi \Lambda} \,\,\,
  \mbox{and} \,\,\, A_{\bar{N}K \Sigma} \,\,,\,
   A_{\bar{N}K \Lambda} \,\,,\, A_{\bar{\Lambda}K \Xi} \,\,,\,
   A_{\bar{\Sigma}K \Xi} \,,
\end{eqnarray}
because no octet operator changes strangeness by $\pm 2$, so there is
no $p \leftrightarrow \Xi^0$ transition amplitude. See the forthcoming
Tables~\ref{su2relations_diag} and \ref{su2relations_trans} for the
explicit results.

To discuss transition matrix elements, we need to specify 
the hadron states carefully. If we do not, then the phases and signs
of transition matrix elements become ambiguous. (This is not 
a problem with masses, or diagonal matrix elements such as 
$\langle p | J | p \rangle $.)

We follow a convention commonly used in chiral perturbation theory%
\footnote{However some papers use different definitions, 
e.g.\ in chapter 18 of \cite{gasiorowicz66} the meson matrix
$M$ is defined the same way as in eq.~(\ref{mesdef}),
but in the baryon matrix $B$ the $\Xi^-$ appears with a minus sign
in comparison to eq.~(\ref{barydef}). Using the Gasiorowicz convention,
\cite{gasiorowicz66}, would give the opposite sign to all
transition matrix elements  to or from the $\Xi^-$.},
e.g.\ \cite{savage96a,walkerloud04a} where the mesons transform
under $SU(3)$ rotations like the $3\times 3$ matrix 
\begin{equation} 
   M = \pmatrix{ \frac{1}{\sqrt{2}} \pi^0 + \frac{1}{\sqrt{6}} \eta & 
                 \pi^+ & K^+                                             \cr 
                 \pi^- &
                 -\, \frac{1}{\sqrt{2}} \pi^0 + \frac{1}{\sqrt{6}} \eta &
                 K^0                                                    \cr 
                 K^- & \bar{K}^0 & -\, \frac{2}{\sqrt{6}} \eta } \,,
\label{mesdef} 
\end{equation} 
and octet baryons like the matrix 
\begin{eqnarray} 
   B &=&  \pmatrix{ \frac{1}{\sqrt{2}} \Sigma^0 
                         + \frac{1}{\sqrt{6}} \Lambda^0 &
                     \Sigma^+ & p                                       \cr
                     \Sigma^- & 
                      -\, \frac{1}{\sqrt{2}} \Sigma^0 
                         + \frac{1}{\sqrt{6}} \Lambda^0 & n             \cr 
                     \Xi^- & \Xi^0 & -\, \frac{2}{\sqrt{6}} \Lambda^0 } \,,
                                                                \nonumber \\
      & &                                                \label{barydef}  \\
    \bar{B} 
      &=&  \pmatrix{ \frac{1}{\sqrt{2}} \bar{\Sigma}^0
                         + \frac{1}{\sqrt{6}} \bar{\Lambda}^0 &
                     \bar{\Sigma}^- & \bar{\Xi}^-             \cr
                     \bar{\Sigma}^+ & 
                     -\, \frac{1}{\sqrt{2}} \bar{\Sigma}^0 
                         + \frac{1}{\sqrt{6}} \bar{\Lambda}^0 & 
                     \bar{\Xi}^0                                   \cr 
                     \bar{p} & \bar{n} & 
                     -\, \frac{2}{\sqrt{6}} \bar{\Lambda}^0 } \,. 
                                                                 \nonumber 
\end{eqnarray}
So for example $\pi^+$, $\pi^0$, $\pi^-$ are represented by the matrices
\begin{eqnarray}
   \pmatrix{ 0 & 1 & 0   \cr 
             0 & 0 & 0   \cr
             0 & 0 & 0}   \,, \qquad
   \pmatrix{ \frac{1}{\sqrt{2}} & 0 & 0   \cr 
             0 & -\frac{1}{\sqrt{2}} & 0   \cr
             0 & 0 & 0}   \,, \qquad
   \pmatrix{ 0 & 0 & 0   \cr 
             1 & 0 & 0   \cr
             0 & 0 & 0}   \,,
\label{pip_matrix}   
\end{eqnarray}
respectively. Under an $SU(3)$ rotation the $M$, $B$ and $\bar{B}$
matrices transform as
\begin{equation} 
   M \to U M U^\dagger, \quad  B \to U B U^\dagger,
      \quad {\rm and } \quad \bar{B} \to U \bar{B} U^\dagger \,.
\label{BMtrans} 
\end{equation} 


\subsection{$SU(2)$ relations}
\label{su2rel}


As discussed previously we use the convention that operator number
$i$, representing an appropriate flavour matrix, has the same effect
on quantum numbers as the absorption of a meson with the index $i$.
So for example, from Table~\ref{ind8} operator $6$ annihilates a $d$ 
quark and creates a $u$, and hence changes a neutron into a proton, i.e.\
\begin{eqnarray}
   \langle p | \bar{u} \gamma d |n \rangle
      \equiv \langle p | J^{\pi^+} | n \rangle
      \equiv \langle B_2 | J^{F_6} | B_1 \rangle \,.
\label{B2J6B1}      
\end{eqnarray}      
In Tables~\ref{su2relations_diag}
\begin{table}[!htb]
   \begin{minipage}{0.40\textwidth}
 \begin{tabular}{ccr}
 $I$ &   &  \\
 \hline
 $0$ & $\langle n | J^\eta | n \rangle$ & \tophat $A_{\bar{N}\eta N}$ \\
 $0$ & $\langle p | J^\eta | p \rangle$ & $A_{\bar{N}\eta N}$ \\
 \hline
 $0$ & $\langle \Sigma^- | J^\eta | \Sigma^- \rangle$ & \tophat $A_{\bar{\Sigma}\eta \Sigma}$ \\
 $0$ & $\langle \Sigma^0 | J^\eta | \Sigma^0 \rangle$ & $A_{\bar{\Sigma}\eta \Sigma}$ \\
 $0$ & $\langle \Sigma^+ | J^\eta | \Sigma^+ \rangle$ & $A_{\bar{\Sigma}\eta \Sigma}$ \\
 \hline
 $0$ & $\langle \Lambda^0 | J^\eta | \Lambda^0 \rangle$ & \tophat $A_{\bar{\Lambda}\eta \Lambda}$ \\
 \hline
 $0$ & $\langle \Xi^- | J^\eta | \Xi^- \rangle$ & \tophat $A_{\bar{\Xi}\eta \Xi}$ \\
 $0$ & $\langle \Xi^0 | J^\eta | \Xi^0 \rangle$ & $A_{\bar{\Xi}\eta \Xi}$ \\
\end{tabular}
   \end{minipage} \hspace*{0.10\textwidth}
   \begin{minipage}{0.40\textwidth}
 \begin{tabular}{ccr}
 $I$ &   &  \\
 \hline
 $1$ & $\langle n |J^{\pi^0} | n \rangle$ & \tophat $- A_{\bar{N}\pi N}$ \\
 $1$ & $\langle p |J^{\pi^0} | p \rangle$ & $  A_{\bar{N}\pi N}$ \\
 $1$ & $\langle n |J^{\pi^-} | p \rangle$ & $\sqrt{2} A_{\bar{N}\pi N}$ \\
 $1$ & $\langle p |J^{\pi^+} | n \rangle$ & $\sqrt{2}  A_{\bar{N}\pi N}$ \\
 \hline
 $1$ & $\langle \Sigma^- |J^{\pi^0} | \Sigma^- \rangle$ & \tophat $-A_{\bar{\Sigma}\pi \Sigma}$ \\
 $1$ & $\langle \Sigma^0 |J^{\pi^0} | \Sigma^0 \rangle$ & $0 \quad$  \\
 $1$ & $\langle \Sigma^+ |J^{\pi^0} | \Sigma^+ \rangle$ & $A_{\bar{\Sigma}\pi \Sigma}$ \\
 $1$ & $\langle \Sigma^- |J^{\pi^-} | \Sigma^0 \rangle$ & $A_{\bar{\Sigma}\pi \Sigma}$ \\
 $1$ & $\langle \Sigma^0 |J^{\pi^-} | \Sigma^+ \rangle$ & $-A_{\bar{\Sigma}\pi \Sigma}$ \\
 $1$ & $\langle \Sigma^0 |J^{\pi^+} | \Sigma^- \rangle$ & $A_{\bar{\Sigma}\pi \Sigma}$ \\
 $1$ & $\langle \Sigma^+ |J^{\pi^+} | \Sigma^0 \rangle$ & $-A_{\bar{\Sigma}\pi \Sigma}$ \\
 \hline
 $1$ & $\langle \Lambda^0 |J^{\pi^0} | \Lambda^0 \rangle$ & \tophat $0 \quad$ \\
 \hline
 $1$ & $\langle \Xi^- |J^{\pi^0} | \Xi^- \rangle$ & \tophat $-A_{\bar{\Xi}\pi \Xi}$ \\
 $1$ & $\langle \Xi^0 |J^{\pi^0} | \Xi^0 \rangle$ & $A_{\bar{\Xi}\pi \Xi}$ \\
 $1$ & $\langle \Xi^- |J^{\pi^-} | \Xi^0 \rangle$ & $-\sqrt{2} A_{\bar{\Xi}\pi \Xi}$ \\
 $1$ & $\langle \Xi^0 |J^{\pi^+} | \Xi^- \rangle$ & $-\sqrt{2} A_{\bar{\Xi}\pi \Xi}$ \\
 \end{tabular} 
   \end{minipage}
\caption{The isospin relations connecting the set of octet
         matrix elements with our standard subsets $A_{\bar B FB}$
         (each independent set separated by horizontal lines).
          Left table: The $I=0$ diagonal relations;
          right table: the $I=1$ transition relations within the same
          isospin multiplet.}
\label{su2relations_diag}
\end{table}
and \ref{su2relations_trans}
\begin{table}[!htb]
   \begin{minipage}{0.40\textwidth}
    \begin{tabular}{ccrl}
 $I$ &   &  \\
 \hline
 $1$ & $\langle \Sigma^- |J^{\pi^-} | \Lambda^0 \rangle$ & \tophat
 $A_{\bar{\Sigma}\pi \Lambda}$ &\\
 $1$ & $\langle \Sigma^0 |J^{\pi^0} | \Lambda^0 \rangle$ & $A_{\bar{\Sigma}\pi \Lambda}$ &\\
 $1$ & $\langle \Sigma^+ |J^{\pi^+} | \Lambda^0 \rangle$ & \tophat $A_{\bar{\Sigma}\pi \Lambda}$ &\\
 \hline \hline
 $\frac{1}{2}$ & $\langle n | J^{K^+} | \Sigma^- \rangle$ & \tophat $A_{\bar{N}K \Sigma}$ &\\
 $\frac{1}{2}$ & $\langle n | J^{K^0} | \Sigma^0 \rangle$ & $-A_{\bar{N}K \Sigma}$
 &\hspace*{-5mm} $/\sqrt{2}$\\
 $\frac{1}{2}$ & $\langle p | J^{K^+} | \Sigma^0 \rangle$ & $A_{\bar{N}K \Sigma}$
 &\hspace*{-5mm} $/\sqrt{2}$\\
 $\frac{1}{2}$ & $\langle p | J^{K^0} | \Sigma^+ \rangle$ & $A_{\bar{N}K \Sigma}$& \\
 \hline
 $\frac{1}{2}$ & $\langle n | J^{K^0} | \Lambda^0 \rangle$ & \tophat $A_{\bar{N}K \Lambda}$& \\
 $\frac{1}{2}$ & $\langle p | J^{K^+} | \Lambda^0 \rangle$ & \tophat $A_{\bar{N}K \Lambda}$& \\
 \hline
 $\frac{1}{2}$ & $\langle \Lambda^0 | J^{K^+} | \Xi^- \rangle$ & \tophat $A_{\bar{\Lambda}K \Xi}$& \\
 $\frac{1}{2}$ & $\langle \Lambda^0 | J^{K^0} | \Xi^0 \rangle$ & $A_{\bar{\Lambda}K \Xi}$& \\
 \hline
 $\frac{1}{2}$ & $\langle \Sigma^- |J^{K^0} |\Xi^- \rangle$ & \tophat $A_{\bar{\Sigma}K \Xi}$& \\
 $\frac{1}{2}$ & $\langle \Sigma^0 |J^{K^+} |\Xi^- \rangle$ & $A_{\bar{\Sigma}K \Xi}$
 &\hspace*{-5mm} $/\sqrt{2}$\\
 $\frac{1}{2}$ & $\langle \Sigma^0 |J^{K^0} |\Xi^0 \rangle$ & $-A_{\bar{\Sigma}K \Xi}$
 &\hspace*{-5mm} $/\sqrt{2}$\\
 $\frac{1}{2}$ & $\langle \Sigma^+ |J^{K^+} |\Xi^0 \rangle$ & $A_{\bar{\Sigma}K \Xi}$&
\end{tabular}
   \end{minipage} \hspace*{0.10\textwidth}
   \begin{minipage}{0.40\textwidth}
 \begin{tabular}{ccrl}
 $I$ &   &  \\
 \hline
 $1$ & $\langle \Lambda^0 |J^{\pi^+} | \Sigma^- \rangle$ & \tophat $A_{\bar{\Lambda}\pi \Sigma}$ &\\
 $1$ & $\langle \Lambda^0 |J^{\pi^0} | \Sigma^0 \rangle$ & $A_{\bar{\Lambda}\pi \Sigma}$ &\\
 $1$ & $\langle \Lambda^0 |J^{\pi^-} | \Sigma^+ \rangle$ & $A_{\bar{\Lambda}\pi \Sigma}$ &\\
 \hline \hline
 $\frac{1}{2}$\tophat & $\langle \Sigma^- | J^{K^-} | n \rangle$ & $A_{\bar{\Sigma}\bar{K} N}$ &\\
 $\frac{1}{2}$ & $\langle \Sigma^0 | J^{\bar{K}^0} | n \rangle$ & $-A_{\bar{\Sigma}\bar{K} N}$ &\hspace*{-5mm} $/\sqrt{2}$\\
 $\frac{1}{2}$ & $\langle \Sigma^0 | J^{K^-} | p \rangle$ & $A_{\bar{\Sigma}\bar{K} N}$ &\hspace*{-5mm} $/\sqrt{2}$\\
 $\frac{1}{2}$ & $\langle \Sigma^+ | J^{\bar{K}^0} | p \rangle$ & $A_{\bar{\Sigma}\bar{K} N}$& \\
 \hline
 $\frac{1}{2}$ & $\langle \Lambda^0 | J^{\bar{K}^0} | n \rangle$ & \tophat $A_{\bar{\Lambda}\bar{K} N}$& \\
 $\frac{1}{2}$ & $\langle  \Lambda^0 | J^{K^-} | p \rangle$ & $A_{\bar{\Lambda}\bar{K} N}$& \\
 \hline
 $\frac{1}{2}$ & $\langle \Xi^- | J^{K^-} | \Lambda^0 \rangle$ & \tophat $A_{\bar{\Xi}\bar{K} \Lambda}$& \\
 $\frac{1}{2}$ & $\langle \Xi^0 | J^{\bar{K}^0} | \Lambda^0 \rangle$ & $A_{\bar{\Xi}\bar{K} \Lambda}$& \\
 \hline
 $\frac{1}{2}$ & $\langle \Xi^- | J^{\bar{K}^0} | \Sigma^- \rangle$ & \tophat $A_{\bar{\Xi}\bar{K} \Sigma}$& \\
 $\frac{1}{2}$ & $\langle \Xi^- | J^{K^-} | \Sigma^0 \rangle$ & $A_{\bar{\Xi}\bar{K} \Sigma}$ &\hspace*{-5mm} $/\sqrt{2}$\\
 $\frac{1}{2}$ & $\langle \Xi^0 | J^{\bar{K}^0} | \Sigma^0 \rangle$ & $-A_{\bar{\Xi}\bar{K} \Sigma}$ &\hspace*{-5mm} $/\sqrt{2}$\\
 $\frac{1}{2}$ & $\langle \Xi^0 | J^{K^-} | \Sigma^+ \rangle$ & $A_{\bar{\Xi}\bar{K} \Sigma}$&
 \end{tabular}
   \end{minipage}
\caption{The isospin relations connecting the transition set of octet
         matrix elements with our standard subsets $A_{\bar B^\prime FB}$
         (each independent set separated by horizontal lines).
          Left table: The `forward' $I = 1$ and $\frac{1}{2}$ relations;
          right table: the inverse relations.}
 \label{su2relations_trans}
\end{table}
we list the isospin relationships between all of the allowed matrix
elements in the octet, and our standard $7 + 5 = 12$ matrix elements.

Making the choice given in eqs.~(\ref{mesdef}, \ref{barydef})
which is conventional in chiral perturbation theory, the isospin
raising and lowering operators do not follow the usual Condon--Shortley 
sign convention. The Wigner--Eckart theorem applies, but the signs
are not always the ones from the standard Clebsch--Gordan coefficients.

To demonstrate this, consider the transformations given in eq.~(\ref{BMtrans})
with $U = \exp( i \alpha_i \lambda^i )$. Infinitesimal transformations
($\alpha_i \to 0$) correspond to commutators of the type $[\lambda^i,B]$ or 
$[\lambda^i,M]$. The isospin operations are constructed from the first
three $\lambda$ matrices
\begin{eqnarray}
   I_3 &=& {\textstyle{1 \over 2}} \lambda^3 \,,              \nonumber \\ 
   I_+ &=& {\textstyle{1 \over 2}} (\lambda^1 + i \lambda^2) \,,        \\
   I_- &=& {\textstyle{1 \over 2}} (\lambda^1 - i \lambda^2) \,.\nonumber
\end{eqnarray} 
$I_3$ has the expected result
\begin{eqnarray} 
   \hat{I}_3 M = {\textstyle{1 \over 2}} [\lambda^3, M]
        &=& \pmatrix{ 0     & \pi^+ & {\textstyle{1 \over 2}} K^+ \cr
                     -\pi^- & 0     & -\,{\textstyle{1 \over 2}} K^0 \cr
                     -{\textstyle{1 \over 2}} K^- & {\textstyle{1 \over 2}}
                                                \bar{K}^0 & 0 } \,, \\[0.7em] 
   \hat{I}_3 B = {\textstyle{1 \over 2}} [\lambda^3, B]
        &=& \pmatrix{ 0 & \Sigma^+ & {\textstyle{1 \over 2}} p \cr 
                      - \Sigma^- & 0 & -\, {\textstyle{1 \over 2}} n \cr
                      -\, {\textstyle{1 \over 2}}\, \Xi^-
                                 & {\textstyle{1 \over 2}}\, \Xi^0 & 0 } \,.
\end{eqnarray}
For example regarding $\pi^-$ as the matrix in eq.~(\ref{pip_matrix}) gives
\begin{eqnarray}
  \hat{I}_3\pi^- =\pmatrix{ 0  & 0 & 0 \cr 
                            -1 & 0 & 0 \cr 
                            0  & 0 & 0 }
                = - \pi^- \,,
\end{eqnarray}
(see Fig.~\ref{meson_baryon_octet}). Similarly for the baryons, for example
$\hat{I}_3 n = -{\textstyle{1 \over 2}} n$, etc...\,.

However $\hat{I}_+$ and $\hat{I}_-$ produce results at odds with the
Condon-Shortley or CS phase convention, which has positive coefficients
for the non-zero matrix elements of the raising and lowering 
operators. 
\begin{equation} 
   \hat{I}_+ M = {\textstyle{1 \over 2}} [ \lambda^1 + i \lambda^2 , M ]
         = \pmatrix{ \pi^- & -\sqrt{2} \pi^0 & K^0 \cr 
                      0    & -\pi^- & 0 \cr 
                      0    & -K^- & 0 } \,.
\end{equation} 
Again using the $\pi^-$ as an example and comparing this result with
eq.~(\ref{mesdef}) we see that we have
\begin{eqnarray}
   \hat{I}_+ \pi^-
         = \pmatrix{  1    & 0  & 0 \cr 
                      0    & -1 & 0 \cr 
                      0    & 0  & 0 } 
         = \sqrt{2}\pi^0 \,.
\end{eqnarray}         
Listing all the relations gives
\begin{eqnarray} 
   \hat{I}_+ \pi^- &=& \sqrt{2} \pi^0 \,, \nonumber \\
   \hat{I}_+ \pi^0 &=& - \sqrt{2} \pi^+ \,, \label{Iplus_meson} \\
   \hat{I}_+ K^0 &=&  K^+ \nonumber \,, \\
   \hat{I}_+ K^- &=& - \bar{K}^0 \,. \nonumber
\end{eqnarray} 
Similarly
\begin{eqnarray} 
   \hat{I}_+ \Sigma^- &=&   \sqrt{2} \Sigma^0  \,, \nonumber \\
   \hat{I}_+ \Sigma^0 &=& - \sqrt{2} \Sigma^+ \,, \label{Iplus_baryon} \\
   \hat{I}_+ n        &=& p \nonumber \,, \\
   \hat{I}_+ \Xi^-    &=& - \Xi^0 \,. \nonumber
\end{eqnarray} 
The action of $\hat{I}_-$ is similar.
Since these relations are not those usually used to calculate 
the Clebsch-Gordon coefficients, we need to tabulate the 
isospin relations within each multiplet. 
The signs of the $\hat{I}_+$ matrix elements follow directly from the 
choice of signs in the chiral perturbation theory representation
of the meson and baryon octets as $3\times 3$ matrices
in eqs.~(\ref{mesdef}, \ref{barydef}). The guiding principle is to 
make the off-diagonal entries there positive. However this tidy choice 
of matrix leads to a non-standard phase convention within isospin multiplets.

In the CS convention all the coefficients in
eqs.~(\ref{Iplus_meson}, \ref{Iplus_baryon}) would be positive. 
Looking at the baryon results, eq.~(\ref{Iplus_baryon}), we see that
the neutron and proton are consistent with that convention, 
while, for example, the $\Xi^-$ and $\Xi^0$ are not. The minus sign tells us
that one of the $\Xi$ states must have the opposite phase to the CS convention.
Since only relative phases are observable, we could choose the $\Xi^0$ 
to have the CS phase, and the $\Xi^-$ to have the flipped phase. 
(Making the other choice would not change the final result.) Similarly
looking at the $\Sigma$ baryons we could choose the $\Sigma^+$
to have the CS phase, and the $\Sigma^-$ and $\Sigma^0$ to have
flipped phase (or vice versa).

One choice of phases that would match 
eqs.~(\ref{Iplus_meson}, \ref{Iplus_baryon}) would be to choose the 
$n$, $p$, $\Sigma^+$ and $\Xi^0$ as standard, and the $ \Sigma^-$, 
$\Sigma^0$ and $\Xi^-$ as flipped, and the equivalent choice for the 
meson currents (i.e.\ $\pi^-$, $\pi^0$, $K^-$ flipped). If we look in
Tables~\ref{su2relations_diag} and \ref{su2relations_trans} we see 
that matrix elements involving an even number of hadrons from the 
flipped group, the Clebsch-Gordon factor is the same as that in the 
usual tables, if an odd number of flipped hadrons are involved, 
the sign is the opposite to that in the usual tables.

As an example of the use of Table~\ref{su2relations_diag}, we show how the 
unbroken $SU(2)$ symmetry can be used to find the transition
amplitude  $\langle p |J^{\pi^+} | n \rangle$ from the corresponding 
diagonal amplitude $\langle p |J^{\pi^0} | p \rangle$. 
From the table
\begin{eqnarray} 
   \langle p |J^{\pi^+} | n \rangle 
      = \sqrt{2}\, A_{\bar{N}\pi N}
      =  \sqrt{2} \; \langle p | J^{\pi^0} | p \rangle \,,
\end{eqnarray}
giving
\begin{eqnarray}
   \langle p | \bar u \gamma d | n \rangle 
    =  \langle p | ( \bar u \gamma u - \bar d \gamma d )| p \rangle \,,
\label{beta_decay}   
\end{eqnarray} 
which is again the simple example showing the relation between
off-diagonal and diagonal currents briefly discussed in
section~\ref{gen_currents}.


\section{Mass dependence of amplitudes} 
\label{mass_amp}


We first consider the simple singlet case (operators with the
$\eta^\prime$ flavour structure, $i = 0$, see Table~\ref{ind8})
and then consider the octet states.


\subsection{Simple cases II: Flavour-singlet operators}
\label{singop}


For matrix elements involving singlet currents,
$\langle B_i | J^{F_0} | B_i\rangle \equiv
\langle B_i | J^{\eta^\prime} | B_i\rangle $,  
we need the $SU(3)$ analysis of $8 \otimes 1 \otimes 8$ tensors. 
These are just the $8 \otimes 8$ matrices already analysed in 
\cite{bietenholz11a}.
The conclusion is thus that matrix elements of flavour singlet
operators follow the same  formulae as the hadron masses.
An example of a flavour singlet operator is the quark component
to the baryon spin, $\Delta\Sigma$.
For example the LO expansion is given by
\begin{eqnarray}
   A_{\bar{N}\eta^\prime N} &=& a_0 +3a_1\delta m_l \,,
                                                          \nonumber \\
   A_{\bar{\Lambda}\eta^\prime \Lambda}
               &=& a_0 +3a_2\delta m_l \,,
                                         \label{singlet_expansions} \\
   A_{\bar{\Sigma}\eta^\prime \Sigma}
               &=& a_0 -3a_2\delta m_l \,,
                                                          \nonumber \\
   A_{\bar{\Xi}\eta^\prime \Xi}
               &=& a_0 -3(a_1-a_2)\delta m_l \,,
                                                          \nonumber
\end{eqnarray}
with higher orders given in \cite{bietenholz11a}.


\subsection{Group theory classification: Flavour-octet operators}

\begin{figure}[!htb]
   \hspace*{0.75in}
   \begin{tabular}{ccc}
      \includegraphics[width=4.75cm,angle=270]{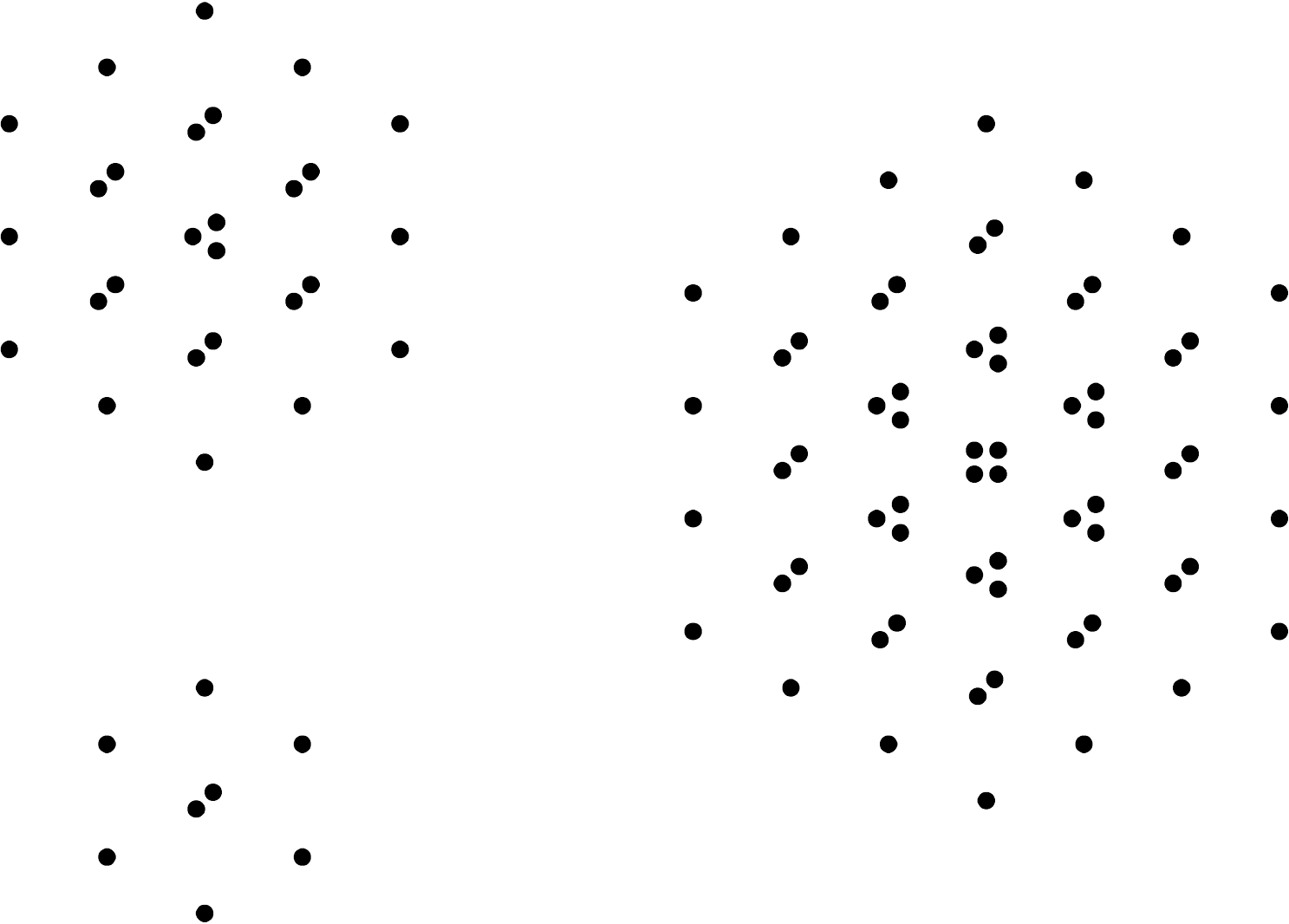}
         &  \hspace*{0.50in}  &
      \includegraphics[width=4.75cm,trim=0.0in 6.0in 0.0in 0.00in]{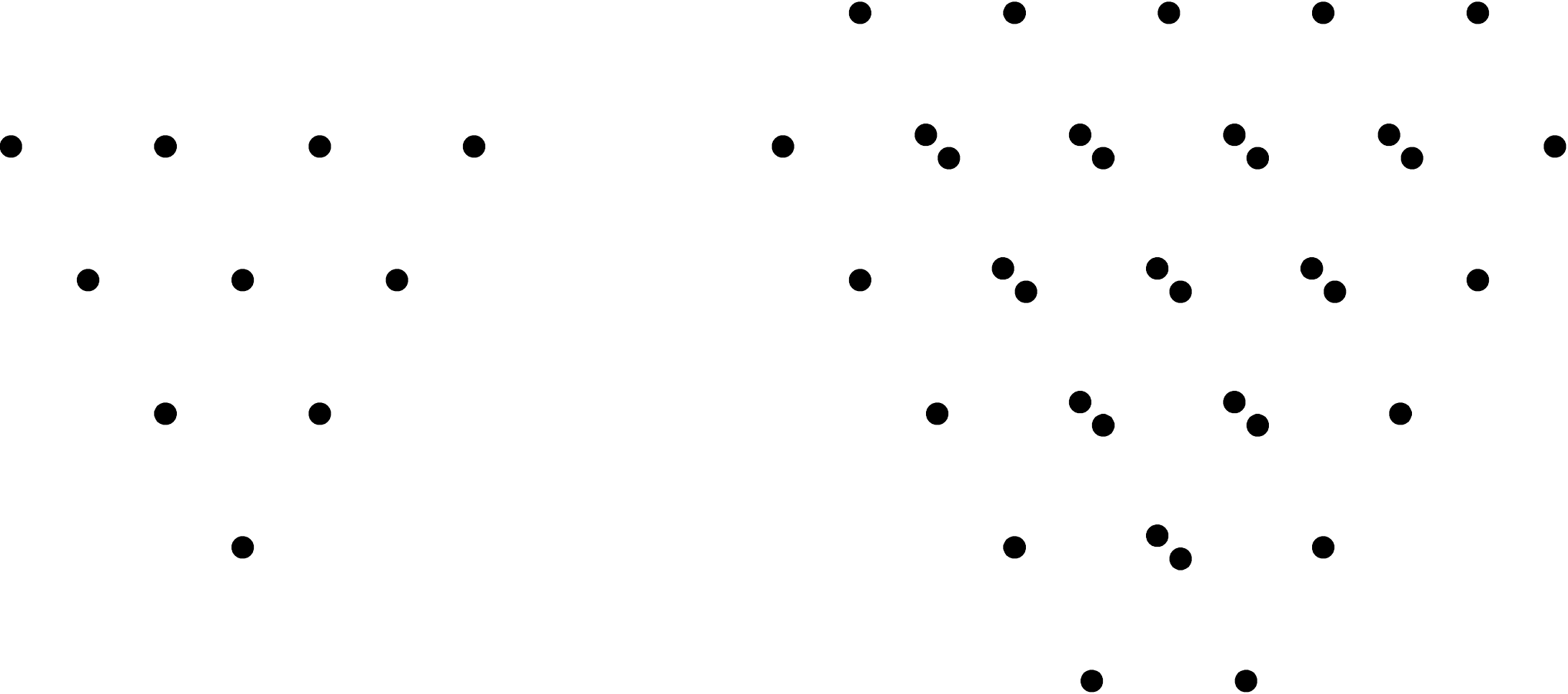}
   \end{tabular}
   \vspace*{0.25in}
   \caption{$I_3$, $Y$ plots for some of the $SU(3)$ multiplets
            which appear in the decomposition of $8 \otimes 8 \otimes 8$.
            The left-hand plot illustrates the octet, $27$-plet and $64$-plet
            representations (clockwise). The right-hand plot shows the 
            $10$ and $35$-plets (left to right). The number of spots in
            the central location gives the number of
            flavour-conserving operators in each multiplet.}
\label{spots}
\end{figure}

To find the allowed mass dependence of octet matrix elements of octet
hadrons we need the $SU(3)$ decomposition of $8 \otimes 8 \otimes 8$. 
Using the intermediate result 
\begin{equation} 
   8 \otimes 8 = 1 \oplus 8 \oplus 8 \oplus 10 \oplus  {\overline {10}} 
                   \oplus 27 \,,
\label{decomp88} 
\end{equation} 
we find
\begin{eqnarray}
   \lefteqn{8 \otimes 8 \otimes 8}
      & &                                       \label{decomp888} \\
      &=& 1 \oplus 1 \oplus 8 \oplus 8 \oplus 8 \oplus 8 \oplus 8 
                                              \oplus 8 \oplus 8 \oplus 8 
            \oplus 27  \oplus 27  \oplus 27  \oplus 27  \oplus 27  \oplus 27
                                              \oplus 64
                                                        \nonumber \\ 
      & & \phantom{1} \oplus 10 \oplus 10 \oplus 10 \oplus 10    
          \oplus {\overline {10}}  \oplus {\overline {10}} 
          \oplus {\overline {10}}  \oplus {\overline {10}}   
          \oplus 35  \oplus 35 \oplus{\overline{35}} \oplus{\overline{35}} \,.
                                                        \nonumber
\end{eqnarray} 
With three unequal quark masses, the $n_f = 1+1+1$ case, 
$I_3$ and $Y$ are both `good' flavour quantum numbers, 
so the tensors in eq.~(\ref{schema}) will satisfy $I_3 = 0$, $Y=0$, 
i.e.\ they will be the central locations (spots) in each multiplet
in Fig.~\ref{spots}.
Thus in a full $n_f = 1+1+1$ flavour calculation (three different quark 
masses) we would see contributions from all the representations in
eq.~(\ref{decomp888}).  

Fortunately in the $n_f = 2+1$ case the good flavour quantum
numbers are $I$ and $Y$, giving us the stronger constraint that 
only tensors with $I=0$, $Y=0$ enter into eq.~(\ref{schema}). 
The $10$, $\overline {10}$, $35$ and $\overline{35}$ do not contain 
any $I=0$, $Y=0$ operators, so they no longer contribute in the $2+1$ case,
which means that we can neglect those representations at present
\cite{gasiorowicz66,pfeifer03a}.
For example for the $Y = 0$ line for the octet, we have an
isospin triplet and singlet of states and similarly for the $27$-plet
(isospin $5$-plet, triplet and singlet) and $64$-plet (isospin $7$-plet,
$5$-plet, triplet and singlet). However for the $10$-plet we have just
an isospin triplet and for the $35$-plet a $5$-plet and triplet.
In both cases there is no $Y=0$ isospin singlet.

We have already seen this phenomenon in \cite{bietenholz11a} for 
the case of the $10$ and $\overline {10}$. The simplest quark-mass
polynomial with $10$, $\overline {10}$ symmetry was 
$(\delta m_s - \delta m_u)(\delta m_s - \delta m_d)(\delta m_u-\delta m_d)$
(see Table~\ref{quadratic}), which vanishes if any two quark masses are equal.
The $10$ and $\overline {10}$ only appeared in two quantities we 
have considered, the violation of the Coleman-Glashow mass relation,
and in $\Sigma^0$ -- $\Lambda^0$ mixing \cite{horsley14a}, both of
which are isospin violating.


\subsection{The $SU(3)$ symmetry-breaking expansions} 
\label{SU3_theory}


\subsubsection{Basis} 


Because $8\times 8 \times 8$ tensors are easier to think about
than $3 \times 3 \times 3 \times 3 \times 3 \times 3 $ tensors
we switch to regarding baryons and mesons as vectors of length~$8$.
We have used the ordering
\begin{equation} 
   \pmatrix{ n \cr p \cr \Sigma^- \cr \Sigma^0 \cr \Lambda^0 
          \cr \Sigma^+ \cr \Xi^- \cr \Xi^0 } \qquad {\rm and} \qquad
 \pmatrix{K^0 \cr K^+ \cr \pi^- \cr \pi^0 \cr \eta \cr \pi^+ 
 \cr K^- \cr \bar{K}^0  } \,.
\end{equation}  
The $8$ generators of $SU(3)$ are now a set of $8 \times 8$ matrices,
chosen so that $\lambda B$ in the matrix-vector notation has the
same effect as $[\lambda, B]$ in the $3 \times 3$ matrix-matrix notation. 
We have
\begin{small}
 \begin{eqnarray} 
 \lambda^1 &=& \pmatrix{ 
 0 & 1 & 0 & 0 & 0 & 0 & 0 & 0 \cr 
 1 & 0 & 0 & 0 & 0 & 0 & 0 & 0 \cr
 0 & 0 & 0 & \sqrt{2} & 0 & 0 & 0 & 0 \cr
 0 & 0 & \sqrt{2} & 0 & 0 & -\sqrt{2} & 0 & 0 \cr
 0 & 0 & 0 & 0 & 0 & 0 & 0 & 0 \cr
 0 & 0 & 0 & -\sqrt{2}  & 0 & 0 & 0 & 0 \cr
 0 & 0 & 0 & 0 & 0 & 0 & 0 & -1 \cr
 0 & 0 & 0 & 0 & 0 & 0 & -1 & 0 }  \,, \nonumber \\[0.8em]
 \lambda^2 &=& \pmatrix{ 
 0 & i & 0 & 0 & 0 & 0 & 0 & 0 \cr 
 -i & 0 & 0 & 0 & 0 & 0 & 0 & 0 \cr
 0 & 0 & 0 & i\sqrt{2} & 0 & 0 & 0 & 0 \cr
 0 & 0 & -i\sqrt{2} & 0 & 0 & -i\sqrt{2} & 0 & 0 \cr
 0 & 0 & 0 & 0 & 0 & 0 & 0 & 0 \cr
 0 & 0 & 0 & i\sqrt{2}  & 0 & 0 & 0 & 0 \cr
 0 & 0 & 0 & 0 & 0 & 0 & 0 & -i \cr
 0 & 0 & 0 & 0 & 0 & 0 & i & 0 }  \,, \nonumber \\[0.8em]
 \lambda^3 &=& \pmatrix{
 -1& 0 & 0 & 0 & 0 & 0 & 0 & 0 \cr
 0 & 1 & 0 & 0 & 0 & 0 & 0 & 0 \cr
 0 & 0 & -2& 0 & 0 & 0 & 0 & 0 \cr
 0 & 0 & 0 & 0 & 0 & 0 & 0 & 0 \cr
 0 & 0 & 0 & 0 & 0 & 0 & 0 & 0 \cr
 0 & 0 & 0 & 0 & 0 & 2 & 0 & 0 \cr
 0 & 0 & 0 & 0 & 0 & 0 & -1& 0 \cr
 0 & 0 & 0 & 0 & 0 & 0 & 0 & 1 }  \,, \nonumber \\[0.8em]
 \lambda^4 &=& \frac{1}{\sqrt{2}} \pmatrix{
 0 & 0 & -\sqrt{2} & 0 & 0 & 0 & 0 & 0 \cr
 0 & 0 & 0 & -1 & -\sqrt{3} & 0 & 0 & 0 \cr
 -\sqrt{2} & 0 & 0 & 0 & 0 & 0 & 0 & 0 \cr
 0 & -1 & 0 & 0 & 0 & 0 & 1 & 0 \cr
 0 & -\sqrt{3} & 0 & 0 & 0 & 0 & \sqrt{3} & 0 \cr
 0 & 0 & 0 & 0 & 0 & 0 & 0 & \sqrt{2} \cr
 0 & 0 & 0 & 1 & \sqrt{3} & 0 & 0 & 0 \cr
 0 & 0 & 0 & 0 & 0 & \sqrt{2} & 0 & 0 } \,, \nonumber \\[0.8em]
 \lambda^5 &=& \frac{1}{\sqrt{2}} \pmatrix{
 0 & 0 & i\sqrt{2} & 0 & 0 & 0 & 0 & 0 \cr
 0 & 0 & 0 & i & i\sqrt{3} & 0 & 0 & 0 \cr
 -i\sqrt{2} & 0 & 0 & 0 & 0 & 0 & 0 & 0 \cr
 0 & -i & 0 & 0 & 0 & 0 & -i& 0 \cr
 0 & -i\sqrt{3} & 0 & 0 & 0 & 0 & -i\sqrt{3} & 0 \cr
 0 & 0 & 0 & 0 & 0 & 0 & 0 & -i\sqrt{2} \cr
 0 & 0 & 0 & i & i\sqrt{3} & 0 & 0 & 0 \cr
 0 & 0 & 0 & 0 & 0 & i\sqrt{2} & 0 & 0 } \,, \nonumber \\[0.8em]
 \lambda^6 &=& \frac{1}{\sqrt{2}} \pmatrix{
 0 & 0 & 0 & 1 &-\sqrt{3} & 0 & 0 & 0 \cr
 0 & 0 & 0 & 0 & 0 & -\sqrt{2} & 0 & 0 \cr
 0 & 0 & 0 & 0 & 0 & 0 & \sqrt{2} & 0 \cr
 1 & 0 & 0 & 0 & 0 & 0 & 0 & -1 \cr
 -\sqrt{3} & 0 & 0 & 0 & 0 & 0 & 0 & \sqrt{3} \cr
 0 & -\sqrt{2} & 0 & 0 & 0 & 0 & 0 & 0 \cr
 0 & 0 & \sqrt{2} & 0 & 0 & 0 & 0 & 0 \cr
 0 & 0 & 0 & -1 & \sqrt{3} & 0 & 0 & 0 
  } \,, \nonumber \\[0.8em]
 \lambda^7 &=& \frac{1}{\sqrt{2}} \pmatrix{
 0 & 0 & 0 & -i &i\sqrt{3} & 0 & 0 & 0 \cr
 0 & 0 & 0 & 0 & 0 & i\sqrt{2} & 0 & 0 \cr
 0 & 0 & 0 & 0 & 0 & 0 & -i\sqrt{2} & 0 \cr
 i & 0 & 0 & 0 & 0 & 0 & 0 & i \cr
 -i\sqrt{3} & 0 & 0 & 0 & 0 & 0 & 0 & -i\sqrt{3} \cr
 0 & -i\sqrt{2} & 0 & 0 & 0 & 0 & 0 & 0 \cr
 0 & 0 & i\sqrt{2} & 0 & 0 & 0 & 0 & 0 \cr
 0 & 0 & 0 & -i & i\sqrt{3} & 0 & 0 & 0 
  } \nonumber \,, \\[0.8em]
 \lambda^8 &=& \sqrt{3} \pmatrix{
 1& 0 & 0 & 0 & 0 & 0 & 0 & 0 \cr
 0 & 1 & 0 & 0 & 0 & 0 & 0 & 0 \cr
 0 & 0 & 0 & 0 & 0 & 0 & 0 & 0 \cr
 0 & 0 & 0 & 0 & 0 & 0 & 0 & 0 \cr
 0 & 0 & 0 & 0 & 0 & 0 & 0 & 0 \cr
 0 & 0 & 0 & 0 & 0 & 0 & 0 & 0 \cr
 0 & 0 & 0 & 0 & 0 & 0 & -1 & 0 \cr
 0 & 0 & 0 & 0 & 0 & 0 & 0 & -1 }  \,. 
 \end{eqnarray}
\end{small} 
 These $8\times 8$ $\lambda$ matrices follow similar relations to 
 the familiar $3\times 3$ matrices, 
 \begin{eqnarray}
    \left[ \lambda^i , \lambda^j \right] = 2 i f^{ijk} \lambda^k
    \,, \qquad 
    {\rm Tr}( \lambda^i \lambda^j ) = 12\, \delta^{ij} \,,
\end{eqnarray}
and
\begin{eqnarray} 
   I_3 = {1 \over 2}\lambda^3 \,, \qquad Y = {1 \over \sqrt{3}}\lambda^8 \,,
 \end{eqnarray} 
 with the difference that the $3 \times 3$ matrices tell us about
 $I_3$ and $Y$ for the individual quarks, but the $8 \times 8$ 
 matrices give the quantum numbers of the octet baryons or octet 
 mesons.


\subsubsection{Transformations}


 Under an $SU(3)$ rotation the tensors on the right-hand side
 of eq.~(\ref{schema}) transform according to 
 \begin{equation} 
    T^\prime_{ijk} = U^\dagger_{ia} T_{abc} U_{bj} U_{ck} \,.
 \end{equation} 
 The change in $T$ under an infinitesimal transformation by
 the generator $\lambda^\alpha$ is
 \begin{equation} 
    \hat{O}^\alpha T \equiv 
      -\lambda^\alpha_{ia} T_{ajk} + T_{ibk} \lambda^\alpha_{bj} 
      + T_{ijc} \lambda^\alpha_{ck} \,.
 \end{equation} 
  The Casimir operator for the $SU(3)$ representation is 
 \begin{equation} 
    \hat{C} T
       = {\textstyle \frac{1}{4} }
              \sum_{\alpha=1,8} \hat{O}^\alpha \hat{O}^\alpha T \,,
 \end{equation}
 while the Casimir for the $SU(2)$ isospin subgroup is
 \begin{equation} 
    \hat{I}^2 T = {\textstyle \frac{1}{4} }
                       \sum_{\alpha=1,3} \hat{O}^\alpha \hat{O}^\alpha T \,.
 \end{equation}
 The $n_f = 2+1$ mass matrix commutes with $\lambda^1,\lambda^2,\lambda^3$
 (the generators of isospin)
 and $\lambda^8$ (hypercharge). We are looking for tensors 
 which obey these symmetries, so we require
 \begin{equation} 
   \hat{O}^\alpha T = 0\,, \quad \alpha=1, 2, 3, 8 \,. 
 \label{zeroeig}
 \end{equation} 
 The Casimir operator has the following eigenvalues for the 
 representations occurring in $8 \otimes 8 \otimes 8$,
 see for example chapter~4 of \cite{pfeifer03a} or chapter~7 
 (exercise 7.12) of \cite{greiner94}
 \begin{equation}
 \begin{tabular}{lcccccccc}
 representation & 1 & 8 & 10 &$\overline{10}$ & 27 & 35 
 &$\overline{35}$&64\cr 
 Casimir eigenvalue & 0 & 3 & 6 & 6 & 8 & 12  & 12 &  15\cr
 \end{tabular} \,.
 \end{equation}
 We now want to construct tensors which are eigenstates of the 
 Casimir operator, and which satisfy the conditions in eq.~(\ref{zeroeig}).
 This is analogous to constructing an eigenvector if we know the 
 eigenvalues. We have a large number of simultaneous linear 
 equations involving the numbers $T_{ijk}$. The solutions 
 tend to be sparse with the conditions in eq.~(\ref{zeroeig}) 
 forcing many entries to be zero.
 We calculate the tensors of a given symmetry with the help 
 of Mathematica, \cite{mathematica}. We begin with a completely general tensor 
 $T_{ijk}$ with $8^3$ entries, and impose the conditions eq.~(\ref{zeroeig}).
 This forces many entries to be zero, 
 as it eliminates all entries in which the flavour quantum 
 numbers of the `outgoing' particle $i$ is not the sum of
 the flavours of $j$ and $k$ (for example
 $\langle \Xi^0 | J^{\pi^+} | p \rangle = 0$ because charge and
 strangeness do not balance). 
 The conditions eq.~(\ref{zeroeig}) are also sufficient to force all 
 the relations in Tables~\ref{su2relations_diag} and \ref{su2relations_trans}
 to hold. 
 After imposing eq.~(\ref{zeroeig}) we have reduced the initial 
 general tensor with $8^3=512$ entries down to a tensor with 
 only 17 independent parameters. 
 From the decomposition of $8 \otimes 8 \otimes 8$ as given in
 eq.~(\ref{decomp888}) we can work out  how many solutions there
 are of each symmetry. 
 The representations $1$, $8$, $27$ and $64$ each have a single state 
 satisfying eq.~(\ref{zeroeig}), while the $10$, $\overline{10}$, $35$
 and $\overline{35}$ have no states compatible with eq.~(\ref{zeroeig}) 
 because they do not have a $Y=0, I^2=0$ central state,
 see Fig.~\ref{spots} and the related discussion.
 The $17$ linearly independent tensors remaining after 
 imposing eq.~(\ref{zeroeig}) can now be further classified as eigenstates 
 of the Casimir operator. Finding these tensors is a simple matter of solving 
 simultaneous equations, analogous to determining an eigenvector
 once the eigenvalue is known.

 As in the case of degenerate eigenvalues, there is a degree of
 choice in choosing which linear combinations of the eigenstates we
 choose as our basis. Often there are interchange operations which
 we can choose to be even or odd. In particular we can choose our
 tensors to be first class or second class depending on the symmetry
 or antisymmetry when the baryons are switched, as discussed in
 section~\ref{gen_currents}.

 We can see this by introducing a reflection matrix $R$ which 
 inverts each octet, leaving the central two states unchanged
 \begin{small}
 \begin{equation} 
 R = \pmatrix{  0 & 0 & 0 & 0 & 0 & 0 & 0 & 1 \cr
                0 & 0 & 0 & 0 & 0 & 0 & 1 & 0 \cr
                0 & 0 & 0 & 0 & 0 & 1 & 0 & 0 \cr
                0 & 0 & 0 & 1 & 0 & 0 & 0 & 0 \cr
                0 & 0 & 0 & 0 & 1 & 0 & 0 & 0 \cr
                0 & 0 & 1 & 0 & 0 & 0 & 0 & 0 \cr
                0 & 1 & 0 & 0 & 0 & 0 & 0 & 0 \cr
                1 & 0 & 0 & 0 & 0 & 0 & 0 & 0 } \,.  
 \end{equation} 
 \end{small}
 For the mesons this is the charge conjugation operation. 
 We note that $R^2 = I$ (the unit matrix), so $R$ can only have
 the eigenvalues $\pm 1$,  hence we can classify states according
 to whether they are even or odd under operations involving $R$. 
 Tensors can be divided into first or second class depending 
 on the symmetry 
 \begin{eqnarray} 
 {\rm first \ class \ \ \ }  & &  T_{ijk} = + T_{kai} R_{aj} \,,
                                                      \nonumber  \\  
 {\rm second \ class} & &  T_{ijk} = - T_{kai} R_{aj}\,,
 \label{FirstSecond}
 \end{eqnarray} 
 in which the baryon order is reversed, and $R$ applied to the current
 (meson) index. Furthermore the definition of first/second class tensors
 in eq.~(\ref{FirstSecond}) agrees with the previous discussion:
 in eqs.~(\ref{V_first}, \ref{V_second}) we interchanged $B$ and $B^\prime$
 and took the transpose of the flavour matrix, $F$. This latter operation
 is easily seen to be equivalent to the reflection, $R$ 
 in eq.~(\ref{FirstSecond}).

 We can further classify tensors by the symmetry when $R$ is applied
 to all three indices
 \begin{eqnarray} 
    d{\rm -like} & &  T_{ijk} = + R_{ia} T_{abc} R_{bj} R_{ck} \,,
                                                      \nonumber  \\ 
    f{\rm -like}  & &  T_{ijk} = -R_{ia} T_{abc} R_{bj} R_{ck} \,.
 \label{f_d_like}
 \end{eqnarray} 

As can be seen from eq.~(\ref{decomp888}) 
there must be two singlet eigenstates, eight octets, six $27$-plets 
and one $64$-plet, $17$ in total. All tensors, $T$, are classified by 
their symmetry properties, according to whether first or 
second class, eq.~(\ref{FirstSecond}), and whether they are 
$f{\rm -like}$ or $d{\rm -like}$, eq.~(\ref{f_d_like}), and 
are given by
\begin{eqnarray}
   \begin{tabular} {c|cc|cc}
   $SU(3)$ & \multicolumn{2}{c}{$T$, $1^{\rm st}$ class} & \multicolumn{2}{c}{$T$, $2^{\rm nd}$ class} \\
           & $d{\rm -like}$ & $f{\rm -like}$ & $d{\rm -like}$ & $f{\rm -like}$  \\
   \hline 
   $1$  & $d$ & $f$ &  &  \\
   $8$  & $r_1$, $r_2$, $r_3$ & $s_1$, $s_2$ & $t_1$, $t_2$ & $u_1$ \\
   $27$ & $q_1$, $q_2$ & $w_1$, $w_2$ &  $x_1$ & $y_1$ \\
   $64$ & $z$ &  &  &  \\
   \end{tabular} \,.
\label{tensor_names}
\end{eqnarray}
Furthermore in Appendix~\ref{non-zero_tensor} we list all non-zero 
elements for all $17$ tensors, together with their values.
For example in eq.~(\ref{Tijk}) we give the non-zero elements
of the tensors $T = r_1$ and $t_1$,
\begin{eqnarray}
   \begin{tabular}{lrl}
  $T$ & $T_{ijk}$  & $ijk$ \\\hline\\[-0.9em]
 $r_1$& $1$ & $ 151 \q 252 \q 353 \q 454 \q 555 \q 656 \q 757 \q 858 $ \cr
 \hline \\[-0.9em]
 $t_1$ & $1$ & $  115 \q 225 \q 335 \q 445\q 665 \q 775 \q 885 $\cr 
    & $-1$ & $ 518 \q 527 \q 536 \q  544 \q 563 \q 572 \q 581 $ \cr
   \end{tabular} \,.
\label{Tijk}
\end{eqnarray}
The values of the non-zero $T_{ijk}$ elements are given in the 
second column, while their position is given in the third block. 
In particular we see that the $r_1$ tensor only has $8$ non-zero entries, 
all identical in value, in the positions $T_{i5i}$, where $i$ can take 
any value from $1$ to $8$. It can easily be checked, for example, 
that the tensors $r_1$ and $t_1$ with non-zero elements as given 
in eq.~(\ref{Tijk}) are first and second class tensors, respectively.

The $r_i$ tensors are $d$-like and can be regarded as responsible 
for the quark-mass dependence of the $d$ coupling (see the $d$-fan in 
section~\ref{FanSection}), while the $s_i$ tensors are $f$-like and act as 
quark-mass-dependent additions to the $f$ coupling (as seen 
in the $f$-fan -- see section~\ref{FanSection}).
 
We are now finally in a position to present the $SU(3)$ 
flavour-symmetry-breaking expansions. As we are considering only the 
isospin limit, eq.~(\ref{mass_isospin}), then Table~\ref{quadratic} 
reduces to Table~\ref{quadratic_dml}.
\begin{table}[!htb] 
   \begin{center} 
   \begin{tabular} {c|cccccc}
   Polynomial & \multicolumn{6}{c}{$SU(3)$}                      \\
   \hline 
   $1$ & $1$ &  &  &  &  &                                       \\
   \hline
   $\delta m_l$ &  & $8$ &  &  &  &                              \\
   \hline
   $\delta m_l^2$ & $1$ &  &  &  & $27$ &  \\
   $\delta m_l^2$ &  & $8$ &  &  & $27$ &  \\
   \hline
   $\delta m_l^3$ & $1$ &  &  &  & $27$ & $64$  \\
   $\delta m_l^3$ &  & $8$ &  &  & $27$ & $64$  \\
\end{tabular}
\caption{All the quark-mass polynomials in the isospin limit
         up to $O(\delta m_l^3)$, classified by symmetry properties.}
\label{quadratic_dml}
\end{center} 
\end{table} 
For example, let us consider 
$\langle p | J^{\pi^+} | n\rangle \equiv \langle B_2 | J^{F_6} | B_1 \rangle$, 
eq.~(\ref{B2J6B1}). From Table~\ref{su2relations_diag}, this is 
$\sqrt{2}A_{\bar{N}\pi N}$. Hence from eq.~(\ref{schema}), and using 
Table~\ref{quadratic_dml} and Appendix~\ref{non-zero_tensor}
(for the non-zero $261$ component of the appropriate tensor)
and using the same notation for the expansion coefficients
as for the tensor gives the LO expansion
\begin{eqnarray}
   \sqrt{2}A_{\bar{N}\pi N}
      = 1 \times (\sqrt{2}f + \sqrt{6}d) 
            + \delta m_l \times (-2\sqrt{2}r_3 + 2\sqrt{2}s_1) \,.
\end{eqnarray}
At higher orders, we also need in addition the non-zero elements
of the $27$- and $64$-plet. Further examples are given in the 
next section in eqs.~(\ref{first_example}, \ref{second_example}).


\section{Coefficient tables} 
\label{coeff_tables}


We use the same notation for the expansion coefficients
as for the tensor. For example the $r_1$ tensor (with components
$T_{i5i}$) has expansion coefficient $r_1$.


\subsection{Leading-order coefficient tables} 
\label{LO_coeff_tables}


The $SU(3)$ singlet and octet coefficients in the mass Taylor expansion
of operator amplitudes are tabulated in Table~\ref{coef_1_8}. 
 \begin{table}[htb]
 \begin{small}
 \begin{center}
 \begin{tabular}{cccrr|rrrrrcrrr}

 & & & \multicolumn{2}{c}{1, $1^{\rm st}$ class}
      & \multicolumn{5}{c}{8, $1^{\rm st}$ class}
      &
      & \multicolumn{3}{c}{8, $2^{\rm nd}$ class}  \\
 & & & \multicolumn{2}{c|}{$O(1)$}
      & \multicolumn{5}{c}{$O(\delta m_l)$}
      &
      & \multicolumn{3}{c}{$O(\delta m_l)$}  \\
 & & & $f$ & $d$ & $d$ &$d$ &$d$ & $f$ &$f$ &  & $d$ & $d$ & $f$ \cr

 \hhline{~~~-------~---}
 & & & & & & & & & & & & & \\

   $I$ & $A_{\bar{B}^\prime FB}$  & & $f$ & $d$ &$r_1$ &$r_2$ &$r_3$ &$s_1$ &$s_2$
 & & $t_1$ & $t_2$ & $u_1$ \\

 \hhline{--~-------~---}

 $0$ &  \tophat ${\bar{N}\eta N}$ & & $\sqrt{3}$ & $-1$
 & 1 & 0 & 0 & 0 & $-1$
 & & 0 & 0 & 0 \\
 $0$ & ${\bar{\Sigma}\eta \Sigma}$ & & 0 & 2
 & 1 & 0 & $2 \sqrt{3}$ & 0 & 0
 & & 0 & 0 & 0 \\
 $0$ & ${\bar{\Lambda}\eta \Lambda}$ & & 0 & $-2$
 & 1 & 2 & 0 & 0 & 0
 & & 0 & 0 & 0 \\
 $0$ & ${\bar{\Xi}\eta \Xi}$ & & $-\sqrt{3}$ & $-1$
 & 1 & 0  & 0 & 0 & 1
 & & 0 & 0 & 0 \\

 \hhline{--~-------~---}

 $1$ & \tophat ${\bar{N}\pi N}$ & & 1 & $\sqrt{3}$
 & 0 & 0 & $-2$ & 2 & 0
 & & 0 & 0 & 0 \\
 $1$ &  ${\bar{\Sigma}\pi \Sigma}$ & & 2 & 0
 & 0 & 0 & 0 &$-2$ &$\sqrt{3}$
 & & 0 & 0 & 0 \\
 $1$ &  ${\bar{\Xi}\pi \Xi}$ & & 1 & $-\sqrt{3}$
 & 0 & 0 & $2$ & 2 & 0
 & & 0 & 0 & 0 \\

 \hhline{==~=======~===}

 $1$ &  ${\bar{\Sigma}\pi \Lambda}$ & & 0 & \tophat 2
 & 0 & 1 & $-\sqrt{3}$ & $0$ & 0
 & & 1 & 0 & 0 \\
 $1$ &  ${\bar{\Lambda}\pi \Sigma}$ & & 0 & 2
 & 0 & 1 & $-\sqrt{3}$ & 0 & 0
 & & $-1$ & 0 & 0 \\

 \hhline{--~-------~---}

 $\frac{1}{2}$ & ${\bar{N} K \Sigma}$ & & $-\sqrt{2}$ & $\sqrt{6}$&
 \tophat
   0 & 0 & $\sqrt{2}$ & $\sqrt{2}$ & 0
 & & 0 & $ \sqrt{2}$ & $ \sqrt{6}$ \\
 $\frac{1}{2}$ & ${\bar{N} K \Lambda}$ & & $-\sqrt{3}$ & $-1$
 & 0 & 1 & 0 & $-\sqrt{3}$ & 1
 & & 1 & $\sqrt{3}$ & $-1$ \\
 $\frac{1}{2}$ & ${\bar{\Lambda} K \Xi}$ & & $\sqrt{3}$ & $-1$
 & 0 & 1 & 0 & $\sqrt{3}$ & $-1$
 & & $-1$ & $-\sqrt{3}$ & $-1$ \\
 $\frac{1}{2}$ & ${\bar{\Sigma} K \Xi}$ & & $\sqrt{2}$ & $\sqrt{6}$
 & 0 & 0 & $ \sqrt{2}$ & $-\sqrt{2}$ & 0
 & & 0 & $-\sqrt{2}$ & $\sqrt{6}$ \\

 \hhline{--~-------~---}

 $\frac{1}{2}$ & ${\bar{\Sigma} \bar{K} N}$ & & $-\sqrt{2}$ & $\sqrt{6}$
 & \tophat 0 & 0 & $ \sqrt{2}$ & $\sqrt{2}$ & 0
 & & 0 & $-\sqrt{2}$ & $-\sqrt{6}$ \\
 $\frac{1}{2}$ & ${\bar{\Lambda} \bar{K} N}$ & & $-\sqrt{3}$ & $-1$
 & 0 & 1 & 0 & $-\sqrt{3}$ & 1
 & & $-1$ & $-\sqrt{3}$ & 1 \\
 $\frac{1}{2}$ & ${\bar{\Xi} \bar{K} \Lambda}$ & & $ \sqrt{3}$ & $-1$
 & 0 & 1 & 0 & $\sqrt{3}$ & $-1$
 & & 1 & $\sqrt{3}$ & 1 \\
 $\frac{1}{2}$ & ${\bar{\Xi} \bar{K} \Sigma}$ & & $\sqrt{2}$ & $\sqrt{6}$
 & 0 & 0 & $\sqrt{2}$ & $-\sqrt{2}$ & 0
 & & 0 & $ \sqrt{2}$ & $-\sqrt{6}$ \\

 \end{tabular}
 \caption{Coefficients in the mass Taylor expansion of $A_{\bar{B}^\prime FB}$
          operator amplitudes: $SU(3)$ singlet and octet, for both 
          first-class and second-class currents. The first row gives
          whether singlet or octet and first or second class, and 
          the second row gives the order in $\delta m_l$. The third row 
          gives whether the associated tensor is $f{\rm -like}$ or 
          $d{\rm -like}$ according to the definition given in 
          eq.~(\protect\ref{f_d_like}). These coefficients are 
          sufficient for the linear expansion of hadronic amplitudes.}
 \label{coef_1_8}
 \end{center}
 \end{small}
 \end{table}
These coefficients are sufficient for the linear expansion of
hadronic amplitudes on the constant $\bar{m}$ line. (If $\bar{m}$ were
not kept constant there would be two more linear terms.)

The table is to be read: for first-class currents the $f$ and $d$ terms
are independent of the quark mass, while the $r_1$, $r_2$, $r_3$ and
$s_1$, $s_2$ coefficients are the leading order (LO) or $\delta m_l$ terms.
For second-class currents, as discussed previously, there are no
leading $f$ and $d$ terms, the expansion starts at $O(\delta m_l)$
for the off-diagonal currents or completely vanishing for the
diagonal currents.

Thus for example to first order in $\delta m_l$ (i.e.\ LO)
we can read off from Tables~\ref{su2relations_diag},
\ref{su2relations_trans} and \ref{coef_1_8}
\begin{eqnarray} 
   \langle p | J^\eta | p \rangle = A_{\bar{N}\eta N} 
      &=& \sqrt{3} f - d + (r_1  - s_2) \delta m_l \,,
                                                         \nonumber  \\
   \langle n| J^{K^+} | \Sigma^- \rangle = A_{\bar{N}K\Sigma}
      &=& -\sqrt{2}f + \sqrt{6}d + (\sqrt{2}r_3+\sqrt{2}s_1)\delta m_l \,,
                                                         \nonumber  \\
   \langle \Sigma^+ | J^\eta | \Sigma^+ \rangle = A_{\bar{\Sigma}\eta \Sigma}
      &=& 2 d  + (r_1 + 2 \sqrt{3} r_3 ) \delta m_l \,,
\label{LO_example}   
\end{eqnarray} 
for first-class currents (for example for the vector current the 
form factors $F_1$ and $F_2$ from eq.~(\ref{1+2class})) and
\begin{eqnarray}
   \langle n| J^{K^+} | \Sigma^- \rangle = A_{\bar{N}K\Sigma}
      &=& ( \sqrt{2}t_2 +\sqrt{6}u_1 ) \delta m_l \,,
                                                          \nonumber \\
   \langle \Sigma^-| J^{K^-} | n \rangle = A_{\bar{\Sigma}\bar{K}N}
      &=& - ( \sqrt{2}t_2 +\sqrt{6}u_1 ) \delta m_l \,,
\end{eqnarray}
for second-class currents (for example for the $F_3$ vector form factor).

A notational comment: we shall usually suppress arguments and indices,
but each coefficient in Table~\ref{coef_1_8}  is a function of 
the  $(\mbox{momentum transfer})^2$, $Q^2$, as well as being renormalised
or not. Thus for example for the renormalised vector current, 
the $f$ coefficient in Table~\ref{coef_1_8} is to be understood as 
$f \to f^{\ind{V\,R}}(\bar{m},Q^2)$.

Note that the clean separation of amplitudes and form factors
into first and second class depends on the fact that we have
defined our amplitudes in ways that treat the parent and daughter
baryons symmetrically. If we had used an unsymmetric definition,
for instance always normalising amplitudes in terms of the
parent baryon's mass, we would find $t_i$ and $u_1$
coefficients appearing in the expansions of quantities
which `should' only involve the symmetric terms.


\subsection{Higher-order coefficient tables} 
\label{higher_order}


For completeness in Table~\ref{coef_27_64} we detail
 \begin{table}[htb]
 \begin{small}   
 \begin{center}
 \begin{tabular}{ccrrrrr|rcrr}

 & & & \multicolumn{4}{c}{27, $1^{\rm st}$ class}
      & \multicolumn{1}{c}{64, $1^{\rm st}$}&
   & \multicolumn{2}{c}{27, $2^{\rm nd}$ class} \\
    & & & \multicolumn{4}{c|}{$O(\delta m_l^2)$}
      & \multicolumn{1}{c}{$O(\delta m_l^3)$} &
      & \multicolumn{2}{c}{$O(\delta m_l^2)$} \\
   &  & & $d$ & $d$ & $f$ & $f$ & $d$ && $d$ & $f$ \cr 

 \hhline{~~~-----~--}
 & & & & & & & & & &  \\

  $I$ & $A_{\bar{B}^\prime FB}$  & & $q_1$ & $q_2$ & $w_1$
 & $w_2$ & $z$ & & $x_1$ & $y_1$  \\

 \hhline{--~-----~--}

 $0$ & \tophat ${\bar{N}\eta N}$ &
 & 9 & 3 & 0 & $3\sqrt{3}$
 & $3\sqrt{3}$
 & & 0 & 0 \\
 $0$ & ${\bar{\Sigma}\eta \Sigma}$ &
 & $-6$& $-10$ & 0 & 0
 & $-\sqrt{3}$
 & & 0 & 0 \\
 $0$ & ${\bar{\Lambda}\eta \Lambda}$ &
 & $-18$ & $18$ & 0 & 0
 & $-9\sqrt{3}$
 & & 0 & 0 \\
 $0$ & ${\bar{\Xi}\eta \Xi}$ &
 & 9 & 3 & 0 & $-3\sqrt{3}$
 & $3\sqrt{3}$
 & & 0 & 0 \\

 \hhline{--~-----~--}

 $1$ & \tophat ${\bar{N}\pi N}$ &
 & $-5\sqrt{3}$ & $\sqrt{3}$ & 4 & $-1$
 & 1
 & & 0 & 0 \\
 $1$ & ${\bar{\Sigma}\pi \Sigma}$ &
 & 0 & 0 &  $-4$ & 2
 & 0
 & & 0 & 0 \\
 $1$ & ${\bar{\Xi}\pi \Xi}$ &
 & $5\sqrt{3}$ & $-\sqrt{3}$ & 4 & $-1$
 & $-1$
 & & 0 & 0 \\

 \hhline{==~=====~==}

 $1$ &  ${\bar{\Sigma}\pi \Lambda}$\tophat&
 & 14 & $-6$ & 0 & 0
 & $-\sqrt{3}$
 & & 4 & 0 \\
 $1$ &  ${\bar{\Lambda}\pi \Sigma}$ &
 & 14 & $-6$ & 0 & 0
 & $-\sqrt{3}$
 & & $-4$ & 0 \\

 \hhline{--~-----~--}

 $\frac{1}{2}$ & ${\bar{N} K \Sigma}$ \tophat &
 & $0$ & $2\sqrt{6}$ & $-3\sqrt{2}$& $2\sqrt{2}$
 & $\sqrt{2}$
 & & $\sqrt{6}$ & $\sqrt{2}$ \\
 $\frac{1}{2}$ & ${\bar{N} K \Lambda}$ &
 & $-6$ & $0 $ & $3 \sqrt{3}$& 0
 & $3\sqrt{3}$
 & & $-3$ & $3\sqrt{3}$ \\
 $\frac{1}{2}$ & ${\bar{\Lambda} K \Xi}$ &
 & $-6$ & $ 0 $ & $-3 \sqrt{3}$& 0
 & $3\sqrt{3}$
 & & $3$ & $3\sqrt{3}$ \\
 $\frac{1}{2}$ & ${\bar{\Sigma} K \Xi}$ &
 & $0$ & $2\sqrt{6}$ & $3\sqrt{2}$& $-2\sqrt{2}$
 & $\sqrt{2}$
 & & $-\sqrt{6}$ & $\sqrt{2}$ \\

 \hhline{--~-----~--}

 $\frac{1}{2}$ & ${\bar{\Sigma} \bar{K} N}$ \tophat &
 & $0$ & $2\sqrt{6}$ & $-3\sqrt{2}$& $2\sqrt{2}$
 & $\sqrt{2}$
 & & $-\sqrt{6}$ & $-\sqrt{2}$ \\
 $\frac{1}{2}$ & ${\bar{\Lambda} \bar{K} N}$ &
 & $-6$ & $ 0$ & $3\sqrt{3}$& 0
 & $3\sqrt{3}$
 & & $3$ & $- 3\sqrt{3}$ \\
 $\frac{1}{2}$ & ${\bar{\Xi} \bar{K} \Lambda}$ &
 & $-6$ & $ 0$ & $-3\sqrt{3}$& 0
 & $3\sqrt{3}$
 & & $-3$ & $-3\sqrt{3}$ \\
 $\frac{1}{2}$ & ${\bar{\Xi} \bar{K} \Sigma}$ &
 & $0$ & $2\sqrt{6}$ & $3\sqrt{2}$& $-2\sqrt{2}$
 & $\sqrt{2}$
 & & $\sqrt{6}$ & $-\sqrt{2}$ \\

 \end{tabular}
 \caption{Additional coefficients in the mass Taylor expansion of
          operator amplitudes: $SU(3)$ 27-plet and 64-plet.
          These additional terms first appear at the quadratic and
          cubic levels respectively. The same notation as for
          Table~\protect\ref{coef_1_8}.}
 \label{coef_27_64}
 \end{center}
 \end{small}
 \end{table}
the additional quadratic and cubic coefficients in the mass
Taylor expansion of the operator amplitudes for the $27$ and $64$-plets.

For first-class currents in Table~\ref{quadratic_dml} the singlet terms
do not contribute at the linear $O(\delta m_l)$ level, but are present at
the quadratic $O(\delta m_l^2)$ and cubic $O(\delta m_l^3)$ levels.
Similarly the octet terms are missing at the $O(1)$ level, but are present
at higher orders. Hence these terms are also present at the higher orders
in the $SU(3)$ flavour-breaking expansion.
There are $5+7 = 12$ amplitudes, and at the $O(\delta m_l^2)$ level
$11$ free parameters, so there is one constraint.
(Alternatively at the $O(\delta m_l^2)$ level one can
have all possibilities which are orthogonal to the $64$-plet,
so there is again just one constraint.)
At the $O(\delta m_l^3)$ level one has $12$ free parameters for the
$12$ amplitudes ($11$ previous and one extra one from the $64$-plet,
the $z$ term). Hence there are now no more constraints available
at this and higher orders in $\delta m_l$.

For second-class currents, there are constraints at the $O(\delta m_l)$
order as we have $5$ amplitudes, but only $3$ expansion coefficients.
However at the next $O(\delta m_l^2)$ level we have additional
$2$ parameters, so there are no more constraints available.
Hence for second-class operators there is no point in going higher
than linear in the quark mass in the $SU(3)$ flavour-breaking expansion.

Thus, for example, from Tables~\ref{coef_1_8} and \ref{coef_27_64}
we would have for the first-class current
\begin{eqnarray} 
   \langle p | J^\eta | p \rangle
      &=& A_{\bar{N}\eta N}
                                                    \nonumber  \\
      &=& \sqrt{3} f - d + (r_1  - s_2) \delta m_l
                                                               \\
      & & + (\sqrt{3}f^{\ind{x}} - d^{\ind{x}}
          + r_1^{\ind{x}} - s_2^{\ind{x}}
          + 6q_1 + 3q_2 + 3\sqrt{3}w_2)\delta m_l^2
                                                    \nonumber  \\
      & & + (\sqrt{3}f^{\ind{xx}} - d^{\ind{xx}}
          + r_1^{\ind{xx}} - s_2^{\ind{xx}}
          + 6q_1^{\ind{x}} + 3q_2^{\ind{x}} + 3\sqrt{3}w_2^{\ind{x}}
          + 3\sqrt{3}z)\delta m_l^3 \,,
                                                   \nonumber
\label{first_example}
\end{eqnarray}
where $f$, $d$ is the leading coefficients, and $f^{\ind{x}}$, $f^{\ind{xx}}$
and $d^{\ind{x}}$, $d^{\ind{xx}}$ are the additional subdominant coefficients
of the same form as the LO singlet, see Table~\ref{quadratic_dml}.
(We use ${\rm x}$ and ${\rm xx}$ superscripts to distinguish them.)
Similarly for $r_1$, $s_2$, $q_1$, $q_2$ and $w_2$ and the octet.
For the second-class current
\begin{eqnarray}
   \langle n| J^{K^+} | \Sigma^- \rangle
      &=& A_{\bar{N}K\Sigma}
                                                               \\
      &=& ( \sqrt{2}t_2 +\sqrt{6}u_1 ) \delta m_l
          + ( \sqrt{2}t_2^{\ind{x}} +\sqrt{6}u_1^{\ind{x}}
             + \sqrt{5}x_1 + \sqrt{2}y_1) \delta m_l^2 \,.
                                                    \nonumber
\label{second_example}
\end{eqnarray}
However as just discussed the $O(\delta m_l^3)$ term for the
first-class currents and the $O(\delta m_l^2)$ term for the 
second-class currents have no constraints between the coefficients and 
hence contain no new information.

From eqs.~(\ref{decomp88}, \ref{decomp888}) and as
previously discussed we see that
there is one $64$-plet in the decomposition of $8\otimes 8\otimes 8$,
but none in $8\otimes 8$ and therefore $64$-plet quantities only
show up at $O(\delta m_l^3)$ as shown in Table~\ref{quadratic_dml}.
In \cite{bietenholz11a} we have seen
that the $64$-plet combination of decuplet baryon masses
is extremely small and we should probably expect that the $64$-plet
combination of amplitudes will also remain very small all the way 
from the symmetric point to the physical point. 
By using Mathematica we construct the $64$-plet flavour tensor, 
and find that it corresponds to the combination
\begin{eqnarray} 
   Q_{64} &\equiv& 2 A_{\bar N \eta N} - A_{\bar \Sigma \eta \Sigma} 
                  - 3 A_{\bar \Lambda \eta \Lambda} + 2 A_{\bar \Xi \eta \Xi}
                  + {\ts \frac{2}{\sqrt{3}}} \left( A_{\bar N \pi N} 
                  - A_{\bar \Xi \pi \Xi} \right) 
                                                         \nonumber  \\
          &  &    - \left( A_{\bar \Sigma \pi \Lambda } 
                  + A_{\bar \Lambda\pi \Sigma} \right) 
                  + 2  \left( A_{\bar \Lambda K \Xi} + A_{\bar N K \Lambda} 
                  + A_{\bar \Lambda \bar K N} + A_{\bar \Xi \bar K \Lambda} \right)
                                                         \nonumber  \\
          & &     + \sqrt{ \ts \frac{2}{3}} 
                   \left(  A_{\bar N K \Sigma} + A_{\bar \Sigma K \Xi} 
                         + A_{ \bar \Xi \bar K \Sigma} 
                         + A_{ \bar \Sigma \bar K N} \right)
                                                         \nonumber  \\
         &=& O(\delta m_l^3) \,,
\label{Q_constraint}                                                         
\end{eqnarray}
and as expected the linear and quadratic terms in $\delta m_l$ vanish.
We also note that this quantity should be zero at the $1$-loop
level in chiral perturbation theory \cite{Jenkins:2009wv}.
 
In the remainder of this article we shall not consider these 
next-to-leading-order (NLO) and next-to-next-leading-order (NNLO)
higher orders further.

 
\section{Amplitudes at the symmetric point}
\label{am_sym_pt}


We now further discuss amplitudes at the symmetric point.
From eq.~(\ref{decomp888}) there are two octets and one singlet in the
decomposition of $8 \otimes 8$, so there will be two singlets in
$8 \otimes 8 \otimes 8$. This means that at the symmetric point
there are two ways to couple an octet operator between octet baryons.
These correspond to the first two columns of Table~\ref{coef_1_8}.
These two couplings are traditionally given the letters $F$ and $D$.
The $F$ coupling has a pattern related to the $SU(3)$ structure
constant $f_{ijk}$ and the $D$ coupling is related to $d_{ijk}$.
In terms of the $3 \times 3$ matrices, the
$F$ coupling is proportional to ${\rm Tr}( M [\bar B, B])$, the 
$D$ coupling to ${\rm Tr}( M \{\bar B, B \})$. 

Let us first look at the pattern of amplitudes 
at the symmetric point (with no breaking of $SU(3)$ flavour symmetry). 
We can read off the corresponding hadronic matrix elements from
Table~\ref{coef_1_8} 
and can construct many matrix element combinations which have to be 
equal at the symmetric point, for example 
 \begin{eqnarray} 
    \frac{\sqrt{3}}{2}  \langle p | J^\eta | p \rangle
       + \frac{1}{2}  \langle p | J^{\pi^0} | p \rangle
       &=& \langle \Sigma^+ |J^{\pi^0} |\Sigma^+ \rangle 
                                                         \nonumber  \\
       &=& - \frac{\sqrt{3}}{2} \langle \Xi^0 | J^\eta | \Xi^0 \rangle  
           + \frac{1}{2} \langle \Xi^0 | J^{\pi^0} | \Xi^0 \rangle
                                                         \nonumber  \\
       &=& 2 f \,,
                                                  \label{fd_ident}  \\
    - \frac{1}{2} \langle p | J^\eta | p \rangle
        +  \frac{\sqrt{3}}{2} \langle p | J^{\pi^0} | p \rangle
       &=&  \langle \Sigma^+ |J^\eta |\Sigma^+ \rangle
                                                         \nonumber  \\
       &=& - \langle \Lambda^0 |J^\eta | \Lambda^0 \rangle 
                                                         \nonumber  \\
       &=& - \frac{1}{2}  \langle \Xi^0 | J^\eta | \Xi^0 \rangle
           - \frac{\sqrt{3}}{2} \langle \Xi^0 | J^{\pi^0} | \Xi^0 \rangle 
                                                         \nonumber  \\
       &=& 2 d \,.
                                                         \nonumber 
\end{eqnarray} 
These relations become more transparent if we write the operators
out in $\bar q \gamma q$ form, following Table~\ref{ind8} giving 
\begin{eqnarray}
   \frac{1}{\sqrt{2}} \langle p | ( \bar u\gamma u - \bar s\gamma s)| p \rangle
      &=& \frac{1}{\sqrt{2}}\langle \Sigma^+ |
                   ( \bar u\gamma u - \bar d\gamma d)|\Sigma^+ \rangle
                                                         \nonumber  \\
      &=& \frac{1}{\sqrt{2}}\langle\Xi^0 | 
                   ( \bar s\gamma s - \bar d\gamma d) | \Xi^0 \rangle 
                                                    \label{fcombo}  \\
      &=& 2 f \,,
                                                         \nonumber
\end{eqnarray}
from the first line of eq.~(\ref{fd_ident}).  
Written out in this form, it is clear why these three matrix elements
have to be the same at the symmetric point. The $u$ content of the proton 
is the same as the $u$ content of the $\Sigma^+$ or the $s$ content
of the $\Xi^0$, because in each case it is the `doubly represented'
valence quark. Likewise the $s$ in the proton is the same as the 
$d$ in the $\Sigma^+$ or the $d$ in the $\Xi^0$ because in each 
case it is the non-valence flavour. So the relations in eq.~(\ref{fcombo})
are simple consequences of flavour permutation (the $S_3$ subgroup of $SU(3)$). 
Similarly, the second line of eq.~(\ref{fd_ident}) implies 
\begin{eqnarray}
   \frac{1}{\sqrt{6}} \langle p | ( \bar u\gamma u + \bar s\gamma s
                                    - 2 \bar d \gamma d)| p \rangle
      &=& \frac{1}{\sqrt{6}}\langle \Sigma^+ |
                                  ( \bar u\gamma u + \bar d\gamma d
                                    -2 \bar s\gamma s)|\Sigma^+ \rangle
                                                         \nonumber  \\
      &=& \frac{1}{\sqrt{6}}\langle\Xi^0 | ( \bar s\gamma s + \bar d\gamma d
                                            -2 \bar u\gamma u)
                                         | \Xi^0 \rangle 
                                                     \label{dcombo} \\ 
      &=& 2\, d \,.
                                                         \nonumber
\end{eqnarray} 
All these matrix elements have the same pattern, `doubly represented
$+$ non-valence $-2 \times$ singly represented', so again we can understand
why they all have to be the same at the symmetric point.
Note that the operator in the $d$ equation, eq.~(\ref{dcombo}), is always
orthogonal to the operator in the $f$ equation, eq.~(\ref{fcombo}).
We could also look at the pattern `doubly represented
$-$ singly represented', which is just a linear
combination of eq.~(\ref{fcombo}) and eq.~(\ref{dcombo}). Thus
\begin{eqnarray}
   \frac{1}{\sqrt{2}} \langle p | ( \bar u\gamma u - \bar d\gamma d)| p \rangle
      &\equiv& \frac{1}{\sqrt{2}}\langle \Sigma^+ |
                          ( \bar u\gamma u - \bar s\gamma s)|\Sigma^+ \rangle
                                                         \nonumber  \\
  &\equiv& \frac{1}{\sqrt{2}}\langle\Xi^0 | ( \bar s\gamma s - \bar u\gamma u)
                               | \Xi^0 \rangle 
                                                  \label{doub-sing} \\ 
      &=& f + \sqrt{3} \, d \,. 
                                                         \nonumber 
\end{eqnarray}

Of course we can not deduce the full structure at the symmetric point 
from flavour permutations alone, identities such as 
\begin{eqnarray}
   A_{\bar{\Sigma}\eta\Sigma} = - A_{\bar{\Lambda}\eta\Lambda}
                        = A_{\bar{\Lambda}\pi\Sigma} \,,
\end{eqnarray}
connecting diagonal matrix elements to transition amplitudes 
require more general $SU(3)$ rotations to establish them.

 
\section{Mass dependence: `fan' plots}
\label{FanSection}


If we move away from the symmetric point, keeping $\bar{m}$ fixed, 
non-singlet tensors can contribute to eq.~(\ref{schema}).
To first order in $\delta m_l$ we only need consider the octets,
so we can then read the mass terms off from Table~\ref{coef_1_8}
with an example being given in eq.~(\ref{LO_example}).
We can examine the violation of $SU(3)$ symmetry caused by the $m_s - m_l$
mass difference by constructing quantities which must all be equal in 
the fully symmetric case, but which can differ in the case of $n_f = 2+1$ 
quark masses. 

We now discuss two so--called `fan' plots -- the $d$-fan plot and the
$f$-fan plot. In Appendix~\ref{doub-sing_rep} we discuss some further
fan plots (called there the doubly represented -- singly represented fan
plots, namely the $P$-fan plot and the $V$-fan plot).

  
\subsection{The $d$-fan}


Using Table~\ref{coef_1_8} we can construct seven quantities, $D_i$, 
which all have the same value ($2d$) at the symmetric point, but which
can differ once $SU(3)$ is broken
\begin{eqnarray}
 D_1 \equiv - ( A_{\bar N \eta N} + A_{\bar \Xi \eta \Xi} ) 
 &=& 2 d - 2 r_1 \delta m_l \,, \nonumber \\
 D_2 \equiv A_{\bar \Sigma \eta \Sigma }
 &=& 2 d + (r_1 + 2 \sqrt{3} r_3) \delta m_l  \,,\nonumber \\
 D_3 \equiv{} - A_{\bar \Lambda \eta \Lambda }
 &=& 2 d - (r_1 + 2 r_2) \delta m_l  \,,\nonumber \\
 D_4 \equiv \frac{1}{\sqrt{3}} ( A_{\bar N \pi  N} - A_{\bar \Xi \pi  \Xi} )
 &=& 2 d -\frac{4}{\sqrt{3}} r_3 \delta m_l  \,, 
                                            \label{D_fan} \\
 D_5 \equiv A_{\bar \Sigma \pi \Lambda }
 &=& 2 d + ( r_2 - \sqrt{3} r_3) \delta m_l  \,,\nonumber \\
 D_6 \equiv \frac{1}{\sqrt{6}}   (A_{\bar N K \Sigma} + A_{\bar \Sigma K \Xi} )
 &=& 2 d + \frac{2}{\sqrt{3}} r_3 \delta m_l \,,\nonumber \\
 D_7 \equiv - ( A_{\bar N K \Lambda} + A_{\bar \Lambda K \Xi} ) 
 &=& 2 d - 2 r_2 \delta m_l \,. \nonumber 
                                                  \nonumber
\end{eqnarray} 
Plotting these quantities gives a fan~plot with seven lines,
but only three slope parameters ($r_1, r_2$ and $r_3$), so the splittings
between these observables are highly constrained. 
Of course, these seven quantities are not a unique choice, other linear
combinations of them could be chosen. 
At the next order (quadratic) in $\delta m_l$ there is one constraint,
from eq.~(\ref{Q_constraint}). In terms of the $D_i$ this reads
\begin{eqnarray} 
   -2 D_1 - D_2 + 3 D_3 + 2 D_4 - 2 D_5 + 4 D_6 - 4 D_7 = O(\delta m_l^3) \,.
\end{eqnarray}
In the $d$-fan we can thus choose six independent quadratic coefficients, 
and fix the seventh from this constraint.

A useful `average $D$' can be constructed from the diagonal amplitudes
\begin{eqnarray}
   X_D \equiv \frac{1}{6} ( D_1 + 2 D_2 + 3 D_4) = 2 d + O(\delta m_l^2) \,,
\label{XD_def}  
\end{eqnarray}
chosen so that the $O(\delta m_l)$ coefficient vanishes.
Other average $D$ quantities are possible if we also incorporate
transition matrix elements. These average quantities can be useful for
helping to set the lattice scale, \cite{Bornyakov:2015eaa}.

It is useful to construct from this fan plots of $D_i/X_D$. However
for our later example of the vector current, $X_D$ vanishes at $Q^2 = 0$
and is always small, so we consider alternatively here
$\tilde{D}_i \equiv D_i/X_F$.


 \subsection{The $f$-fan}


Again using Table~\ref{coef_1_8} we can construct five quantities $F_i$, 
which all have the same value ($2f$) at the symmetric point, but which
can differ  once $SU(3)$ is broken. 
\begin{eqnarray}
 F_1 \equiv \frac{1}{\sqrt{3}} ( A_{\bar N \eta N} - A_{\bar \Xi \eta \Xi} ) 
 &=& 2 f - \frac{2}{\sqrt{3}} s_2 \delta m_l \,, \nonumber \\
 F_2 \equiv  ( A_{\bar N \pi  N} + A_{\bar \Xi \pi  \Xi} )
 &=& 2 f + 4 s_1 \delta m_l \,, \nonumber \\
 F_3 \equiv A_{\bar \Sigma \pi \Sigma }
 &=& 2 f + (-2 s_1 + \sqrt{3} s_2) \delta m_l \,,
                                                 \label{F_fan} \\ 
 F_4 \equiv \frac{1}{\sqrt{2}}   ( A_{\bar \Sigma K \Xi} -A_{\bar N K \Sigma} )
 &=& 2 f - 2 s_1 \delta m_l \,, \nonumber \\
 F_5 \equiv \frac{1}{\sqrt{3}}   ( A_{\bar \Lambda K \Xi}  -A_{\bar N K \Lambda}) 
 &=& 2 f + \frac{2}{\sqrt{3}} (\sqrt{3} s_1 - s_2) \delta m_l \,.
                                                        \nonumber
\end{eqnarray} 
Plotting these quantities gives a fan~plot with $5$ lines, but only two
slope parameters ($s_1$ and $s_2$), so the splittings between these 
observables are again highly constrained. 
At quadratic and higher level there are no constraints
between the coefficients for the $f$-fan. 

Again a useful `average $F$' can be constructed from the diagonal amplitudes
\begin{eqnarray}
   X_F \equiv \frac{1}{6} (3 F_1 + F_2 + 2 F_3) = 2 f + O(\delta m_l^2) \,,
\label{XF_def}  
\end{eqnarray}
and again we can we can construct fan plots of $\tilde{F}_i \equiv F_i/X_F$.
  
The $f$-fan has the nice property that, to linear order, 
there is no error from dropping quark-line-disconnected 
contributions. This is because $r_1$ is the only parameter
with a quark-line-disconnected piece, and none of the 
$r_i$ parameters appear in the $f$-fan.
We shall prove and expand on this point in the following sections
by considering the connected and disconnected expansions separately.

 
\section{Quark-line-connected and -disconnected diagrams}
\label{con+disc}

\begin{figure}[htb]
   \hspace*{0.50in}
   \begin{tabular}{cc}
      \includegraphics[width=5.00cm]{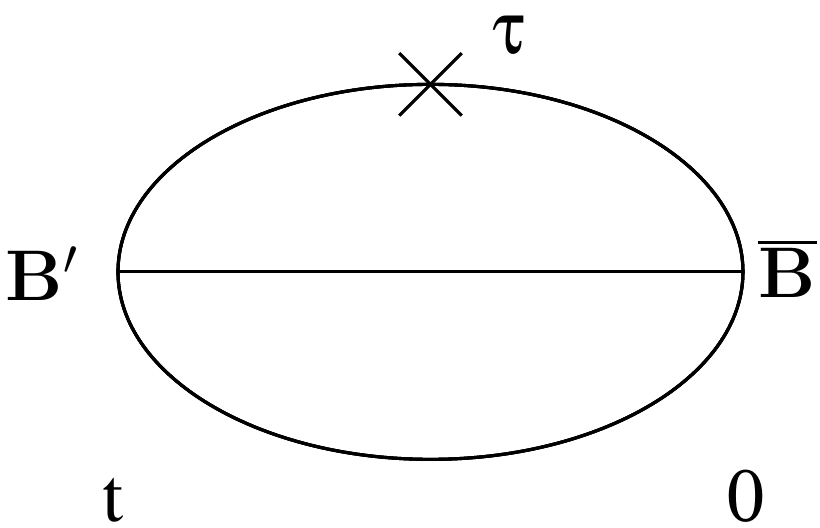}     &
      \hspace{1.0cm}
      \includegraphics[width=5.00cm]{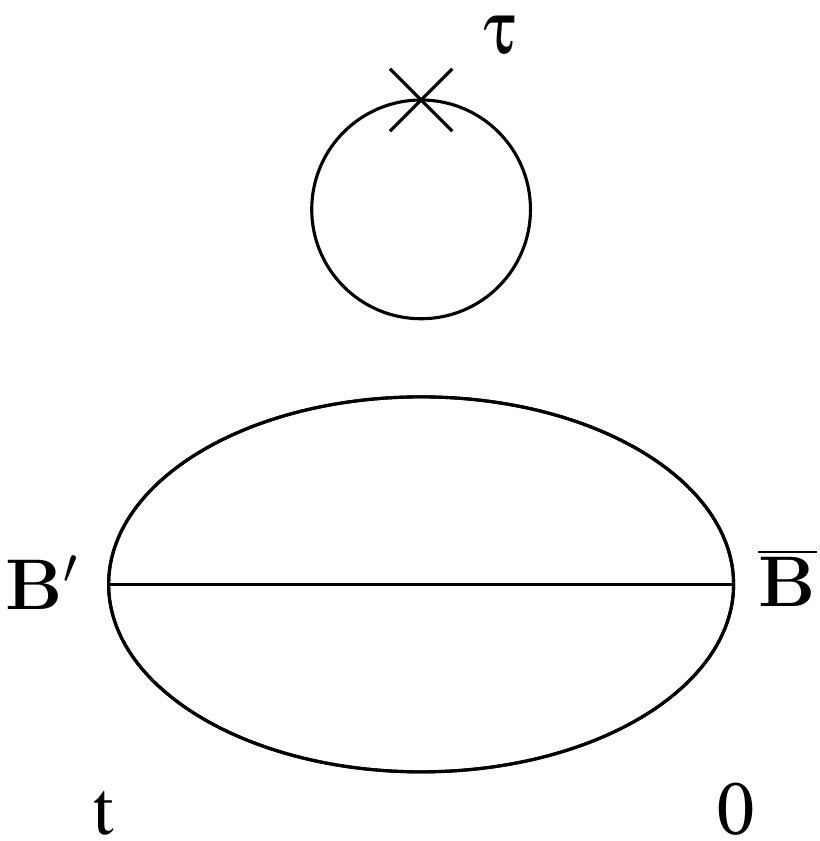}     
   \end{tabular}
   \caption{The three point quark correlation function for a baryon.
            The cross represents the current insertion.
            Left panel: the quark-line-connected piece; right panel:
            the quark-line-disconnected piece.}
   \label{fig_B_3pt}
\end{figure}

In lattice QCD for the three point function and its associated matrix 
element (see section~\ref{lat_comp} for some further details) we have two
classes of diagrams to compute: quark-line connected
(left panel of Fig.~\ref{fig_B_3pt}) and quark-line disconnected
(the right panel of Fig.~\ref{fig_B_3pt}). We first write
\begin{eqnarray}
   \langle B^\prime|J^F|B \rangle =
      \langle B^\prime|J^F|B \rangle^{\ind con } +
      \langle B^\prime|J^F|B \rangle^{\ind dis} \,,
\label{con+dis_splitting}      
\end{eqnarray}      
corresponding to the left and right panels of Fig.~\ref{fig_B_3pt}
respectively. Note that an alternative notation for the 
quark-line-connected piece is the valence matrix element
$\langle B^\prime|J^F|B \rangle^{\ind{con}} \equiv
\langle B^\prime|J^F|B \rangle^{\ind{val}}$. 
However we shall usually just say connected matrix element.

The quark-line-disconnected diagrams cannot occur for
transition matrix elements, $B^\prime \not= B$, but can for 
diagonal matrix elements $B^\prime = B$. 
From Table~\ref{ind8} we see that disconnected diagonal matrix elements
can only happen for the currents $J^{\pi^0}$, $J^\eta$ and $J^{\eta^\prime}$ 
(indices $4$, $5$ and $0$ respectively). 
As we are only considering mass degenerate
$u$ and $d$ quarks then for the $J^{\pi^0}$ operators, the $u$-loop
and $d$-loop quark-line-disconnected pieces always cancel.
Thus apart from the singlet operator $J^{\eta^\prime}$, this leaves
only the $J^\eta$ operator to consider. At the symmetric point,
the disconnected contribution to $J^{\eta}$ will cancel.
If one moves to $m_s \ne m_l$, then disconnected $\eta$ contributions
will become non-zero, as twice the strange loop will not
be equal to the $u$ loop + $d$ loop.
However, at leading order, this effect is going to be the same
for all baryons, so it has the pattern only of $r_1$ in
Table~\ref{coef_1_8}. Hence $r_1$ must have a disconnected piece.

More explicitly first consider the flavour diagonal amplitudes.
In each baryon the disconnected $u$ and $d$ terms
are equal (as $m_u = m_d$), so
\begin{eqnarray}
   \langle p| J^{\pi^0} | p \rangle^{\ind{dis}}\,,
   \quad
   \langle \Sigma^+| J^{\pi^0} | \Sigma^+\rangle^{\ind{dis}}\,,
   \quad
   \langle \Xi^0| J^{\pi^0} |\Xi^0\rangle^{\ind{dis}} \,,
\end{eqnarray}
all vanish. Hence
\begin{eqnarray}
   f^{\ind{dis}} + \sqrt{3}d^{\ind{dis}} = 0\,,
   \quad
   f^{\ind{dis}} = 0\,,
   \quad
   f^{\ind{dis}} - \sqrt{3}d^{\ind{dis}} = 0
\end{eqnarray}
and
\begin{eqnarray}
   -r_3^{\ind{dis}} + s_1^{\ind{dis}} = 0\,,
   \quad
   -2s_1^{\ind{dis}} + \sqrt{3}s_2^{\ind{dis}} = 0\,,
   \quad
   r_3^{\ind{dis}} + s_1^{\ind{dis}} = 0
\end{eqnarray}
giving
\begin{eqnarray}
   f^{\ind{dis}}\,,\,d^{\ind{dis}}\,,\,r_3^{\ind{dis}}\,,\,
   s_1^{\ind{dis}}\,,\,s_2^{\ind{dis}} = 0 \,.
\label{dis_zero_text}
\end{eqnarray}
This was briefly considered for the axial current in \cite{horsley18a}
but the results here are more general than given there.

Consider now the transition amplitudes. As stated previously disconnected 
terms cannot cause a transition that changes flavour. In particular 
considering $K$ current transitions they must all be connected, 
so from Table~\ref{coef_1_8} this again shows that all the above 
coefficients in eq.~(\ref{dis_zero_text}) have no disconnected 
piece, together with the additional result
\begin{eqnarray}
   r_2^{\ind{dis}} = 0 \,,
\label{dis_zero_r2}
\end{eqnarray}   
which means that indeed only $r_1^{\ind dis}$ contributes. Thus in future
we need only distinguish between connected and disconnected contributions
for the $r_1$ coefficient. Differences between the disconnected pieces
in different baryons will therefore first contribute at quadratic order
in the $SU(3)$ flavour-symmetry-breaking expansion. 

We shall now develop and make these considerations more explicit in 
the following section.


\section{Mass dependence: flavour-diagonal matrix elements}
\label{matrix_els}


In the previous sections we have developed $SU(3)$ flavour-breaking
expansions for $\langle B^\prime| J^F | B\rangle$, which are sufficient
for transition matrix elements. However for diagonal matrix elements 
we need the additional expansion $\langle B| J^{\eta^\prime} | B \rangle$
as discussed in section~\ref{singop}. This will now enable all diagonal
matrix elements to be given for each individual quark flavour.

From Table~\ref{ind8} we see that the diagonal flavour states are
given by $\pi^0$ (index $4$) and $\eta$ (index $5$), together
with the singlet flavour state, $\eta^\prime$ (index $0$). These
can be inverted to give $\bar{u}\gamma u$,
$\bar{d}\gamma d$ and $\bar{s}\gamma s$ in terms of
$J^{\eta^\prime}$, $J^{\pi^0}$ and $J^\eta$ as
\begin{eqnarray}
   \bar{u}\gamma u
      &=& {1 \over \sqrt{3}}J^{\eta^\prime} + {1 \over \sqrt{2}}J^{\pi^0}
                                + {1 \over \sqrt{6}}J^{\eta} \,,
                                                         \nonumber \\
   \bar{d}\gamma d
      &=& {1 \over \sqrt{3}}J^{\eta^\prime} - {1 \over \sqrt{2}}J^{\pi^0}
                                + {1 \over \sqrt{6}}J^{\eta} \,,
                                                    \label{ubargu} \\
   \bar{s}\gamma s
      &=& {1 \over \sqrt{3}}J^{\eta^\prime} - {\sqrt{2 \over 3}}J^{\eta} \,.
                                                         \nonumber
\end{eqnarray}   
As discussed previously in section~\ref{singop}, the additional
expansion for the singlet current $J^{\eta^\prime}$ is the same as 
the mass expansion presented in \cite{bietenholz11a}.
We shall only consider LO here (higher orders are also given in 
\cite{bietenholz11a}). We take the expansion
as already given in eq.~(\ref{singlet_expansions}).

Using eq.~(\ref{ubargu}), together with eq.~(\ref{singlet_expansions})
and Tables~\ref{su2relations_diag} and \ref{coef_1_8}
allows us to give the $SU(3)$ flavour-breaking 
expansion for flavour diagonal matrix elements.
In Appendix~\ref{LO_flav_diag_ME} we give this expansion to LO for
the representative octet baryons $p$, $\Sigma^+$, $\Lambda^0$
and $\Xi^0$ (the others $n$, $\Sigma^-$, $\Sigma^0$ and $\Xi^-$
can be similarly determined).

While it appears from eq.~(\ref{singlet_expansions}) that we now
have extra coefficients $a_0$, $a_1$ and $a_2$ that have to be determined,
this can be somewhat ameliorated when the quark-line-connected and
-disconnected matrix elements are considered. There was a general
discussion in section~\ref{con+disc}. We now consider this in more detail
by considering separate expansions for both the connected and
disconnected pieces. So the previous equations are doubled,
as given in eq.~(\ref{con+dis_splitting}).
For example 
\begin{eqnarray}
   \langle p|\bar{u}\gamma u|p \rangle =
      \langle p|\bar{u}\gamma u|p \rangle^{\ind con} +
                  \langle p|\bar{u}\gamma u|p \rangle^{\ind dis} \,,
\end{eqnarray}
corresponding to the left and right panels of Fig.~\ref{fig_B_3pt}
respectively. There are now some additional constraints.

For completeness we list the disconnected matrix element
results in Appendix~\ref{LO_flav_diag_ME_dis}, using $a_0^{\ind dis}$,
$a_1^{\ind dis}$, $a_2^{\ind dis}$ and
eqs.~(\ref{dis_zero_text}, \ref{dis_zero_r2}).


\subsection{Connected terms}


For $p(uud)$, $\Sigma^+(uus)$ and $\Xi^0(ssu)$ there are no connected
pieces for $\langle p| \bar{s}\gamma s | p\rangle$,
$\langle\Sigma^+|\bar{d}\gamma d |\Sigma^+\rangle$ and
$\langle\Xi^0|\bar{d}\gamma d |\Xi^0\rangle$. Thus there are now
conditions on $a_0^{\ind con}$, $a_1^{\ind con}$ and $a_2^{\ind con}$
from the previous expansion parameters. We find
\begin{eqnarray}
   a_0^{\ind{con}}   &=& \sqrt{6}f - \sqrt{2}d \,,
                                                        \nonumber \\
   3a_1^{\ind{con}}  &=& \sqrt{2}r_1^{\ind{con}} - \sqrt{2}s_2 \,,
                                                        \label{a_con} \\
   3a_2^{\ind{con}}  &=& {1\over\sqrt{2}}r_1^{\ind{con}}
                      + \sqrt{6}r_3 + \sqrt{6}s_1
                      - {3\over\sqrt{2}}s_2 \,.
                                                        \nonumber
\end{eqnarray}
(These consistently satisfy all the previous equations.)
Using these expressions for $a_0^{\ind{con}}$, $a_1^{\ind{con}}$
and $a_2^{\ind{con}}$ gives for the octet baryons $p$, $\Sigma^+$, $\Lambda^0$
and $\Xi^0$
\begin{eqnarray}
   \langle p|\bar{u}\gamma u | p\rangle^{\ind{con}}
      &=& 2\sqrt{2}f
                 + \left( \sqrt{3\over 2}r_1^{\ind{con}} - \sqrt{2}r_3
                     + \sqrt{2}s_1 - \sqrt{3\over 2}s_2
                   \right) \delta m_l \,,
                                         \label{delta_q_expan_con_p} \\
   \langle p|\bar{d}\gamma d | p\rangle^{\ind{con}}
      &=& \sqrt{2}\left( f -\sqrt{3}d\right)
            + \left( \sqrt{3\over 2}r_1^{\ind{con}} + \sqrt{2}r_3
                      - \sqrt{2}s_1 - \sqrt{3\over 2}s_2
                   \right) \delta m_l \,,
                                                          \nonumber
\end{eqnarray}
\begin{eqnarray}
   \langle \Sigma^+|\bar{u}\gamma u | \Sigma^+\rangle^{\ind{con}}
      &=& 2\sqrt{2}f
                 + \left(-2\sqrt{2}s_1 + \sqrt{6}s_2
                   \right) \delta m_l \,,
                                                          \nonumber  \\
   \langle \Sigma^+|\bar{s}\gamma s | \Sigma^+\rangle^{\ind{con}}
      &=& \sqrt{2}\left( f -\sqrt{3}d\right)
                                         \label{delta_q_expan_con_Sigma} \\
      & &   + \left( -\sqrt{3\over 2}r_1^{\ind{con}} -3\sqrt{2}r_3
                      - \sqrt{2}s_1 + \sqrt{3\over 2}s_2
              \right) \delta m_l \,,
                                                          \nonumber
\end{eqnarray}   
\begin{eqnarray}
   \langle \Lambda^0|\bar{u}\gamma u | \Lambda^0\rangle^{\ind{con}}
      &=& \langle \Lambda^0|\bar{d}\gamma d | \Lambda^0\rangle^{\ind{con}}
                                                             \nonumber \\
      &=& \sqrt{2}\left(f - {2\over \sqrt{3}}d\right)
                                                             \nonumber \\
      & & + \left(  \sqrt{2\over 3}r_1^{\ind{con}}
                   + \sqrt{2\over 3}r_2
                   + \sqrt{2}r_3
                   + \sqrt{2}s_1 - \sqrt{3\over 2}s_2
                   \right) \delta m_l \,,
                                          \label{delta_q_expan_con_Lam} \\
   \langle \Lambda^0|\bar{s}\gamma s | \Lambda^0\rangle^{\ind{con}}
      &=& \sqrt{2}\left(f + {1\over \sqrt{3}}d\right)
                                                             \nonumber \\
      & &  + \left(- {1\over\sqrt{6}}r_1^{\ind{con}}
                   - {4\over\sqrt{6}}r_2
                   + \sqrt{2}r_3
                   + \sqrt{2}s_1 - \sqrt{3\over 2}s_2
                   \right) \delta m_l \,,
                                                          \nonumber
\end{eqnarray}   
and
\begin{eqnarray}
   \langle \Xi^0|\bar{u}\gamma u | \Xi^0\rangle^{\ind{con}}
      &=& \sqrt{2}(f - \sqrt{3}d)
                 + \left(2\sqrt{2}r_3 + 2\sqrt{2}s_1
                   \right) \delta m_l \,,
                                                          \nonumber   \\
   \langle \Xi^0|\bar{s}\gamma s | \Xi^0\rangle^{\ind{con}}
      &=& 2\sqrt{2}f
                  + \left( -\sqrt{3\over 2}r_1^{\ind{con}} +\sqrt{2}r_3
                      + \sqrt{2}s_1 - \sqrt{3\over 2}s_2
                   \right) \delta m_l \,.
\label{delta_q_expan_con_Xi}
\end{eqnarray}   
Without $\Lambda^0$ there are six equations, together with six parameters,
so no constraint. Adding the $\Lambda^0$ gives two more equations and
one extra parameter, so this is now constrained.
In addition off-diagonal matrix elements would also give more constraints.


\subsection{The electromagnetic current}
\label{em_current}


Using the previous results of this section, we can also give the
results for the electromagnetic current, eq.~(\ref{em_current_def}).
Using this equation we find, for example, that
for the octet baryons $p$, $\Sigma^+$, $\Lambda^0$ and $\Xi^0$
\begin{eqnarray}
   \langle p|J_{\rm em} | p\rangle^{\ind{con}}
      &=& \sqrt{2}f + \sqrt{2 \over 3}d
          + \left( {1 \over \sqrt{6}}r_1^{\ind{con}} - \sqrt{2}r_3
                     + \sqrt{2}s_1 - {1 \over \sqrt{6}}s_2
                   \right) \delta m_l \,,
                                                          \nonumber  \\
   \langle \Sigma^+|J_{\rm em} | \Sigma^+\rangle^{\ind{con}}
      &=& \sqrt{2}f + \sqrt{2 \over 3}d
          + \left( {1 \over \sqrt{6}}r_1^{\ind{con}} + \sqrt{2}r_3
                     - \sqrt{2}s_1 - \sqrt{3 \over 2}s_2
                   \right) \delta m_l \,,
                                                          \nonumber  \\
   \langle \Lambda^0|J_{\rm em} | \Lambda^0\rangle^{\ind{con}}
      &=& - \sqrt{2 \over 3}d
            + \left( {1 \over \sqrt{6}}r_1^{\ind{con}}
            + \sqrt{2 \over 3}r_3
              \right) \delta m_l \,,
                                               \label{EM_expansion}  \\
   \langle \Xi^0|J_{\rm em} | \Xi^0\rangle^{\ind{con}}
      &=& - 2\sqrt{2 \over 3}d
          + \left( {1 \over \sqrt{6}}r_1^{\ind{con}} + \sqrt{2}r_3
                     + \sqrt{2}s_1 + {1 \over \sqrt{6}}s_2
                   \right) \delta m_l \,,
                                                          \nonumber
\end{eqnarray}
for the quark-line-connected terms, and for the quark-line-disconnected
terms
\begin{eqnarray}
   \langle p|J_{\rm em} | p\rangle^{\ind{dis}}
     = \langle \Lambda^0|J_{\rm em} | \Lambda^0\rangle^{\ind{dis}}
     = \langle \Sigma^+|J_{\rm em} | \Sigma^+\rangle^{\ind{dis}}
     = \langle \Xi^0|J_{\rm em} | \Xi^0\rangle^{\ind{dis}}
     = {1 \over \sqrt{6}} r_1^{\ind{dis}} \delta m_l \,.
\label{EM_dis}     
\end{eqnarray}
Similar expansions hold for the $n$, $\Sigma^0$, $\Sigma^-$ and
$\Xi^-$ electromagnetic matrix elements.


\section{Renormalisation and $O(a)$ improvement for the vector current}
\label{renorm+improv}



\subsection{General comments}
\label{general_comments}


The computed matrix elements are bare (or lattice) quantities
and must be renormalised and $O(a)$ improved. We would expect that
the effect of the $O(a)$ improvement terms is simply to modify the
$SU(3)$ flavour-breaking expansion coefficients. In this section
we shall show that this expectation is indeed correct. Again,
for illustration, we shall only consider the diagonal sector
($B^\prime = B$) of the vector current here.
By using the results and notation in
\cite{Bhattacharya:2005rb} (see also \cite{Gerardin:2018kpy}) we have
for on-shell improvement
\begin{eqnarray}
   V_\mu^{\pi^0\,\ind{R}}
      &=& Z_V \left[ 1 + ( b_V + 3{\bar{b}}_V)\bar{m}
                       + b_V \delta m_l \right] {\cal V}_\mu^{\pi^0} \,,
                                                              \nonumber  \\
   V_\mu^{\eta\,\ind{R}}
      &=& Z_V \left[ \left( 1 + ( b_V + 3\bar{b}_V)\bar{m}
                       - b_V \delta m_l \right) {\cal V}_\mu^{\eta}
                     + \sqrt{2}(b_V + 3f_V) \delta m_l
                              {\cal V}_\mu^{\eta^\prime} \right] \,,
                                                              \nonumber  \\
   V_\mu^{\eta^\prime\,\ind{R}}
      &=& Z_V r_V \left[ \left(1 + (d_V + 3\bar{d}_V)\bar{m}\right)
                       {\cal V}_\mu^{\eta^\prime}
                     + 2\sqrt{2}d_V \delta m_l
                              {\cal V}_\mu^{\eta} \right] \,.
\end{eqnarray}
where ${\cal V}$ for the local vector current denotes
\begin{eqnarray}
   {\cal V}_\mu^F = V_\mu^F + ic_V \partial_\nu T_{\mu\nu}^F \,,
\end{eqnarray}
with $T^F_{\mu\nu} = \bar{q}F\sigma_{\mu\nu}q$ and
$\partial_\mu\phi(x) = [\phi(x+\hat{\mu}) - \phi(x-\hat{\mu})]/2$.
This additional term only plays a role in non-forward matrix elements.
Note that all the improvement coefficients
$b_V$, $d_V$, $\bar{b}_V$, $\bar{d}_V$ and $c_V$
are just functions of the coupling constant, $g_0$%
\footnote{There is a further improvement coefficient,
$g_0^2 \to \hat{g}_0^2 = g_0^2 \, ( 1 + b_g \bar{m} )$, 
where $b_g$ is a function of $g_0^2$. Little is known about
the value of $b_g$, however perturbatively it is very small,
so we shall ignore it here. Note that as we always consider
$\bar{m} = \mbox{const}.$, then the value of $g_0^2$ is
only slightly shifted by a constant.}.
Thus we do not have to be precisely at the correct (physical) $\bar{m}$
to determine the coefficients. The $r_V$ parameter accounts
for the fact that the singlet renormalisation is different to the
non-singlet renormalisation, $Z_V(g_0)$. $r_V$ also depends on the
chosen scheme and scale.
Tree level gives for the relevant coefficients
\begin{eqnarray}
   b_V(g_0) =  1 + O(g_0^2) \,, \qquad
   f_V(g_0) =  O(g_0^2) \,, \qquad
   c_V(g_0) = O(g_0^2) \,,
\label{tree_improvement}
\end{eqnarray}
(together with $Z_V(g_0) =  1 + O(g_0^2)$
and $d_V(g_0) =  O(g_0^2)$)
where $\bar{b}_V(g_0)$, $\bar{d}_V(g_0)$, being connected with
the sea contributions are $\sim O(g_0^4)$, and are usually taken
as negligible. Furthermore we can write
\begin{eqnarray}
   V_\mu^{\pi^0\,\ind{R}}
      &=& \hat{Z}_V
             \left[ 1 + \hat{b}_V \delta m_l \right] {\cal V}_\mu^{\pi^0} \,,
                                                              \nonumber  \\
   V_\mu^{\eta\,\ind{R}}
      &=& \hat{Z}_V
             \left[ (1 -  \hat{b}_V \delta m_l) {\cal V}_\mu^{\eta}
                      + \sqrt{2}(\hat{b}_V+3\hat{f}_V) \delta m_l
                              {\cal V}_\mu^{\eta^\prime} \right] \,,
                                                              \nonumber  \\
   V_\mu^{\eta^\prime\,\ind{R}}
      &=& \hat{Z}_V \hat{r}_V
             \left[ {\cal V}_\mu^{\eta^\prime}
                     + 2\sqrt{2}\hat{d}_V \delta m_l
                         {\cal V}_\mu^{\eta} \right] \,,
\label{VR}             
\end{eqnarray}
where for constant $\bar{m}$ we have absorbed these $\bar{m}$ terms
into  the renormalisation constant and improvement coefficients.
For example we have%
\footnote{Similarly
  $\hat{r}_V = r_V(1 + (d_V + 3\bar{d}_V)\bar{m})
               ( 1 + (b_V + 3\bar{b}_V)\bar{m} )^{-1}$ 
  and
  $\hat{d}_V = d_V (1 + (d_V + 3\bar{d}_V)\bar{m})^{-1}$.}
\begin{eqnarray}
   \hat{Z}_V &=& Z_V(1 + (b_V + 3\bar{b}_V)\bar{m}) \,, 
                                                             \nonumber  \\
   \hat{b}_V &=& b_V ( 1 + (b_V + 3\bar{b}_V)\bar{m} )^{-1} \,, 
                                                             \nonumber  \\
   \hat{f}_V &=& f_V ( 1 + (b_V + 3\bar{b}_V)\bar{m} )^{-1} \,.
\label{ZVprime_def}  
\end{eqnarray}
We take eq.~(\ref{VR}) as our definition of the improvement coefficients,
as the $SU(3)$ flavour-breaking expansion coefficients are already
functions of $\bar{m}$. To avoid confusion with the previous $SU(3)$
flavour-breaking expansion coefficients we have denoted them with a caret.
Note that in any case we have also numerically that
$|\bar{m}\delta m_l| \ll 1$ and $\bar{m}^2 \ll 1$ so the
improvement coefficients are effectively unchanged.


\subsubsection{$V_\mu^{\pi^0\,{\ind R}}$}


Let us first consider $V_\mu^{\pi^0\,\ind{R}}$ in eq.~(\ref{VR}),
together with (for example) $\langle p| V_4^{\pi^0} | p \rangle^{\ind R}$,
$\langle \Sigma^+| V_4^{\pi^0} | \Sigma^+ \rangle^{\ind R}$,
$\langle \Xi^0| V_4^{\pi^0} | \Xi^0 \rangle^{\ind R}$.
From the expansion for $F = \pi^0$ given in Table~\ref{coef_1_8}
for $A_{\bar{N}\pi N}$, $A_{\bar{\Sigma}\pi \Sigma}$ and $A_{\bar{\Xi}\pi \Xi}$
we see that as expected the effects of the expansion coefficients
simply change their value slightly
\begin{eqnarray}
   s_1 &\to& s_1^\prime = s_1 + {1\over 2} f \hat{b}_V \,,
                                                              \nonumber  \\
   s_2 &\to& s_2^\prime = s_2 + \sqrt{3} f \hat{b}_V \,,
                                                              \nonumber  \\
   r_3 &\to& r_3^\prime = r_3 - {\sqrt{3} \over 2} d \hat{b}_V \,.
\label{stobV}                                                              
\end{eqnarray}  
Furthermore, as a reminder, from eq.~(\ref{dis_zero_text}) the
disconnected pieces for $f$, $d$, $r_2$, $r_3$, $s_1$, $s_2$ all vanish,
which implies that $\hat{b}_V$ also has no disconnected piece.
In particular this means that the results for $V_\mu^{\pi^0\,\ind{R}}$
remain valid when just considering the connected matrix elements.


\subsubsection{$V_\mu^{\eta\,{\ind R}}$}


We can repeat the process for $V_\mu^{\eta\,{\ind R}}$, which gives in addition
to the results of eq.~(\ref{stobV}), the further results
\begin{eqnarray}
   r_1 \to r_1^\prime
       &=& r_1 + d\hat{b}_V + \sqrt{2}a_0(\hat{b}_V+3\hat{f}_V) \,,
                                                              \nonumber  \\
   r_2 \to r_2^\prime
       &=& r_2 + d\hat{b}_V \,.
\label{etar_stobV}
\end{eqnarray}
In addition splitting $r_1$ into $r_1^{\ind con}$ and $r_1^{\ind dis}$ pieces
gives upon using $a_0^{\ind con}$ from eq.~(\ref{a_con})
\begin{eqnarray}
   r_1^{\ind con}
      & \to & r_1^{{\ind con}\,\prime}
              = r_1^{\ind con}
                + 2\sqrt{3}f(\hat{b}_V+3\hat{f}_V^{\ind con})
                - d( \hat{b}_V+ 6\hat{f}_V^{\ind con}) \,,
                                                             \nonumber \\
   r_1^{\ind dis}
      & \to & r_1^{{\ind dis}\,\prime}
              =  r_1^{\ind dis} +3\sqrt{2}a_0^{\ind dis}\hat{f}_V^{\ind dis} \,.
\label{r1+improv}
\end{eqnarray}


\subsubsection{$V_\mu^{\eta^\prime\,\ind{R}}$}


Lastly, considering $V_\mu^{\eta^\prime\,\ind{R}}$, we find
\begin{eqnarray}
   a_1 \to a_1^\prime
      &=& a_1 + 2\sqrt{2 \over 3}
              \left(f - {1 \over \sqrt{3}}d\right) \hat{d}_V \,,
                                                              \nonumber  \\
   a_2 \to a_2^\prime
      &=& a_2 - {4\over 3}\sqrt{2}d\,\hat{d}_V \,.
\label{etap_stobV}                                                         
\end{eqnarray}  


\subsubsection{Concluding remarks}


As expected, all improvement coefficients are terms in the
$SU(3)$ symmetry flavour-breaking expansion, and indeed upon inclusion
leads to slightly modified expansion coefficients, as given in
eqs.~(\ref{stobV}, \ref{etar_stobV}, \ref{etap_stobV}).
We anticipate that the additional improvement term, $\hat{c}_V$,
is also of this form.


\subsection{Determination of $\hat{Z}_V$ and $\hat{b}_V$, $\hat{f}_V^{\ind con}$}


There is an exact global symmetry of the lattice action,
$q \to e^{-i\alpha_q}q$, valid for each quark separately. Using Noether's
theorem this leads to an exactly conserved vector current, CVC.
Practically the operator counts the number of $u$ quarks and the number
of $d$ quarks in the baryon. The local current considered here is not
exactly conserved, so that $V_{\ind{CVC}} = V + O(a)$. We can use this
to define the renormalisation constant and several improvement terms.
(A similar method was used for two flavours and quenched QCD in,
e.g., \cite{Bakeyev:2003ff}.) Thus we shall see that imposing CVC
is equivalent to determining some improvement coefficients.

Practically here we restrict our considerations to the forward
matrix elements for $V_4$ at $Q^2 = 0$ (no momentum transfer,
so there is no additional $\hat{c}_V$ term).


\subsubsection{$V_4^{\pi^0\,{\ind R}}$}
\label{V4pi}


First for the CVC, we consider the representative matrix elements
\begin{eqnarray}
   \langle p| V_4^{\pi^0} | p \rangle^{\ind{R}}
      &=& A^{\ind{R}}_{\bar{N}\pi N} = {1 \over \sqrt{2}}(2-1) \,,
                                                              \nonumber  \\
   \langle \Sigma^+| V_4^{\pi^0} | \Sigma^+ \rangle^{\ind{R}}
      &=& A^{\ind{R}}_{\bar{\Sigma}\pi \Sigma} = {1 \over \sqrt{2}}(2-0) \,,
                                                              \nonumber  \\
   \langle \Xi^0| V_4^{\pi^0} | \Xi^0 \rangle^{\ind{R}}
      &=& A^{\ind{R}}_{\bar{\Xi}\pi \Xi} = {1 \over \sqrt{6}}(1-0) \,.
\label{forcing_CVC}   
\end{eqnarray}     
Using this together with $V_4^{\pi^0}$ in eq.~(\ref{VR}) gives
\begin{eqnarray}
   f = {1 \over \sqrt{2}\hat{Z}_V} \,, \qquad d = 0 \,.
\label{fd_F1q20}  
\end{eqnarray}
One possibility is thus to determine $f$ from $X_F$ at $Q^2 = 0$,
see eq.~(\ref{XF_def}) as
\begin{eqnarray}
   \hat{Z}_V = {\sqrt{2} \over X_F} \,.
\label{ZV_determination}  
\end{eqnarray}
Also from eq.~(\ref{stobV}) and due to the lack of $O(\delta m_l)$ terms
in eq.~(\ref{forcing_CVC}) we have $s_1^{\prime} = 0$, $s_2^{\prime} = 0$ 
and $r_3^{\prime} = 0$ or
\begin{eqnarray}
   s_1 = - {1 \over 2} f \hat{b}_V\,, \qquad
   s_2 = - \sqrt{3} f \hat{b}_V\,,
                                      \qquad r_3 = 0 \,.
\label{bV_determination}   
\end{eqnarray}
Using $\tilde{s}_i = s_i/X_F$, which to leading order is $s_i/(2f)$,
gives directly the  $\hat{b}_V$ improvement coefficient.


\subsubsection{$V_4^{\eta\,{\ind R}}$}


Additionally using the equivalent results from eq.~(\ref{forcing_CVC})
but now for $V_4^{\eta\,{\ind R}}$ namely
\begin{eqnarray}
   \langle p| V_4^{\eta} | p \rangle^{\ind{R}}
      &=& A^{\ind{R}}_{\bar{N}\eta N} = {1 \over \sqrt{6}}(2+1-0) \,,
                                                              \nonumber  \\
   \langle \Sigma^+| V_4^{\eta} | \Sigma^+ \rangle^{\ind{R}}
      &=& A^{\ind{R}}_{\bar{\Sigma}\eta \Sigma} = {1 \over \sqrt{6}}(2+0-2) \,,
                                                              \nonumber  \\
   \langle \Xi^0| V_4^{\eta} | \Xi^0 \rangle^{\ind{R}}
      &=& A^{\ind{R}}_{\bar{\Xi}\eta \Xi} = {1 \over \sqrt{2}}(1+0-4) \,,
\label{forcing_CVC_eta}   
\end{eqnarray}     
not only gives consistency with the previous results
eqs.~(\ref{fd_F1q20}, \ref{ZV_determination}),
but in addition we have $r_1^{{\ind con}\,\prime}=0$, $r_2^\prime = 0$
or from eqs.~(\ref{etar_stobV}, \ref{r1+improv})
\begin{eqnarray}
   r_1^{\ind con}
      = -2\sqrt{3}f\left(\hat{b}_V+3\hat{f}_V^{\ind con}\right) \,, \qquad
   r_2 = 0 \,.
\label{fV_determination}      
\end{eqnarray}
Again using $\tilde{r}_1^{\ind con} = r_1^{\ind con}/X_F = r_1^{\ind con}/(2f)$
automatically eliminates $f$.
We observe that once $\hat{Z}_V$, $\hat{b}_V$ (and $\hat{f}_V^{\ind con}$)
have been determined then by using eq.~(\ref{ZVprime_def}) and 
varying $\bar{m}$, then it is in principle possible to determine 
$\hat{\bar{b}}_V$.


\subsubsection{The Ademollo--Gatto theorem}


The Ademollo--Gatto theorem \cite{ademolla64a} (see also
\cite{anderson93a,cabibbo03b}) in the context of our flavour-breaking
expansions states that the $O(\delta m_l)$ terms vanish for the 
$F_1^{\bar{B}^\prime F B}$ form factor at $Q^2 = 0$ and $B^\prime \not= B$.
This means that $r_2$, $r_3$, $s_1$, $s_2$ vanish at $Q^2 = 0$ (or the 
primed versions if we include the improvement coefficients). This agrees 
with the results of this section.


\section{Lattice computations of form factors}


\subsection{General discussion}
\label{lat_comp}


We now need to determine the matrix elements from a lattice simulation
which computes two- and three-point correlation functions.
For completeness as well as form factors with $B = B^\prime$,
we are developing a formalism for semileptonic decays, $B \not= B^\prime$
so we first consider the general method here.

The baryon two-point correlation function is given by
\begin{eqnarray}
   C^B_\Gamma(t;\vec{p})
      = \sum_{\alpha\beta}\Gamma_{\beta\alpha}
           \left\langle B_\alpha(t;\vec{p}) \bar{B}_\beta(0;\vec{p})
           \right\rangle \,,
\label{2pt_cf}
\end{eqnarray}
while the three-point correlation function generalises this and 
is given by
\begin{eqnarray}
   C^{B^\prime B}_\Gamma(t,\tau;\vec{p},\vec{p}^{\,\prime};J)
      = \sum_{\alpha\beta}\Gamma_{\beta\alpha}
           \left\langle B^\prime_\alpha(t;\vec{p}^{\,\prime})
                        J(\tau;\vec{q}) \bar{B}_\beta(0;\vec{p})
           \right\rangle \,,
\label{3pt_cf}
\end{eqnarray}
with $J$ at time $\tau$ either the vector, axial or tensor current,
and where the source is at time $0$, the sink operator is at time $t$
and
\begin{eqnarray}
   \Gamma \equiv \Gamma^{\ind{unpol}} 
       = {\textstyle{1 \over 2}} ( 1 + \gamma_4)\,, \quad \mbox{or} \,\,\,
   \Gamma \equiv \Gamma^{\ind{pol}} 
       =  {\textstyle{1 \over 2}}
             ( 1 + \gamma_4)i\gamma_5\vec{\gamma}\cdot\vec{n} \,,
\end{eqnarray}
where $\vec{n}$ is the polarisation axis.

To eliminate overlaps of the source and sink operators
with the vacuum, we build ratios of $3$-point to $2$-point
correlation functions. More explicitly let us set
\begin{eqnarray}
   \lefteqn{R_\Gamma(t,\tau;\vec{p},\vec{p}^{\,\prime}; J)}
      & &                                                \nonumber  \\
      &=& { C_\Gamma^{B^\prime B}(t,\tau; \vec{p},\vec{p}^{\,\prime};J) \over
                   C_{\Gamma^{\ind{unpol}}}^{B^\prime}(t;\vec{p}^{\,\prime}) }
           \sqrt{ C_{\Gamma^{\ind{unpol}}}^{B^\prime}(\tau;\vec{p}^{\,\prime})
                  C_{\Gamma^{\ind{unpol}}}^{B^\prime}(t;p^{\,\prime})
                  C_{\Gamma^{\ind{unpol}}}^B(t-\tau;\vec{p}) \over
                  C_{\Gamma^{\ind{unpol}}}^B(\tau;\vec{p})
                  C_{\Gamma^{\ind{unpol}}}^B(t;\vec{p})
                  C_{\Gamma^{\ind{unpol}}}^{B^\prime}(t-\tau;\vec{p}^{\,\prime}) } \,.
\label{ratio_3o2}
\end{eqnarray}
This is designed so that any smearing for the source and sink
operators is cancelled in the ratios,
e.g.\ \cite{capitani98a,Gockeler:2003ay};
of course smearing the baryon operators improves the overlap
with the lowest-lying state, so the relevant overlaps for
the two- and three-point correlation functions must match.

Inserting complete sets of unit-normalised states in eq.~(\ref{ratio_3o2})
and for $0 \ll \tau \ll t \ll {\textstyle{1 \over 2}} T$ gives
\begin{eqnarray}
   R_\Gamma(t,\tau;\vec{p},\vec{p}^{\,\prime}; J)
      = \sqrt{ E_B(\vec{p}) E_{B^\prime}(\vec{p}^{\,\prime}) \over
                \left(E_B(\vec{p})+M_B\right)
                \left(E_{B^{\prime}}(\vec{p}^{\,\prime})+M_{B^\prime}\right) } \,
                F(\Gamma, {\cal J})\,,
\label{R_F}
\end{eqnarray}
with
\begin{eqnarray}
   F(\Gamma,{\cal J})
      = {1 \over 4} \mbox{tr}\, \Gamma
          \left(\gamma_4 
               - i{\vec{p}^{\,\prime}\cdot\vec{\gamma} 
                               \over E_{B^\prime}(\vec{p}^{\,\prime})}
               + {M_{B^\prime} \over E_{B^\prime}(\vec{p}^{\,\prime})}\right)
           {\cal J}
         \left(\gamma_4 - i{\vec{p}\cdot\vec{\gamma} \over E_B(\vec{p})} 
                        + {M_B \over E_B(\vec{p})} \right)
\label{F_fromR}
\end{eqnarray}
(with ${\cal J}$ being given from the Euclideanised version of
eq.~(\ref{current_FF})).
The transferred (Euclidean) momentum from the initial, $B$, to final,
$B^\prime$ state is given by
$Q = (i(E_{B^\prime}(\vec{p}^{\,\prime}) - E_B(\vec{p})),\vec{p}^{\,\prime} - \vec{p})$
so
\begin{eqnarray}
   Q^2 = -(M_{B^\prime}-M_B)^2 + 2\left(E_{B^\prime}(\vec{p}^{\,\prime}) E_B(\vec{p})
          - M_{B^\prime}M_B - \vec{p}\cdot\vec{p}^{\,\prime}\right) \,.
\label{q2_trans}       
\end{eqnarray}
To illustrate the previous $SU(3)$ flavour symmetry-breaking results,
we shall now consider here only the vector current.
Furthermore in general for arbitrary momenta geometry, the kinematic factors
can be complicated; in this article we shall only be considering
the simpler case $\vec{p}^{\,\prime} = \vec{0}$. The technical reason is
that in the lattice evaluation, it requires less numerical inversions
and is hence computationally cheaper. (Physically, of course it is
more natural to start with a stationary baryon, but computationally
of course it does not matter.)
Evaluating $Q^2$ in this frame, eq.~(\ref{q2_trans}), shows that
for flavour diagonal matrix elements form factors $Q^2$ is always 
positive, while for semileptonic decays for small momentum it can 
also be negative. For the vector current with
$\vec{p}^{\,\prime} = \vec{0}$ this gives%
\footnote{We use the Euclideanisation conventions given in \cite{best97a}.
          In particular
          $V_4 = V^{\ind{({\cal M})}0}$,
          $V_i = -iV^{\ind{({\cal M})}i}$ with
          $\gamma_4 = \gamma^{\ind{({\cal M})}0}$,
          $\gamma_i = -i\gamma^{\ind{({\cal M})}i}$,
          $\gamma_5 = - \gamma_{\ind{5}}^{\ind{({\cal M})}}$,
          $\sigma_{\mu\nu} = i/2 [\gamma_\mu, \gamma_\nu]$.}.
\begin{eqnarray}
   R_{\Gamma^{\ind{unpol}}}(t,\tau;\vec{p},0; V_4)
      &=& \sqrt{ E_{\vec{p}}^B+M_B \over 2E_{\vec{p}}^B } \,
          \left[ F_1^{\bar{B}^\prime FB}
                 - {E_{\vec{p}}^B-M_B \over M_B + M_{B^\prime}} F_2^{\bar{B}^\prime FB}
          \right.
                                                              \nonumber  \\
      & & \left. \hspace*{1.50in}
                 - {E_{\vec{p}}^B-M_{B^\prime} \over M_B + M_{B^\prime}}
                       F_3^{\bar{B}^\prime FB} 
          \right] \,,
                                                              \nonumber  \\
   R_{\Gamma^{\ind{unpol}}}(t,\tau;\vec{p},0; V_i)
      &=& - { i p_i \over \sqrt{ 2E_{\vec{p}}^B (E_{\vec{p}}^B+M_B)} } \,
          \left[ F_1^{\bar{B}^\prime FB} 
                 - {E_{\vec{p}}^B-M_{B^\prime} \over M_B + M_{B^\prime}}
                       F_2^{\bar{B}^\prime FB} 
          \right.
                                                            \nonumber   \\
      & & \left. \hspace*{1.50in}
                 - {E_{\vec{p}}^B+M_B \over M_B + M_{B^\prime}}
                       F_3^{\bar{B}^\prime FB} 
          \right] \,,
                                                     \label{vector_rat} \\
   R_{\Gamma^{\ind{pol}}}(t,\tau;\vec{p},0; V_i)
      &=&  { (\vec{p}\times\vec{n})_i 
                \over \sqrt{ 2E_{\vec{p}}^B (E_{\vec{p}}^B+M_B)} } \,
          \left[ F_1^{\bar{B}^\prime FB} + F_2^{\bar{B}^\prime FB} 
          \right] \,, 
                                                              \nonumber  \\
   R_{\Gamma^{\ind{pol}}}(t,\tau;\vec{p},0; V_4)
      &=& 0 \,.
                                                              \nonumber
\end{eqnarray}
In particular for $\vec{p} = 0$ then the only non-zero ratio is
\begin{eqnarray}
   R_{\Gamma^{\ind{unpol}}}(t,\tau;0,0; V_4)
      = F_1^{\bar{B}^\prime FB} - {M_B - M_{B^\prime} \over M_B + M_{B^\prime}}
                  F_3^{\bar{B}^\prime FB} \,,
\label{vector_qmax}
\end{eqnarray}
so we see that in this case for $B^\prime \not= B$ then
we cannot disentangle $F_1^{\bar{B}^\prime FB}$ from $F_3^{\bar{B}^\prime FB}$.
However to LO (i.e.\ $O(\delta m_l)$
effects in the matrix elements) and as $M_B - M_{B^\prime} \propto \delta m_l$ 
then from eq.~(\ref{vector_qmax}) we can write
\begin{eqnarray}
   R_{\Gamma^{\ind{unpol}}}(t,\tau;0,0; V_4)
      = F_1^{\bar{B}^\prime FB} + O(\delta m_l^2) \,,
\label{diag_q2eq0F1}   
\end{eqnarray}
for all $B$ and $B^\prime$, where the $O(\delta m_l^2)$ term
is not present when $B^\prime = B$.


\subsection{Lattice details}
\label{lattice_details}


As a demonstration of the method we apply the formalism outlined
in the previous sections to the form factors published in
\cite{Shanahan:2014uka,Shanahan:2014cga}. Further details of
the numerical simulations can be found there.
The simulations have been performed using $n_f = 2+1$, $O(a)$
improved clover fermions \cite{cundy09a} at $\beta \equiv 10/g_0^2$ of $5.50$
and on $32^3\times 64$ lattice sizes, \cite{bietenholz11a}.
Errors given here are primarily statistical (using $\sim O(1500)$
configurations).

As discussed previously and particularly in section~\ref{choice_qm}
our strategy is to keep the bare quark-mass constant.
Thus once the $SU(3)$ flavour degenerate sea quark mass, $m_0$,
is chosen, subsequent sea quark-mass points $m_l$, $m_s$ are then
arranged in the various simulations to keep $\bar{m}$
($ = m_0$) constant. This then ensures that all the expansion
coefficients given previously do not change. In \cite{bietenholz11a},
masses were investigated and it was seen that a linear fit provides
a good description of the numerical data on the unitary line over
the relatively short distance from the $SU(3)$ flavour symmetric
point down to the physical pion mass. This proved useful in helping
us in choosing the initial point on the $SU(3)$ flavour symmetric
line to give a path that reaches (or is very close to) the physical point.

The bare unitary quark masses in lattice units are given by
\begin{eqnarray}
   m_q = {1 \over 2} 
            \left ({1\over \kappa_q} - {1\over \kappa_{0c}} \right)
            \qquad \mbox{with} \quad q = l, s \,,
\label{kappa_bare}
\end{eqnarray}
and where vanishing of the quark mass along the $SU(3)$ flavour
symmetric line determines $\kappa_{0c}$. We denote the $SU(3)$ flavour
symmetric kappa value, $\kappa_0$, as being the initial point on the path
that leads to the physical point. $m_0$ is given in eq.~(\ref{kappa_bare})
by replacing $\kappa_q$ by $\kappa_0$. Keeping 
$\bar{m} = \mbox{constant} = m_0$ then gives
\begin{equation}
   \delta m_q = {1 \over 2}
                 \left( {1 \over \kappa_q} - {1 \over \kappa_0}
                 \right) \,.
\label{delta_mq_kappa}
\end{equation}
We see that $\kappa_{0c}$ has dropped out of eq.~(\ref{delta_mq_kappa}),
so we do not need its explicit value here. Along the
unitary line the quark masses are restricted and we have
\begin{eqnarray}
   \kappa_s = { 1 \over { {3 \over \kappa_0} - {2 \over \kappa_l} } } \,.
\label{kappas_mbar_const}
\end{eqnarray}
So a given $\kappa_l$ determines $\kappa_s$ here. 
This approach is much cleaner than the more conventional
approach of keeping (the renormalised) strange quark mass constant,
as this necessitates numerically determining the bare strange quark mass.
In addition the $O(a)$ improvement of the coupling constant is much simpler,
in our approach as it only depends on $\bar{m}$,  \cite{bietenholz11a}.
Thus here, the coupling constant remains constant and hence the lattice 
spacing does not change as the quark mass is changed. In the more 
conventional approach this can be problematical as you must in principle 
monitor the changing of the coupling constant as the quark masses vary.

An appropriate $SU(3)$ flavour symmetric $\kappa_0$ value chosen here for this
action was found to be $\kappa_0 = 0.120900$, \cite{bietenholz11a}.
The constancy of flavour-singlet quantities along the unitary line
to the physical point \cite{bietenholz11a}, leads directly from
$X_\pi$ to an estimate for the pion mass of $\sim 465\,\mbox{MeV}$
at our chosen $SU(3)$ flavour symmetric point and from $X_N$ an estimation
of the lattice spacing of $a_N(\kappa_0=0.120900) = 0.074\,\mbox{fm}$.

Specifically as indicated in Table~\ref{table:latticeparameters}
\begin{table}[htb]
   \begin{center}
   \begin{tabular}{ccc}
      $\kappa_{l}$ & $\kappa_{s}$ & $M_{\pi}\,\mbox{MeV}$ \\
      \hline
      $0.120900$ & $0.120900$ & $465$ \\ 
      $0.121040$ & $0.120620$ & $360$ \\
      $0.121095$ & $0.120512$ & $310$ \\ 
   \end{tabular}
   \end{center}
   \caption{Outline of the ensembles used here on the $32^3\times 64$
            lattices together with the corresponding pion masses.}
\label{table:latticeparameters}
\end{table}
we have generated configurations, \cite{Shanahan:2014uka,Shanahan:2014cga},
at the $(\kappa_l, \kappa_s)$ values listed, all with
$\kappa_0 = 0.120900$. 

Eqs.~(\ref{vector_rat}, \ref{diag_q2eq0F1}) are used to determine
from the ratio, $R$, the appropriate form factor.
As described in \cite{Shanahan:2014uka,Shanahan:2014cga},
we bin $Q^2$ to directly compare each configuration and
using the bootstrapped lattice configurations, we set up a weighted
least squares to extract the linear fit parameters and weighted errors
at each $Q^2$ value. The lattice momenta used here in this study in
units of $2\pi/32$ are given by
$a\vec{q} = (0,0,0)$, $(1,0,0)$, $(1,1,0)$, $(1,1,1)$, $ (2,0,0)$, $(2,1,0)$,
$(2,1,1)$, $(2,2,0)$ together with all permutations (where different)
and all possible $\pm$ values.


\section{Results}
\label{results}


We now illustrate some of the features that we have described in
previous sections, using our lattice calculations and the ensembles in
Table~\ref{table:latticeparameters}.


\subsection{X plots}
\label{X_plots_num}

\begin{figure}[!tbp]
   \begin{tabular}{c}
   \includegraphics[width=14.50cm]{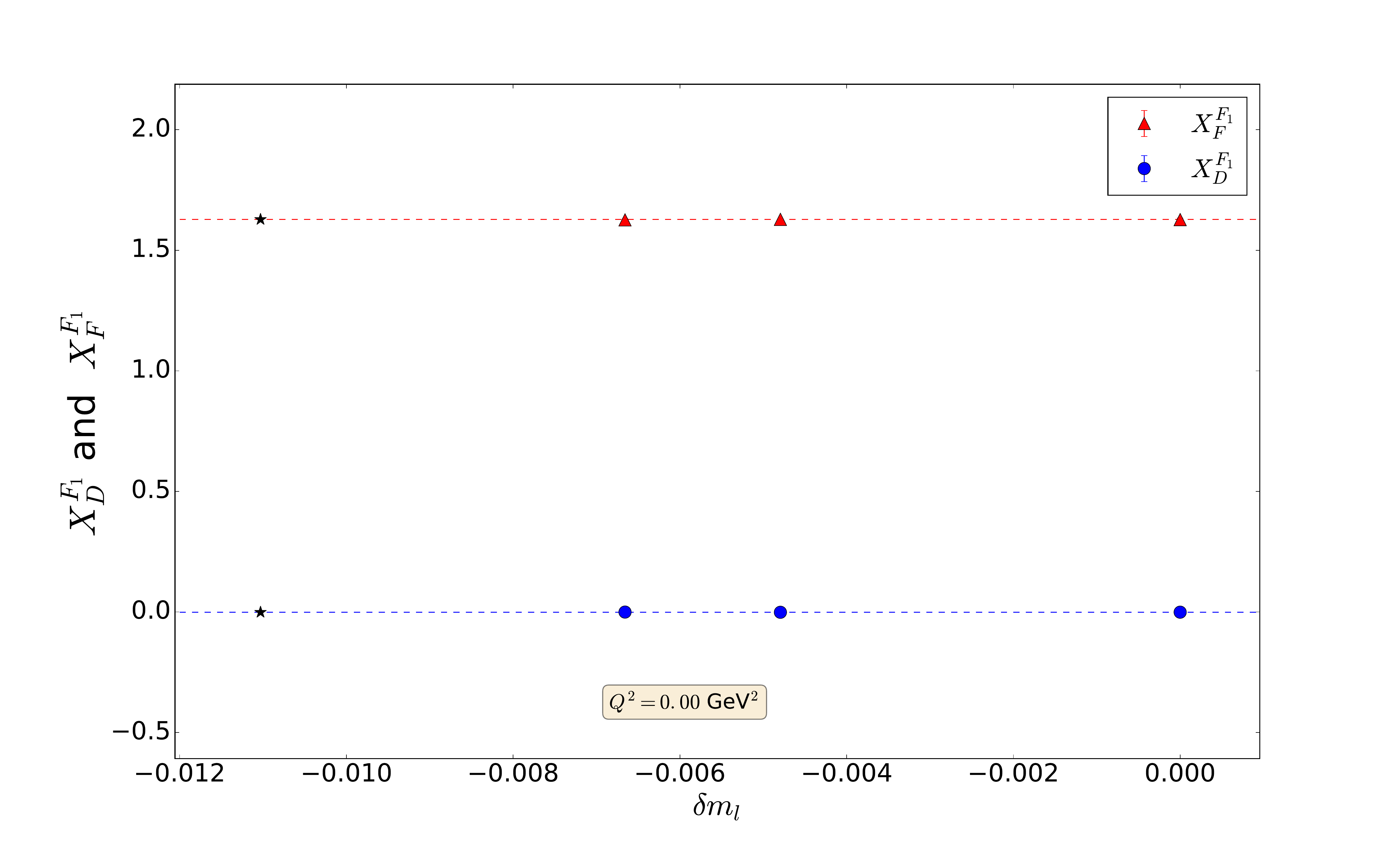} \\
   \includegraphics[width=14.50cm]{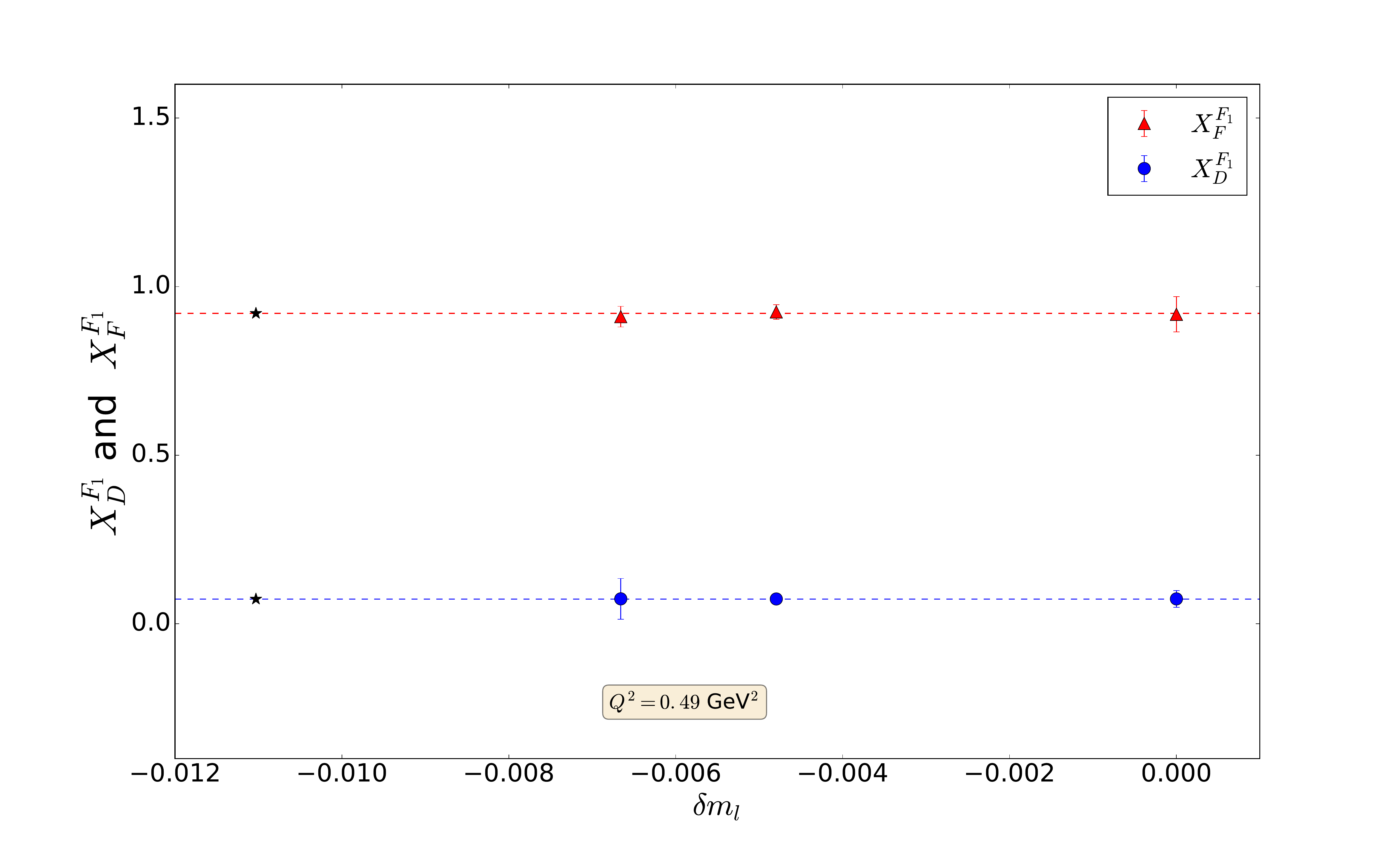}
   \end{tabular}
   \caption{$X^{F_1\,{\ind con}}_D$ and $X^{F_1}_F$ for $F_1$ at $Q^2 = 0$,
         top panel and for $Q^2=0.49\,\mbox{GeV}^2$, lower panel.
         The lower filled circles in each plot are $X^{F_1\,{\ind con}}_D$, the
         upper filled triangles are $X^{F_1}_F$. The dashed lines are
         constant fits and the stars represent the physical point.}
\label{XDF_F1_Q2}
\end{figure}

We first consider the lattice quantities $X^{F_1\,{\ind con}}_D$,
$X^{F_1}_F$ and $X^{F_2\,{\ind con}}_D$, $X^{F_2}_F$.
As discussed previously we only consider
diagonal form factors to construct the $X$s, i.e.\
the equations: $D_1^{\ind{con}}$, $D_2^{\ind{con}}$ and $D_4$
in eq.~(\ref{D_fan}) and $F_1$, $F_2$ and $F_3$
in eq.~(\ref{F_fan})%
\footnote{We note that care needs to be taken to distinguish the $F_i$
          corresponding to a form factor and the $F_i$ defined in
          eq.~(\ref{F_fan}).}.
Using the method of section~\ref{lattice_details} allows us
to create the appropriate $D_1^{\ind con}$, $D_2^{\ind con}$ and
$D_4$ defined in  eq.~(\ref{D_fan}) and hence
$X^{F_1\,{\ind con}}_D$, $X^{F_2\,\ind con}_D$ in eq.~(\ref{XD_def}) or
$F_1$, $F_2$ and $F_3$ in eq.~(\ref{F_fan}) and thus again
$X^{F_1}_F$, $X^{F_2}_F$ in eq.~(\ref{XF_def}).
In Fig.~\ref{XDF_F1_Q2} we consider $X^{F_1\,{\ind con}}_D$ and $X^{F_1}_F$
for the $F_1$ form factor for $Q^2 = 0$ and $0.49\,\mbox{GeV}^2$~%
\footnote{This corresponds to a lattice momentum of
$a\vec{q} = (2\pi)/32\,(1,1,0)$.}.
First, as we expect they are constant and show little sign of
$O(\delta m_l^2)$ or curvature effects.
Although not so relevant on this plot, as an indication of
how far we must extrapolate in the quark mass from the symmetric point
to the physical point, we also give this, using the previous
determination, \cite{horsley14a}, of $\delta m_l^* = -0.01103$.
Note also as shown in eq.~(\ref{fd_F1q20}) for
$Q^2= 0$, $X^{F_1\,\ind{con}}_D$ vanishes as $d=0$, which we also
see on the plot.

\begin{figure}[!tbp]
   \begin{tabular}{c}
   \includegraphics[width=14.50cm]{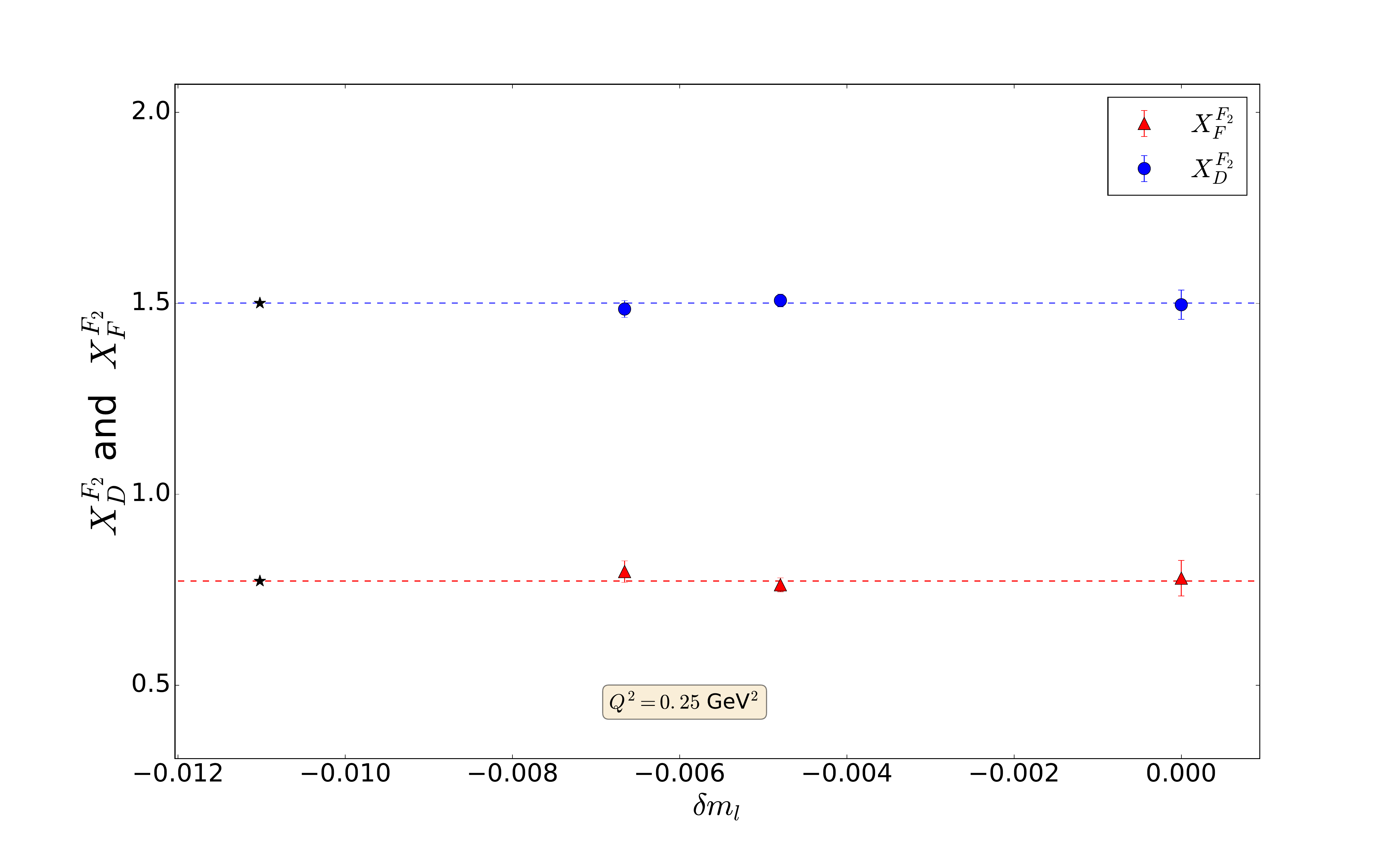} \\
   \includegraphics[width=14.50cm]{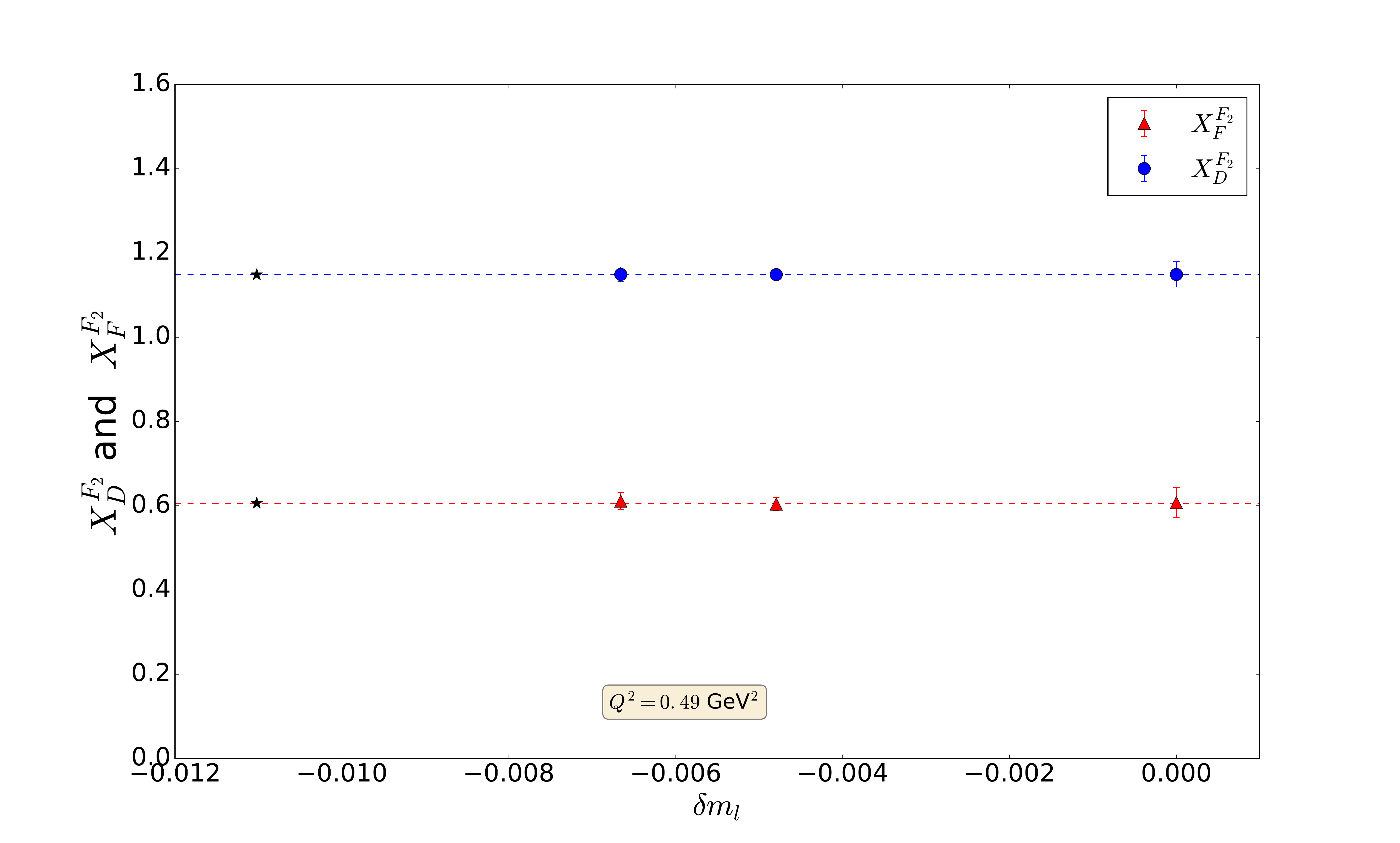}
   \end{tabular}
\caption{$X^{F_2\,{\ind con}}_D$ and $X^{F_2}_F$ for $F_2$
         at $Q^2 = 0.25\,\mbox{GeV}^2$, top panel and for 
         $Q^2=0.49\,\mbox{GeV}^2$, lower panel. The same notation as 
         for Fig.~\protect\ref{XDF_F1_Q2}.}
\label{XFD_F2_Q2}
\end{figure}

This constancy of $X$ does not depend on the form factor used.
In Fig.~\ref{XFD_F2_Q2} we show similar plots, but now for
the $F_2$ form factors: $X^{F_2\,{\ind con}}_D$ and $X^{F_2}_F$,
for $Q^2 = 0.25$~%
\footnote{This corresponds to a lattice momentum of
$a\vec{q} = (2\pi)/32\, (1,0,0)$.}
and $0.49\,\mbox{GeV}^2$. Again these are all constant, within our
statistics. (We can only determine $X^{F_2\,{\ind con}}_D$ at $Q^2 = 0$ 
via an extrapolation, so we show $Q^2 = 0.25\,\mbox{GeV}^2$ instead.)

\begin{figure}[!tbp]
   \begin{tabular}{c}
   \includegraphics[width=14.50cm]{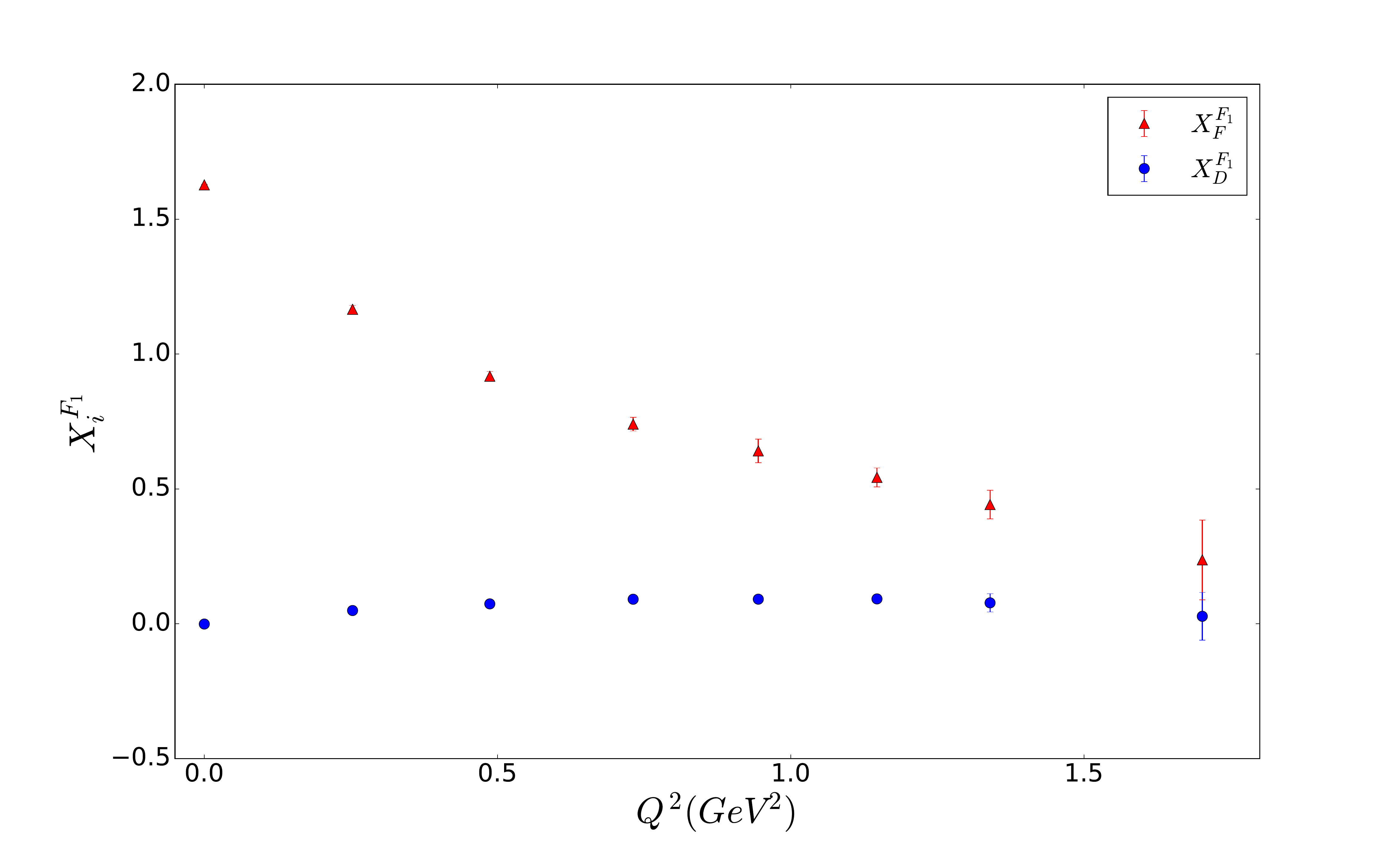} \\
   \includegraphics[width=14.50cm]{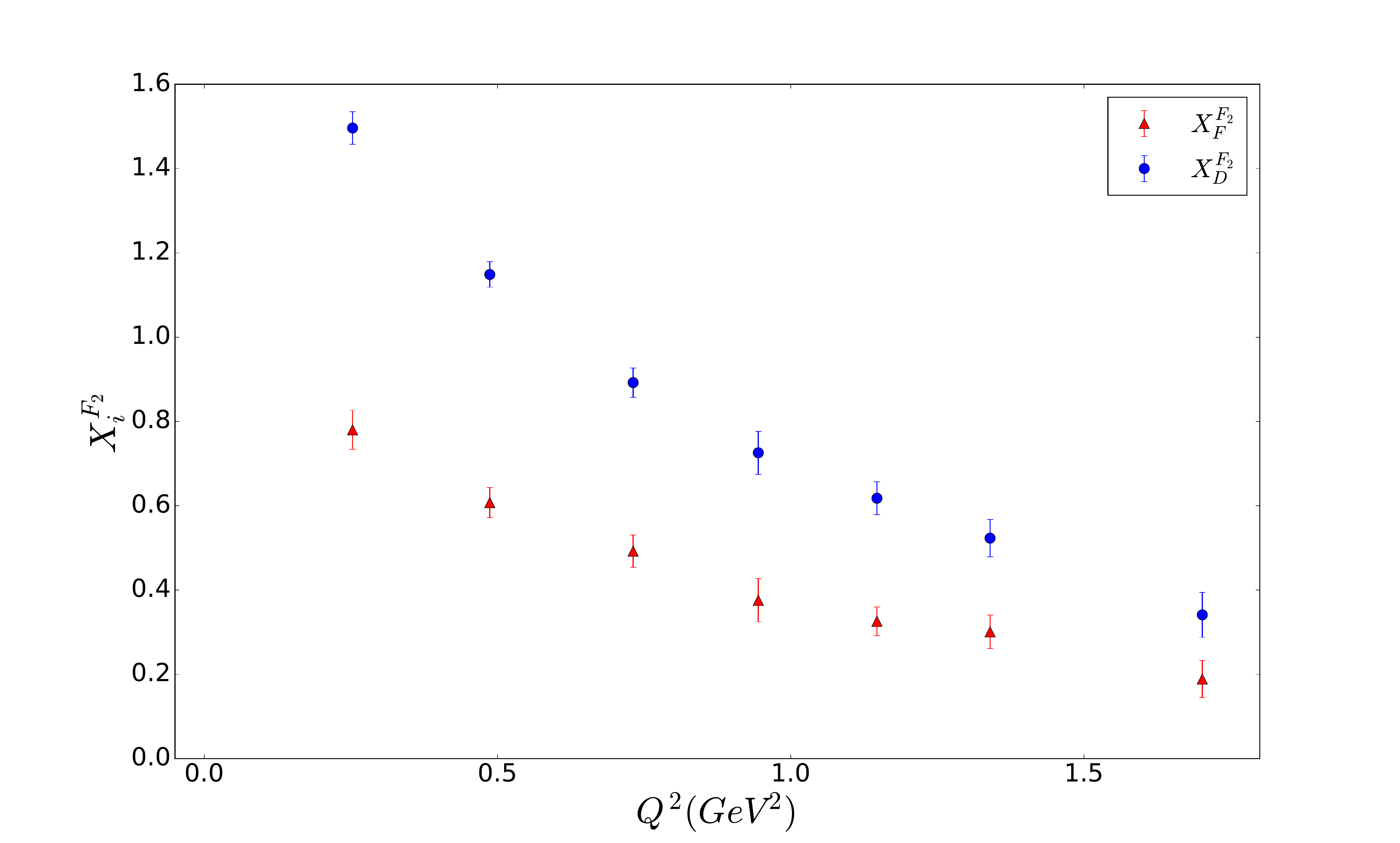}
   \end{tabular}
\caption{Top panel: $X^{F_1}_F$ (filled circles) and $X^{F_1\,{\ind con}}_D$ 
         (filled triangles) versus $Q^2$. Lower panel:
         Similarly for $F_2$.}
\label{X_FD_Q2}
\end{figure}

Finally we can plot the dependence of $X$ on $Q^2$.
In Fig.~\ref{X_FD_Q2} we show  $X^{F_1\,\ind con}_D$
and $X^{F_1}_F$ and similarly for  $X^{F_2}$ versus $Q^2$ (using the 
previously determined fitted values). This gives the $Q^2$ dependence 
of $d$ and $f$ respectively. For  $X^{F_1}_F$, $d$ is initially zero and 
remains small for larger $Q^2$, while $f$ drops monotonically.
We expect $d$ and $f$ to drop like $\sim 1/Q^2$ for large $Q^2$
for all the form factors.


\subsection{Fan plots}
\label{fan_plots_num}

\begin{figure}[!tbp]
   \begin{tabular}{c}
   \includegraphics[width=14.50cm]{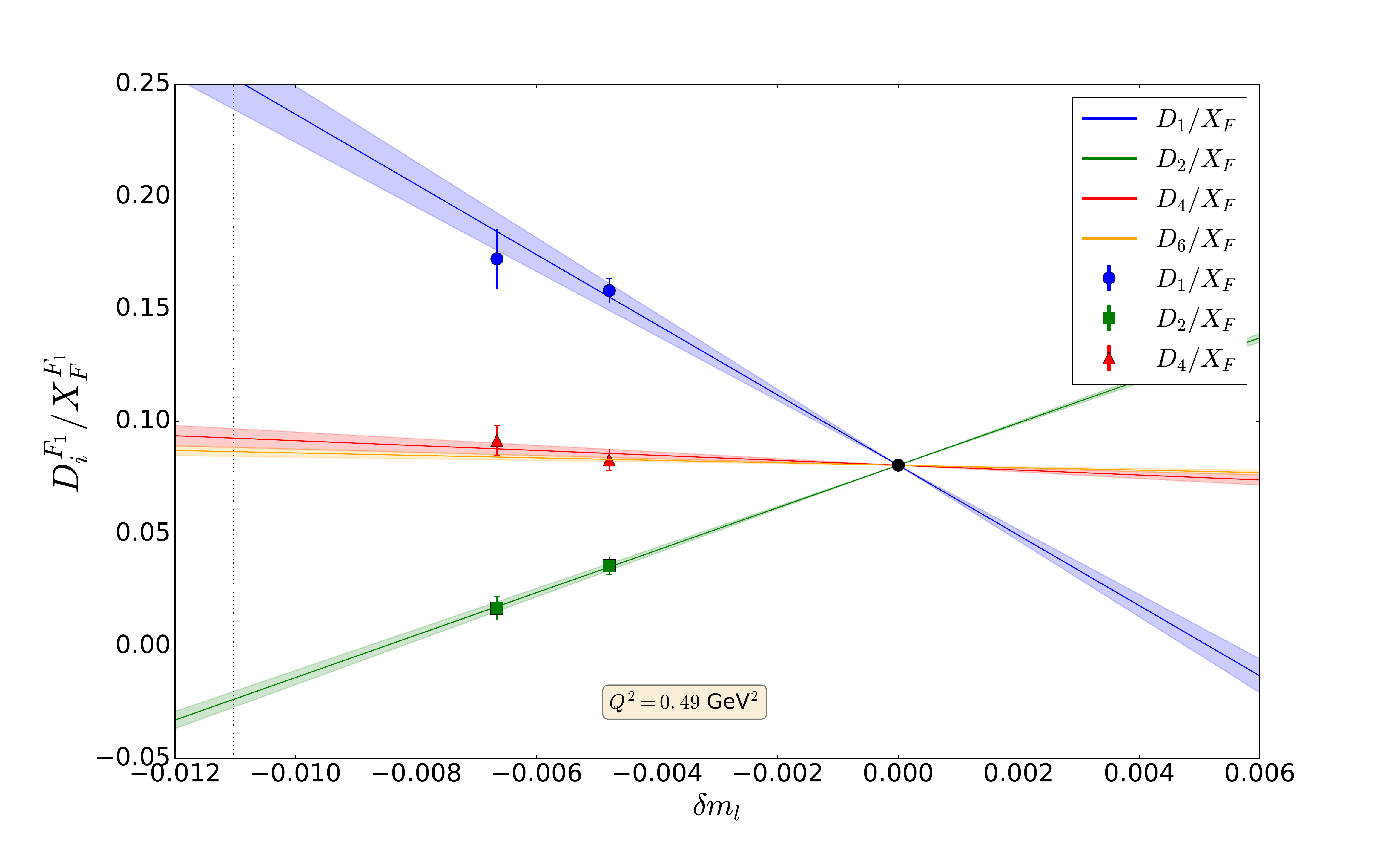} \\
   \includegraphics[width=14.50cm]{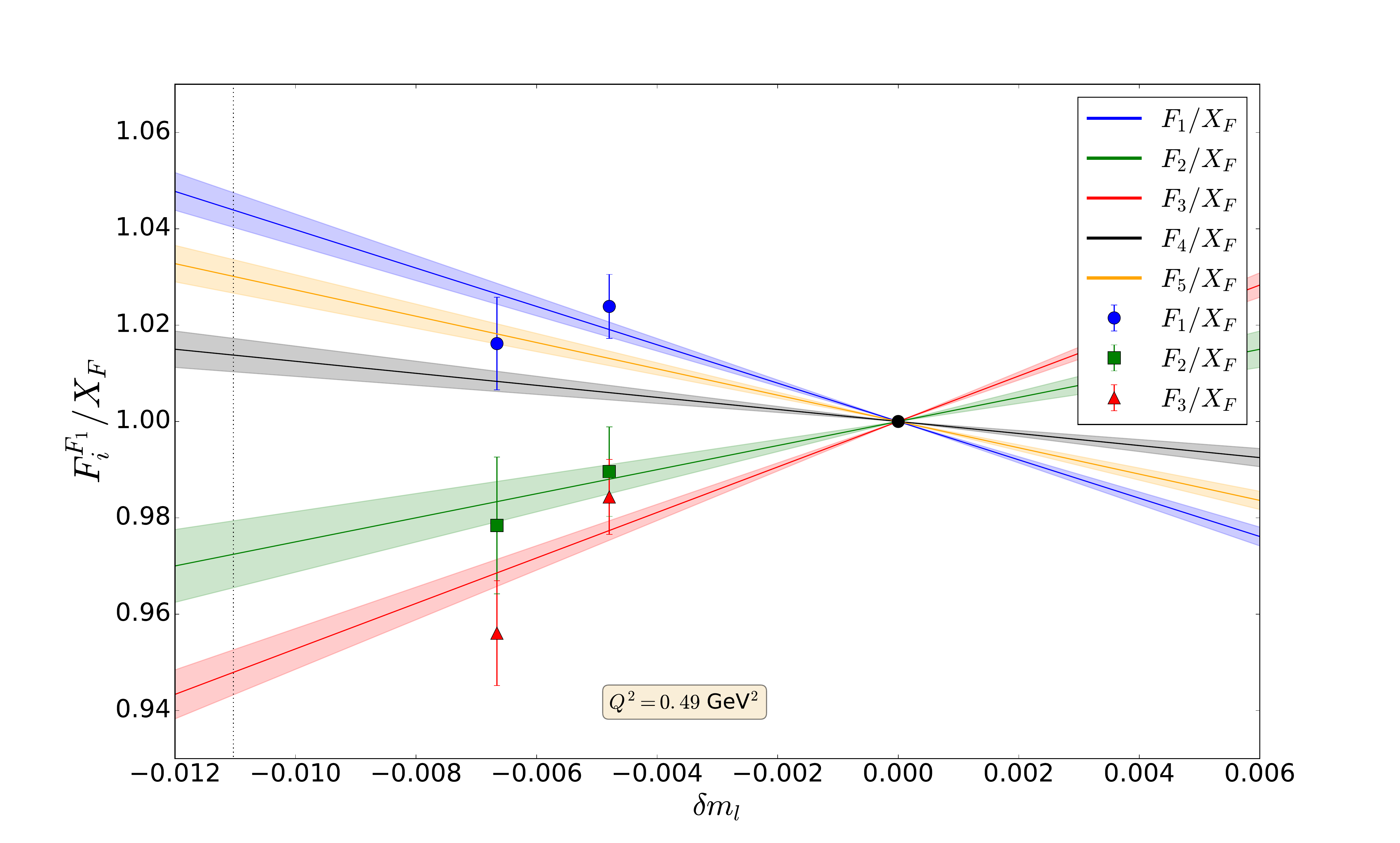}
   \end{tabular}
\caption{Top panel: $\tilde{D}_i^{F_1} \equiv D_i^{F_1}/X^{F_1}_F$
         for $i = 1$ (filled circles), $2$ (filled squares) and 
         $4$ (filled triangles) for $Q^2 = 0.49\,\mbox{GeV}^2$. 
         The three fits are from
         eq.~(\protect\ref{D_fan}), the line for $i=6$ is also
         shown. The vertical dotted line represents the physical point.
         Lower panel: $\tilde{F}_i^{F_1} \equiv F_i^{F_1}/X^{F_1}_F$ again at
         $Q^2 = 0.49\,\mbox{GeV}^2$ for $i = 1$ (filled circles), $2$
         (filled squares) and $3$ (filled triangles),
         together with fits from eq.~(\protect\ref{F_fan}) normalised
         by $X^{F_1}_F$. The line for $i=5$ is also shown.}
\label{fan_DF1_qeq0p49}
\end{figure}

We now turn to `fan' plots, as defined by eqs.~(\ref{D_fan}) and
(\ref{F_fan}). Note that again we only consider lattice quantities,
the improved operator would have small changes to the $SU(3)$ 
flavour-breaking expansion, as discussed in section~\ref{general_comments}.
Again we only consider diagonal form factors in
these equations: $D_1^{\ind con}$, $D_2^{\ind con}$ and $D_4$
in eq.~(\ref{D_fan}) and $F_1$, $F_2$ and $F_4$
in eq.~(\ref{F_fan}). We construct the system of linear equations in
eq.~(\ref{D_fan}) with parameters $r_1^{\ind con}$, $r_3$ and
$d$ for the $d$-fan and eq.~(\ref{F_fan}) with parameters
$s_1$, $s_2$ and $f$ for the $f$-fan.
In Fig.~\ref{fan_DF1_qeq0p49} we show $\tilde{D}_i^{F_1} = D_i^{F_1}/X_F$
for $i = 1$, $2$ and $4$ and $\tilde{F}_i^{F_1} = F_i^{F_1}/X_F$ for 
$i = 1$, $2$ and $3$. Note that as $d$ vanishes for the $F_1$ form factor
at $Q^2 = 0$, and even away from $Q^2 = 0$ it remains small, 
see the lower panel of Fig.~\ref{XDF_F1_Q2}, then dividing by $X^{F_1}_D$ 
is not possible or very noisy, so we use $X^{F_1}_F$. Although for 
$X^{F_2}_D$ this is not the case (as seen in Fig.~\ref{XFD_F2_Q2}) however
for consistency we still use  $X^{F_2}_F$. The only change in these
cases is that the value at the symmetric point is no longer one.

\begin{figure}[!tbp]
   \begin{tabular}{c}
   \includegraphics[width=14.50cm]{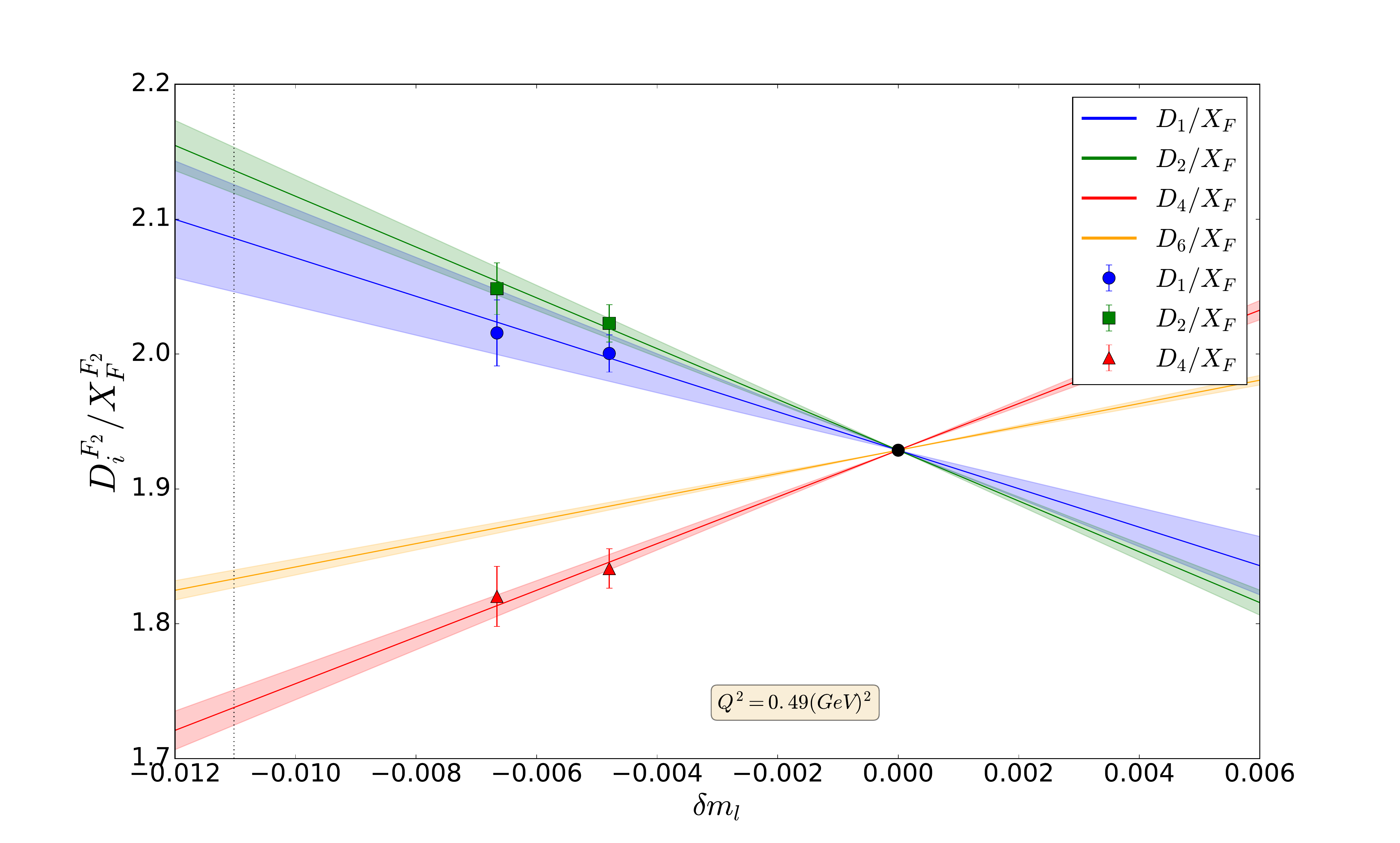} \\
   \includegraphics[width=14.50cm]{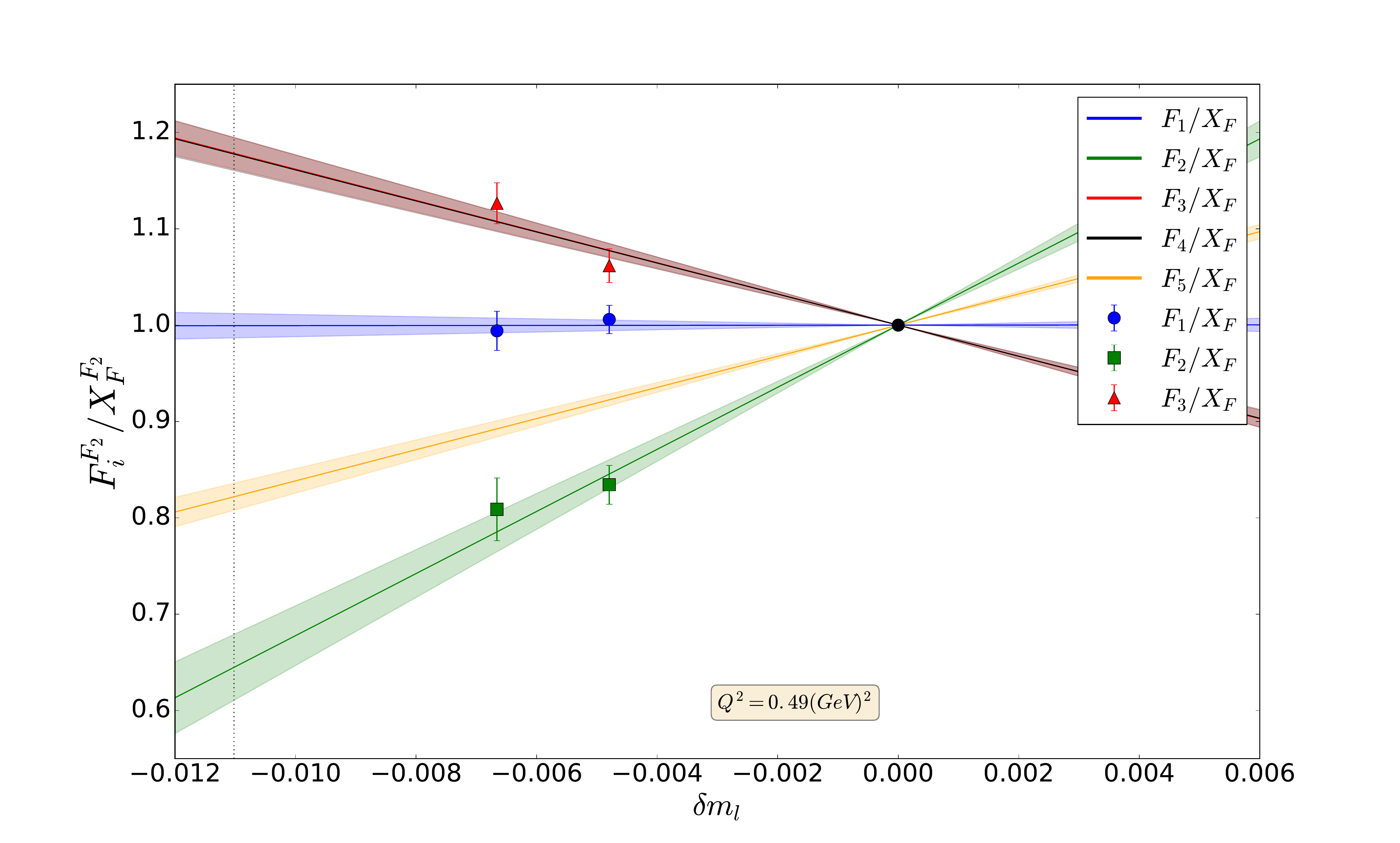}
   \end{tabular}
\caption{Top panel: $\tilde{D}_i^{F_2}$ for $i = 1$ (filled circles),
         $2$ (filled squares) and $4$ (filled triangles) for
         $Q^2 = 0.49\,\mbox{GeV}^2$. The three fits are from
         eq.~(\protect\ref{D_fan}) normalised by $X_D^{F_2}$,
         also shown is the $i=6$ line. The vertical dotted line
         represents the physical point.
         Lower panel: $\tilde{F}_i^{F_2}$ for
         $i = 1$ (filled circles), $2$ (filled squares) and $3$
         (filled triangles), also for $Q^2 = 0.49\,\mbox{GeV}^2$,
         together with fits from eq.~(\protect\ref{F_fan}) normalised
         by $X_F^{F_2}$. Also shown are the lines $i=4$ (upper line),
         $5$ (lower line).}
\label{fan_DF2_qeq0p49}
\end{figure}

The lines shown in Fig.~\ref{fan_DF2_qeq0p49} correspond to linear fits
to the $D_i^{F_1\,{\ind con}}$ using eq.~(\ref{D_fan}) (upper plot) and 
$F_i^{F_1\,{\ind con}}$ using eq.~(\ref{F_fan}) (lower plot). The fits to 
$D_i^{F_1\,{\ind con}}$ determine $r_1^{\ind con}$, $r_3$ using three fits
and are hence constrained. Furthermore determining these two parameters
also allows us to plot the off-diagonal hyperon decays for $i=6$, 
which is also shown. Similarly for $F_i^{F_1}$, we first determine 
the constrained fit parameters 
$\tilde{s}_1 = s_1/X_F$, $\tilde{s}_2= s_2/X_F$
and then plot the off-diagonal hyperon decays for $i=4$, $5$.

Similarly in Fig.~\ref{fan_DF2_qeq0p49} we show the equivalent
results for $F_2$. As previously we have normalised the parameters,
$\tilde{r}_1^{\ind con} = r_1^{\ind con}/X_F$,
$\tilde{r}_3= r_3/X_F$ and $\tilde{s}_1 = s_1/X_F$, 
$\tilde{s}_2= s_2/X_F$. Again we have some constraints. 
In addition off-diagonal hyperon decays for $i=6$, $d$-fan plot and 
$i=4$, $5$, $f$-fan plot are also shown.

\begin{figure}[!tbp]
   \begin{tabular}{c}
      \includegraphics[width=14.50cm]{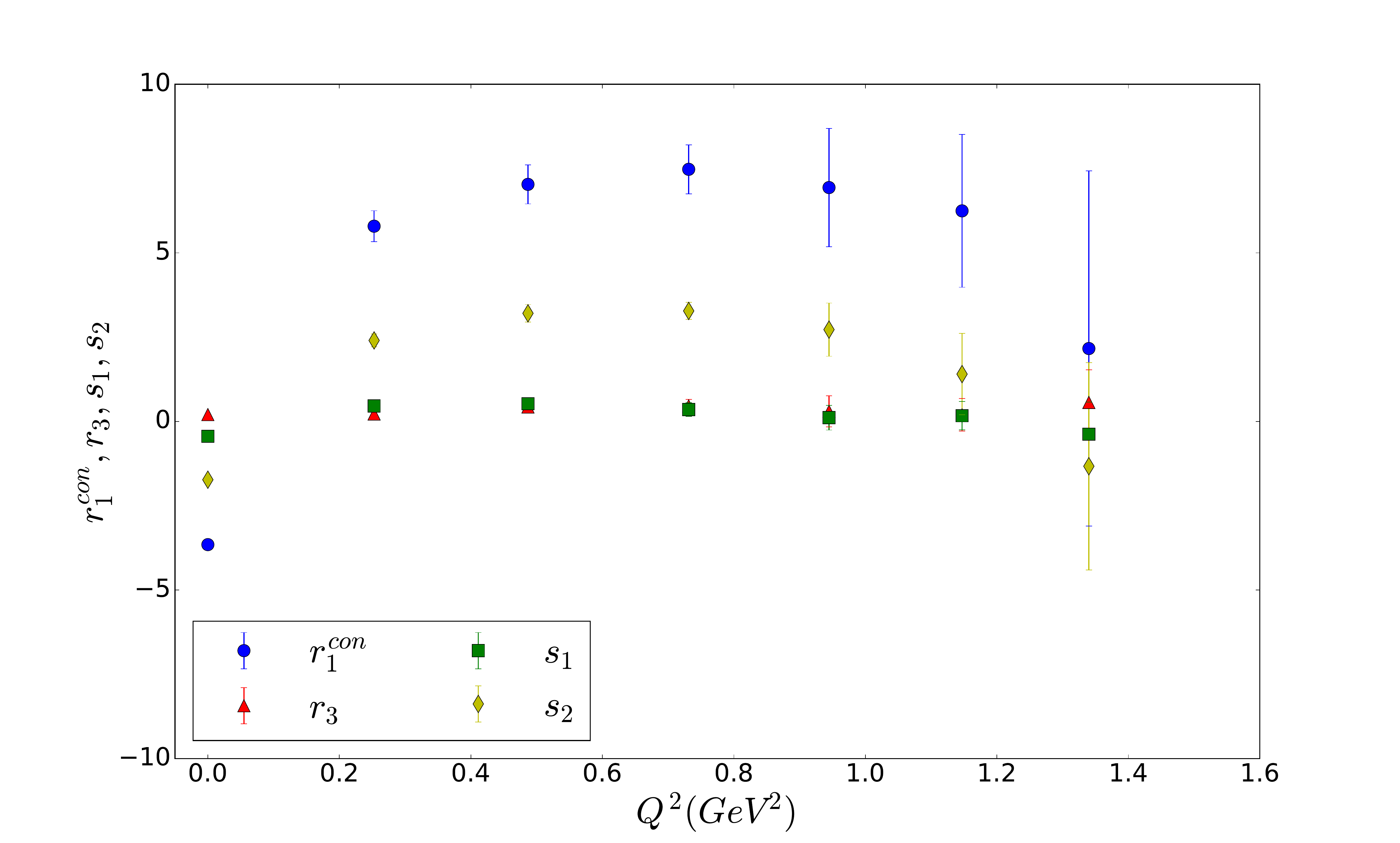} \\
      \includegraphics[width=14.50cm]{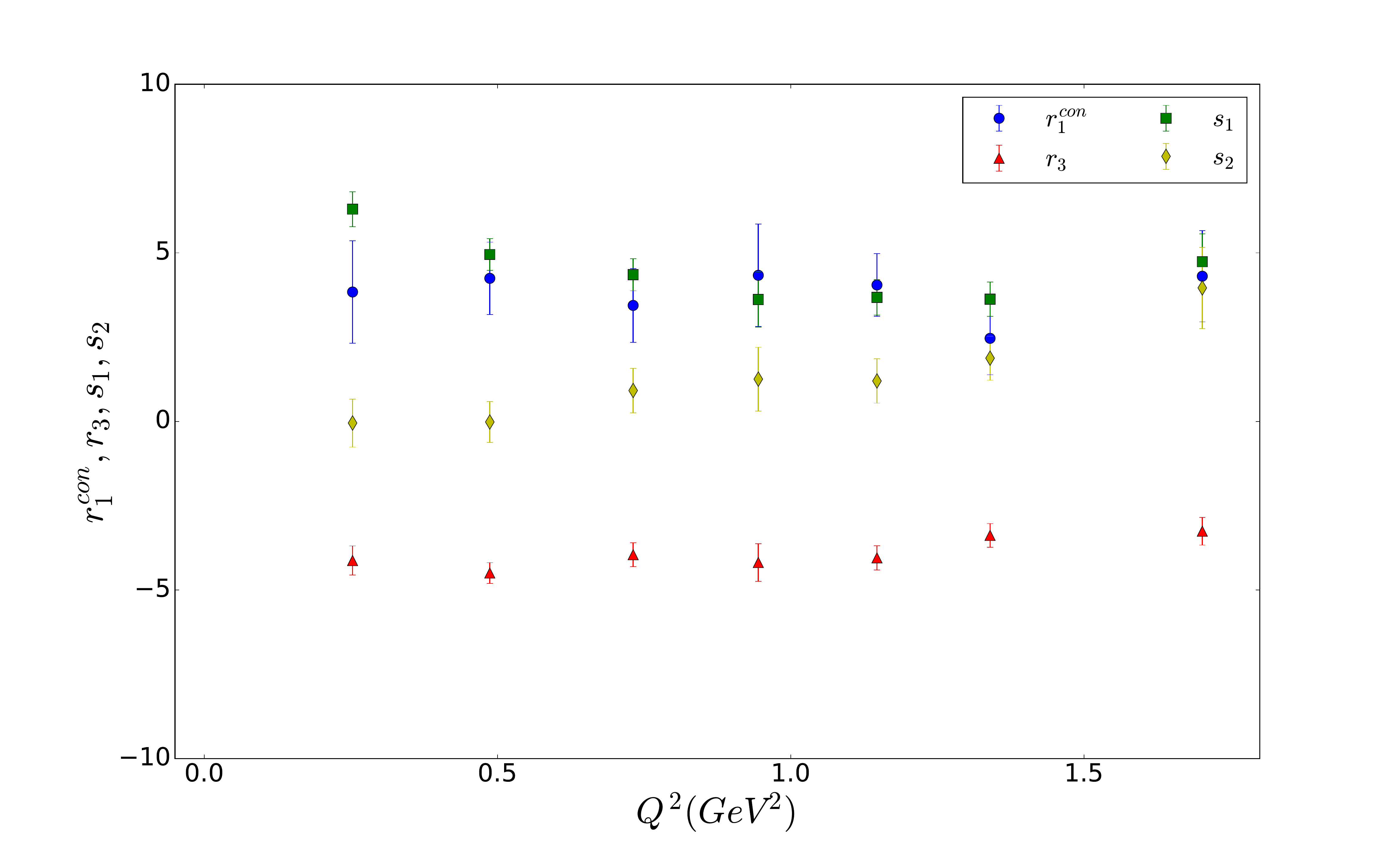}
   \end{tabular}
   \caption{Top panel: $r_1^{\ind con}$ (filled circles), 
            $r_3$ (filled triangles), $s_1$ (filled squares) and 
            $s_2$ (filled diamonds) expansion coefficients for the vector 
            $F_1^{\ind con}$ form factor as a function of $Q^2$.
            Lower panel: Similarly for the $F_2$ form factor.}
\label{r1r3s1s2vQ2}
\end{figure}

From these fan plots at various $Q^2$ we can determine the
dependence of the expansion coefficients as a function of $Q^2$.
In Fig.~\ref{r1r3s1s2vQ2} 
we show the expansion coefficients $r_1^{\ind con}$, $r_3$, $s_1$, $s_2$
for the $F_1^{\ind con}$ and $F_2$ form factors as function
of $Q^2$. As discussed previously in section~\ref{general_comments},
at $Q^2 = 0$ the expansion coefficients for  $F_1^{\ind con}$
vanish, which determines the improvement coefficients $b_V$, $f_V^{\ind con}$.
Thus in the top panel of Fig.~\ref{r1r3s1s2vQ2} the negative values 
of the $r_1^{\ind con}$, $s_1$, $s_2$ are a clear indication of the 
nature of the improvement coefficients. For rather small $Q^2$, 
these all change sign rather quickly and also their order inverts. 
We have (approximately) $|r_3|$, $|s_1| \approx 0$ and $|r_1^{\ind con}|$ 
is a factor of $2$--$4$ larger than $|s_2|$.
For $F_2$ the expansion coefficients tend to be flatter.
Also $s_2 \approx 0$, indicated in Fig.~(\ref{fan_DF2_qeq0p49})
by the small difference between $\tilde{F}_3^{F_2}$ and $\tilde{F}_4^{F_2}$.


\subsection{Estimating $\hat{Z}_V$ and $\hat{b}_V$, $\hat{f}_V^{\ind con}$}
\label{renorm+imp_res}


$X_F^{F_1}$ at $Q^2 = 0$ determines the renormalisation constant
$\hat{Z}_V$ via eq.~(\ref{ZV_determination}).
The constant fit described in eq.~(\ref{XF_def}) and shown in
Fig.~\ref{XDF_F1_Q2}, see also Fig.~\ref{X_FD_Q2}, leads to
$f=0.814(1)$ or
\begin{eqnarray}
  \hat{Z}_V = 0.869(1) \,.
\label{ZhatV+result}  
\end{eqnarray}
Our previous non-perturbative estimates of $Z_V$ at $\beta = 5.50$ are given
in \cite{Cundy:2008mb,Constantinou:2014fka} of $0.863(4)$, $0.857(1)$
respectively, and are quite close to $\hat{Z}_V$ in eq.~(\ref{ZhatV+result}).
Note that the different determinations can have $O(a)$ differences.
Also $\hat{Z}_V$ has been measured rather than $Z_V$. The difference
is $\sim 1 + b_V\bar{m}$. Here we have $b_V \sim O(1)$ and
$\bar{m} \sim 0.01$ (using the $\kappa_{0c}$ found in
\cite{bietenholz11a}), so there a further possible difference
(and reduction from the $\hat{Z}_V$ value) of $\sim 1\%$.

From Fig.~\ref{r1r3s1s2vQ2}, the $Q^2 = 0$ value
for $r_3$ is $0.06(2)$, which compared to other values
is compatible with zero. The $Q^2 = 0$ values for $s_1$, $s_2$ are
$s_1 = -0.479(22)$ and $s_2 = -1.643(44)$, respectively. The ratio is
$s_2/s_1 = 3.42$, which is in good agreement with the
theoretical value for the ratio from eq.~(\ref{bV_determination})
of $2\sqrt{3} \sim 3.46$.
Similarly, using eq.~(\ref{bV_determination}), we find
a weighted average of
\begin{eqnarray}
  \hat{b}_V = 1.174(21) \,,
\end{eqnarray}
which is about a $15\%$ increase from the tree-level value.
Although a strict comparison with other determinations
of this improvement coefficient is not possible,
it is interesting to note that
compared to other computations , e.g.\ \cite{Gerardin:2018kpy}
and for $n_f = 0$, $2$, \cite{Bakeyev:2003ff} the value determined
here is much closer to its tree-level value eq.~(\ref{tree_improvement}).
This suggests that improvement coefficients are small, including
possibly $\hat{c}_V$.

Using the value of $\hat{b}_V$ from $s_1$, $s_2$ and using
eq.~(\ref{fV_determination}) together with $r_1^{\ind con} = -3.65(8)$
gives a weighted average of
\begin{eqnarray}
   \hat{f}_V^{\ind con} = 0.041(4) \,.
\end{eqnarray}  
As expected this is quite small.


\subsection{Electromagnetic form factor results}


With a knowledge of $f$, $d$ and $r_1^{\ind con}$, $r_3$, $d$, $s_1$, $s_2$
we can find the electromagnetic Dirac form factor $F_1^{\ind con}(Q^2)$ 
and Pauli form factor $F_2^{\ind con}(Q^2)$ using the electromagnetic 
current $J_{{\rm em}\,\mu}^{\ind{con}}$ (see section~\ref{em_current}) 
and results of eq.~(\ref{EM_expansion}). Also we shall use $\hat{Z}_V$, 
$\hat{b}_V$ and $\hat{f}_V^{\ind con}$ (i.e.\ equivalent to CVC) 
from section~\ref{renorm+imp_res}.

\begin{figure}[!tbp]
   \begin{center}
     \includegraphics[width=14.50cm]{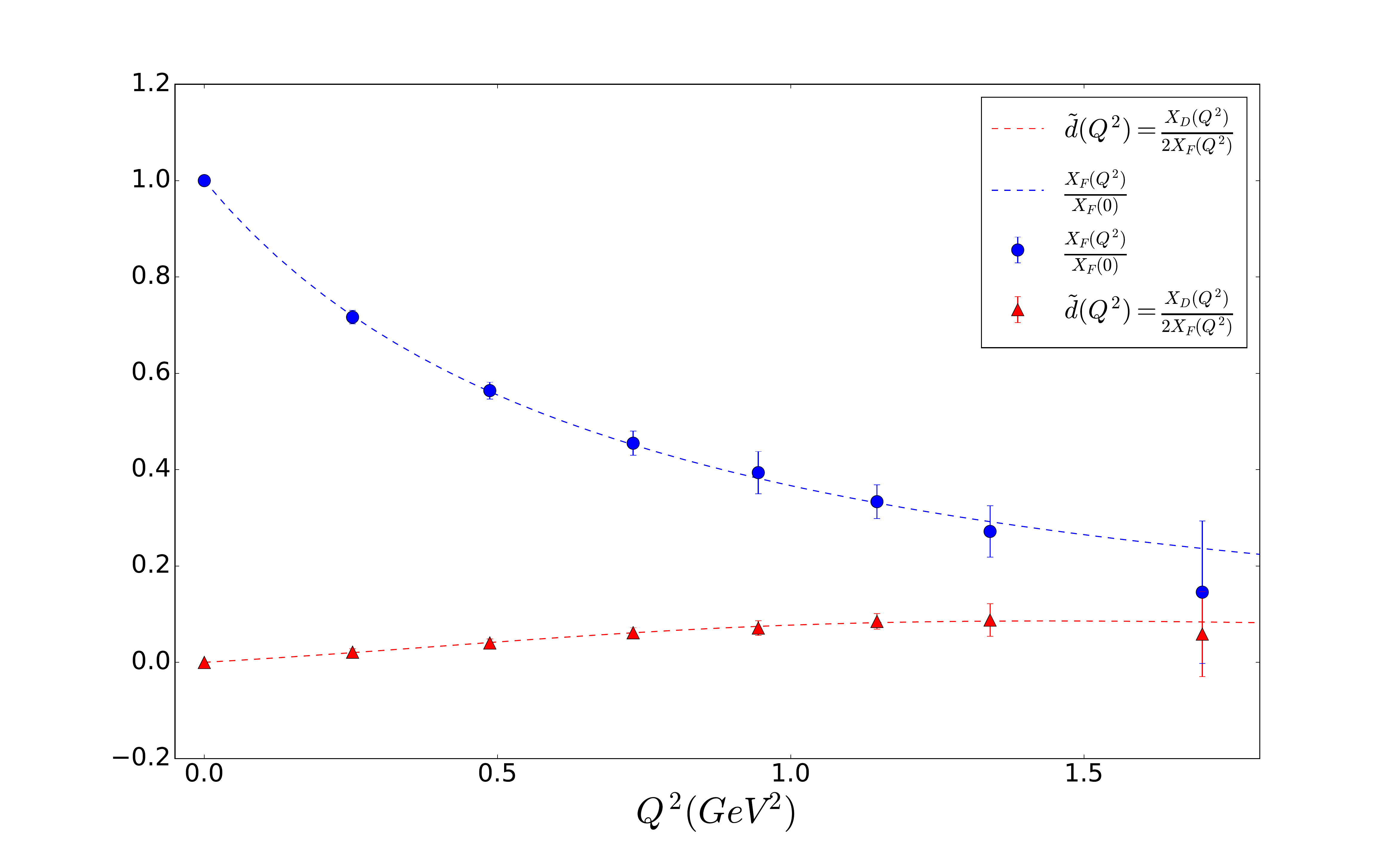}
   \end{center}
   \caption{$X_F(Q^2)/X_F(0)$ (filled circles) and
            $\tilde{d}(Q^2)$ (filled triangles) for $F_1^{\ind con}$
            against $Q^2$. The interpolation formulae used are given 
            in eq.~(\protect\ref{fit_fun}).}
\label{EM_FF}
\end{figure}

\begin{figure}[!tbp]
   \begin{center}
     \includegraphics[width=14.50cm]{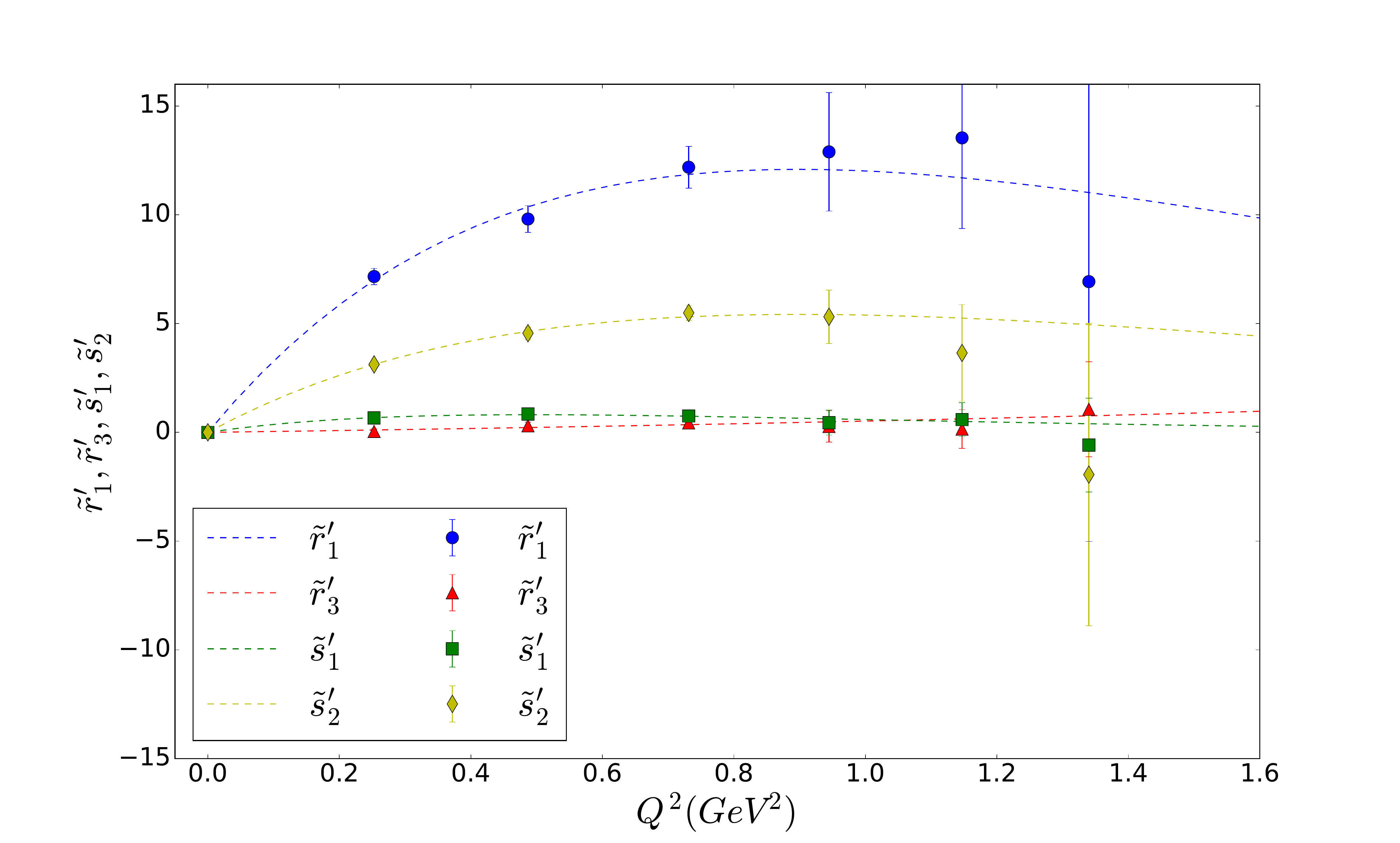}
   \end{center}
   \caption{$\tilde{r}_1^{\ind{con}\,\prime}$ (filled circles), 
            $\tilde{s}_2^{\,\prime}$ (filled diamonds), 
            $\tilde{s}_1^{\,\prime}$ (filled squares) and
            $\tilde{r}_3^{\,\prime}$ (filled triangles) against $Q^2$
            together with interpolation formulae also given by 
            eq.~(\protect\ref{fit_fun}).}
\label{EM_FF_expan}
\end{figure}

It is interesting to determine the various contributions to
the form factors from the expansion coefficients. For illustrative
purposes, we shall just consider $F_1^{\ind con}$ here and for $p$ and $\Xi^0$.
From eq.~(\ref{EM_expansion}) we can write
\begin{eqnarray}
   \langle p|J_{\rm em} | p\rangle^{\ind{con}\,{\ind R}}
      &=& {X_F(Q^2,\bar{m}) \over X_F(0, \bar{m})}
            \left[ 1 + {2\over \sqrt{3}}\tilde{d}(Q^2,\bar{m})
              + \tilde{\epsilon}_p^{\,\prime}(Q^2,\bar{m})\delta m_l \right] \,,
                                                           \nonumber \\
   \langle \Xi^0|J_{\rm em} | \Xi^0\rangle^{\ind{con}\,{\ind R}}
      &=& - {X_F(Q^2,\bar{m}) \over X_F(0, \bar{m})}
             \left[ {4\over \sqrt{3}}\tilde{d}(Q^2,\bar{m})
                     - \tilde{\epsilon}_{\Xi^0}^{\,\prime}(Q^2,\bar{m})\delta m_l)
             \right] \,,
\label{FF_phen}
\end{eqnarray}
with
\begin{eqnarray}
    \tilde{\epsilon}_p^{\,\prime}
      &=& {1 \over \sqrt{3}} (\tilde{r}_1^{\ind{con}\,\prime}
                               - \tilde{s}_2^{\,\prime})
           + 2(\tilde{s}_1^{\,\prime}
                             - \tilde{r}_3^{\,\prime}) \,,
                                                           \nonumber \\
    \tilde{\epsilon}_{\Xi^0}^{\,\prime}
      &=& {1 \over \sqrt{3}} (\tilde{r}_1^{\ind{con}\,\prime}
                               + \tilde{s}_2^{\,\prime})
           + 2(\tilde{s}_1^{\,\prime}
                             + \tilde{r}_3^{\,\prime}) \,,
\label{phen_expan}
\end{eqnarray}
where, for example,
$\tilde{r}_1^{\ind{con}\,\prime}
   = r_1^{\ind{con}\,\prime}(Q^2,\bar{m})/X_F(Q^2,\bar{m})$ and similarly for
the other expansion coefficients. The prime includes the improvement
terms, see eqs.~(\ref{stobV}, \ref{etar_stobV}). In this form, we can
investigate the contributions to the form factors.
In Fig.~\ref{EM_FF} we show the results for the terms of 
eq.~(\ref{FF_phen}): $X_F(Q^2)/X_F(0)$ and $\tilde{d}$.
In Fig.~\ref{EM_FF_expan} we show $\tilde{r}_1^{\ind{con}\,\prime}$,
$\tilde{s}_2^{\,\prime}$, $\tilde{r}_3^{\,\prime}$ and $\tilde{s}_1^{\,\prime}$.
All the interpolation formulae (fits) are of the form
\begin{eqnarray}
   {AQ^2 \over 1 + BQ^2 + C(Q^2)^2} \,.
\label{fit_fun}
\end{eqnarray}
From Fig.~\ref{EM_FF} and the leading term in eq.~(\ref{FF_phen})
for the proton form factor, the dominant contribution comes from
$X_F(Q^2)/X_F(0)$ -- the $f$ term, while there is a small contribution 
from the $d$ term (as $\tilde{d}$). Furthermore from Fig.~\ref{EM_FF_expan}
we see that for the $\tilde{\epsilon}$ coefficients, $\tilde{r}_3^{\,\prime}$
and $\tilde{s}_1^{\,\prime}$ are essentially negligible and most of the 
contribution comes from $\tilde{r}_1^{\ind{con}\,\prime}$ and 
$\tilde{s}_2^{\,\prime}$.

\begin{figure}[!tb]
   \includegraphics[width=14.50cm]{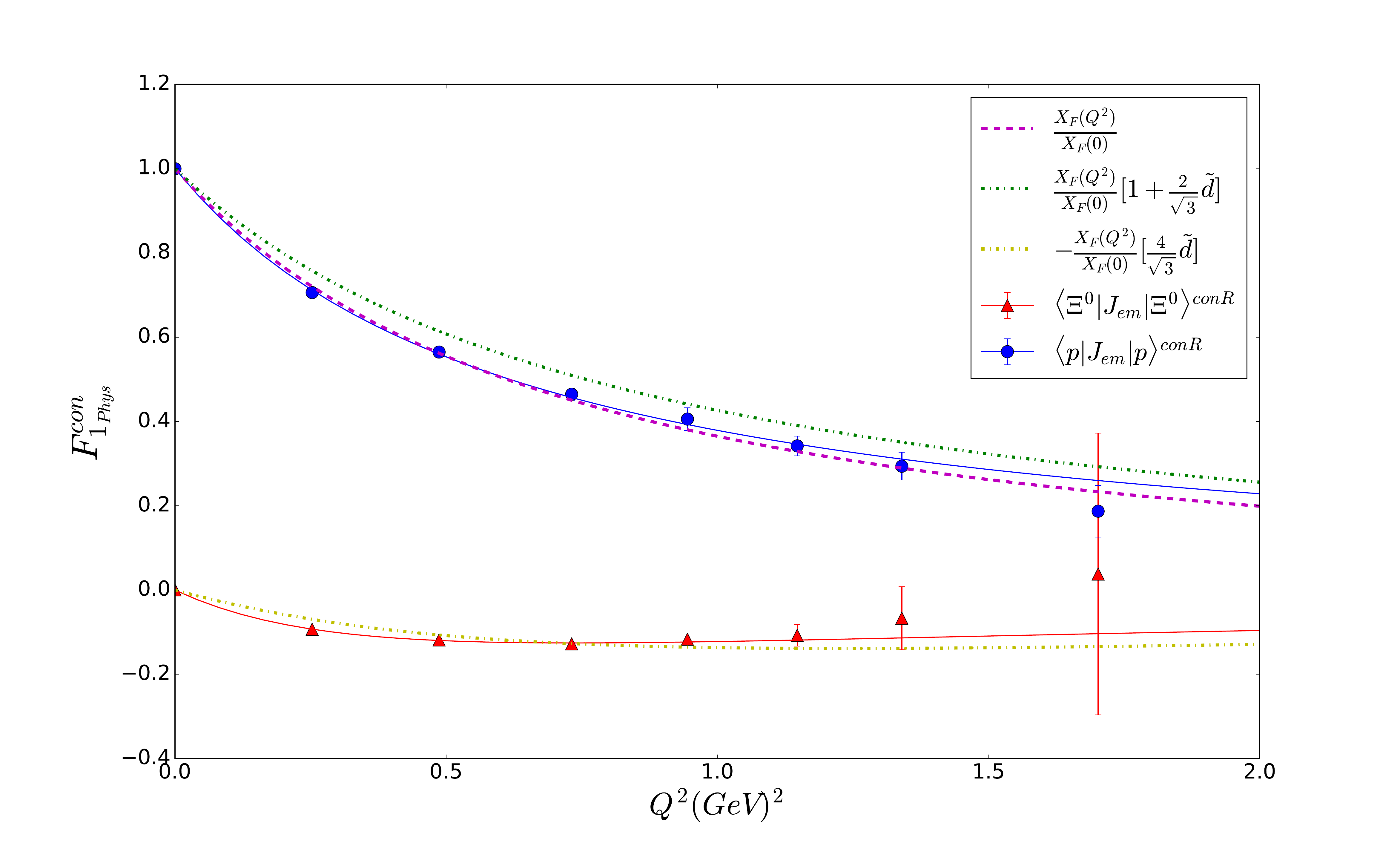}
   \caption{$F_1^{{\ind con}\,{\ind R}}$ for the proton (filled circles) and 
            $\Xi^0$ (filled triangles) at the physical point.
            The dashed line is $X_F(Q^2)/X_F(0)$. The dashed-dotted lines 
            are the  complete leading terms, for the proton:
            $X_F(Q^2,\bar{m})/X_F(0,\bar{m})
             (1 + 2/\sqrt{3}\tilde{d}(Q^2,\bar{m}))$
            and for $\Xi^0$: $X_F(Q^2,\bar{m})/X_F(0,\bar{m})\times 
            4/\sqrt{3}\tilde{d}(Q^2,\bar{m}))$, while the full lines 
            are the complete expressions in eq.~(\protect\ref{FF_phen}).}
\label{F_1_Jem}
\end{figure}

We illustrate this for the $F_1$ form factor for the $p$ and $\Xi^0$. 
In Fig.~\ref{F_1_Jem} we show 
$F_1^{{\ind con}\,{\ind R}}$ for these baryons at the physical point 
$\delta m_l^* = -0.01103$. i.e.\ a small and negative value. The dashed line
is $X_F(Q^2)/X_F(0)$, The dashed-dotted lines are the complete
leading terms: 
$X_F(Q^2,\bar{m})/X_F(0,\bar{m})(1 + 2/\sqrt{3}\tilde{d}(Q^2,\bar{m}))$
for $p$ and 
$X_F(Q^2,\bar{m})/X_F(0,\bar{m})\times4/\sqrt{3}\tilde{d}(Q^2,\bar{m}))$
for the $\Xi^0$, while the full lines are the complete expressions in
eq.~(\ref{FF_phen}). 

We see that for the proton the $f$ term (represented by 
$X_F(Q^2,\bar{m})/X_F(0,\bar{m})$) gives a result very close
to the numerical result; the addition of the $\tilde{d}$ term
pulls it slightly away in the $+$ve direction. The inclusion
of the $O(\delta m_l)$ term, being $-$ve pushes it back.
However the additional terms to the $f$ term contributes
very little (only a few percent) to the final result. 
For the $\Xi^0$ the $O(\delta m_l)$ term improves the agreement.


\section{Conclusions and outlook}
\label{conclusions}


In this article we have outlined a programme for investigating the
quark-mass behaviour of matrix elements, for $n_f = 2+1$ quark flavours
starting from a point on the $SU(3)$ flavour symmetric line when
the $u$, $d$ and $s$ quarks have the same mass and then following
a path keeping the singlet quark-mass constant. This is an extension
of our original programme for masses, \cite{bietenholz10a,bietenholz11a},
using a generalisation of the techniques developed there. 

When flavour $SU(3)$ is unbroken all baryon matrix elements of
a given operator octet can be expressed in terms of just two couplings
($f$ and $d$), as is well known. We find that when $SU(3)$ flavour symmetry
is broken, at LO and NLO, the expansions are constrained 
(but not at further higher orders).
By this we mean that there are a large number of relations between
the expansion coefficients. Our main results for the expansions are 
contained in sections~\ref{LO_coeff_tables} and \ref{higher_order}.
Although we concentrated on the $n_f = 2+1$ case, in which symmetry 
breaking is due to mass differences between the strange and light quarks
our methods are also applicable to isospin-breaking effects coming 
from a non-zero $m_d-m_u$ along the lines of \cite{horsley12a,horsley14a}.

The results here parallel those for the mass case. Firstly, for example
we have constructed `singlet-like' matrix elements -- collectively called
$X$ here -- where the LO term vanishes. As noted in
\cite{bietenholz11a} these can be extrapolated to the physical point,
using a one-parameter constant fit. In this article we constructed
several of these $X$ functions, and indeed can isolate the
constant as either the $f$ or $d$ coupling. Secondly again in analogy
to the mass expansions we constructed `fan' plots, each element of
which is a linear combination of matrix elements, where at the $SU(3)$
flavour symmetric point all the elements have a common value, and
then radiate away from this point as the quark masses change.
This is slightly more complicated than for the mass case as we
now have two couplings, $f$ and $d$. Indeed the `fan' plot expansions
can be constructed involving either $f$ or $d$ alone at the $SU(3)$
flavour symmetric point (more generally we have some combination
of them).

Technically important for lattice determinations of matrix elements is the
difference between quark-line-connected and quark-line-disconnected
terms in the calculation of the three-point correlation functions.
(The quark-line-disconnected terms are small, but difficult to
compute using lattice methods, due to large gluon fluctuations.)
Applying the $SU(3)$ flavour breaking expansion to these cases separately,
we have identified which expansion coefficient(s) have contributions
coming from the quark-line-disconnected terms. We found that at LO
there is just one expansion coefficient which has a 
quark-line-disconnected piece.

As numerically we are using Wilson clover improved fermions, then
for $O(a^2)$ continuum expansions, improvement coefficients need
to be determined. The general structure for $n_f = 2+1$ flavours
of fermions has been determined, see e.g.\ \cite{Bhattacharya:2005rb}.
We showed here these coefficients are equivalent to modifications
to the expansion parameters. 
Using the subsidiary condition that the relation between the local
and conserved vector current is $O(a)$ allowed us to determine two
improvement terms (together with the renormalisation constant).

To demonstrate how the expansions work, we discussed numerical
results using the vector current and diagonal matrix elements.
However these can be extended to include transition hyperon decays
(a phenomenological review is given in \cite{cabibbo03b}). These
would allow an alternative method to the standard $K_{\ell 3}$ decays
of determining  $|V_{us}|$, e.g. \cite{cabibbo03a,cabibbo03b,mateu05a}.
Earlier quenched and $n_f=2$ results for $\Sigma^-\to n\ell\nu$ and 
$\Xi^0\to\Sigma^+\ell\nu$ can be found in \cite{guadagnoli06a,sasaki08a}, 
and $n_f=2+1$ results have been obtained in \cite{sasaki12a,Sasaki:2017jue}.
The latter reference also investigates the possibility of non-linear
effects in the quark-mass, which in the $SU(3)$ symmetry 
flavour-breaking expansion means including terms from Table~\ref{coef_27_64}.

Future theoretical developments include extending the formalism to
partially quenched quark masses, when the valence quark mass, $\delta\mu_q$,
does not have to be the same as the sea or unitary quark mass.
Then eq.~(\ref{dml_def}) is replaced by $\delta\mu_q = \mu_q - \bar{m}$.
In this case the generalisation of eq.~(\ref{zerosum21}) does not hold.
This allows the determination of the expansion coefficients over
a larger quark mass range than is possible using the unitary
quark masses (and allows, for example, the charm quark to
be included, \cite{Horsley:2017azo}).
Furthermore expansions for `fake' hadrons would
be useful. Possible are a `nucleon' with three mass degenerate
strange quarks and a `Lambda' with two mass degenerate strange
quarks. Although they are not physical states, they can be
measured on the lattice, and do not introduce any more
$SU(3)$ mass flavour-breaking expansion coefficients, so
simply add more constraints to the coefficient determination.
An example of this for the baryon octet masses is given in
\cite{horsley14a}.

Another extension of the $SU(3)$ mass flavour breaking method
is to the baryon decuplet with $10\otimes 8 \otimes 10$ tensors,
and also to the meson octet. While the latter extension
is straightforward, there are some extra constraints,
as due to charge conjugation, the particles in the
meson octet are related to each other.

Furthermore generalised currents can be evaluated between quark
states. This leads to a $SU(3)$ mass flavour-breaking expansion
involving $3\otimes 8 \otimes 3$ tensors. This will help
when considering the non-perturbative $RI^\prime - MOM$ scheme which
defines the renormalisation constants (and improvement constants)
by considering the generalised currents between quark states.
Useful would also be to consider the axial current improvement coefficients
using a partially conserved axial-vector current (PCAC) along the 
lines of \cite{Bhattacharya:2005rb}.

Finally, a more distant prospect is to include QED corrections
to the matrix elements, \cite{QCDSF}, along the lines of our previous 
studies of the $SU(3)$ flavour-breaking expansion for masses,
\cite{Horsley:2015eaa,Horsley:2015vla,Horsley:2019wha}.


\section*{Acknowledgements}


The numerical configuration generation (using the BQCD lattice 
QCD program \cite{Haar:2017ubh})) and data analysis 
(using the Chroma software library \cite{edwards04a}) was carried out
on the IBM BlueGene/Q and HP Tesseract using DIRAC 2 resources 
(EPCC, Edinburgh, UK), the IBM BlueGene/Q (NIC, J\"ulich, Germany) 
and the Cray XC40 at HLRN (The North-German Supercomputer 
Alliance), the NCI National Facility in Canberra, Australia 
(supported by the Australian Commonwealth Government) 
and Phoenix (University of Adelaide).
We would like to thank Ashley Cooke for useful
discussions at an early stage in this project.
RH was supported by STFC through grant ST/P000630/1.
HP was supported by DFG Grant No. PE 2792/2-1.
PELR was supported in part by the STFC under contract ST/G00062X/1.
GS was supported by DFG Grant No. SCHI 179/8-1.
RDY and JMZ were supported by the Australian Research Council Grants
FT120100821, FT100100005, DP140103067 and DP190100297. 
We thank all funding agencies.



\appendix

\section*{Appendix}


\section{Non-zero tensor elements}
\label{non-zero_tensor}


 The non-zero elements of the tensors $T_{ijk}$ are listed in 
 Tables~\ref{fd_appendix} -- \ref{zx1y1_appendix}.
 \begin{table}[!htb]
 \begin{small}  
 \begin{center} 
 \begin{tabular}{crl}
 tensor & value\ \ & \ \ position \\ \hline \\[-0.9em]
 &$2 $ \q & $ 334 \q 463 \q 646 $ \cr
 &$-2 $ \q & $ 343 \q 436 \q 664 $ \cr
 &$\sqrt{3} $ \q & $ 151 \q 252 \q 518 \q 527 \q 775 \q 885 $ \cr
 $f$
 &$-\sqrt{3} $ \q & $ 115 \q 225 \q 572 \q 581 \q 757 \q 858 $ \cr
 &$\sqrt{2} $ \q & $ 132 \q 261 \q 317 \q 628 \q 783 \q 876 $ \cr
 &$-\sqrt{2} $ \q & $ 123 \q 216 \q 371 \q 682 \q 738 \q 867 $ \cr
 &$1 $ \q & $ 114 \q 242 \q 427 \q 481 \q 774 \q 848 $ \cr
 &$-1 $ \q & $ 141 \q 224 \q 418 \q 472 \q 747 \q 884 $ \cr
 \hline \\[-0.9em]
 &$\sqrt{6} $ \q & $ 123 \q 132 \q 216 \q 261 \q 317 \q 371 $ \cr
 & \q & $ 628 \q 682 \q 738 \q 783 \q 867 \q 876 $ \cr
 &$2 $ \q & $ 335 \q 353 \q 445 \q 454 \q 536 \q 544 \q 563 \q 656 \q 665 $ \cr
 $d$
 &$-2 $ \q & $ 555 $ \cr
 &$\sqrt{3} $ \q & $ 224 \q 242 \q 427 \q 472 \q 747 \q 774 $ \cr
 &$-\sqrt{3} $ \q & $ 114 \q 141 \q 418 \q 481 \q 848 \q 884 $ \cr
 &$-1 $ \q & $ 115 \q 151 \q 225 \q 252 \q 518 \q 527 $ \cr
 & \q & $ 572 \q 581 \q 757 \q 775 \q 858 \q 885 $ \cr
 \hline \\[-0.9em]
 \end{tabular} \end{center}
 \end{small}
 \caption{Flavour-singlet first-class non-zero elements of the $f$ and 
          $d$ tensors.}
 \label{fd_appendix}
 \end{table}
 
 \begin{table}[!htb]
 \begin{small}
 \begin{center} 
 \begin{tabular}{crl}
 tensor & value\ \ & \ \ position \\ \hline \\[-0.9em]
 $r_1$
 &$1 $ \q & $ 151 \q 252 \q 353 \q 454 \q 555 \q 656 \q 757 \q 858 $ \cr
 \hline \\[-0.9em]
 $r_2$
 &$2 $ \q & $ 555 $ \cr
 &$1 $ \q & $ 115 \q 225 \q 335 \q 445 \q 518 \q 527 \q 536 $ \cr
 & \q & $ 544 \q 563 \q 572 \q 581 \q 665 \q 775 \q 885 $ \cr
 \hline \\[-0.9em]
 &$2 \sqrt{3} $ \q & $ 353 \q 454 \q 656 $ \cr
 &$-2 \sqrt{2} $ \q & $ 132 \q 261 \q 738 \q 867 $ \cr
 &$2 $ \q & $ 141 \q 848 $ \cr
 $r_3$
 &$-2 $ \q & $ 242 \q 747 $ \cr
 &$-\sqrt{3} $ \q & $ 335 \q 445 \q 536 \q 544 \q 563 \q 665 $ \cr
 &$\sqrt{2} $ \q & $ 123 \q 216 \q 317 \q 371 \q 628 \q 682 \q 783 \q 876 $ \cr
 &$1 $ \q & $ 224 \q 427 \q 472 \q 774 $ \cr
 &$-1 $ \q & $ 114 \q 418 \q 481 \q 884 $ \cr
 \hline \\[-0.9em]
 &$2 \sqrt{2} $ \q & $ 132 \q 261 $ \cr
 &$-2 \sqrt{2} $ \q & $ 738 \q 867 $ \cr
 &$2 $ \q & $ 242 \q 343 \q 436 \q 664 \q 848 $ \cr
 &$-2 $ \q & $ 141 \q 334 \q 463 \q 646 \q 747 $ \cr
 $s_1$
 &$\sqrt{3} $ \q & $ 518 \q 527 \q 775 \q 885 $ \cr
 &$-\sqrt{3} $ \q & $ 115 \q 225 \q 572 \q 581 $ \cr
 &$\sqrt{2} $ \q & $ 123 \q 216 \q 371 \q 682 $ \cr
 &$-\sqrt{2} $ \q & $ 317 \q 628 \q 783 \q 876 $ \cr
 &$1 $ \q & $ 224 \q 418 \q 472 \q 884 $ \cr
 &$-1 $ \q & $ 114 \q 427 \q 481 \q 774 $ \cr
 \hline \\[-0.9em]
 &$\sqrt{3} $ \q & $ 334 \q 463 \q 646 $ \cr
 $s_2$
 &$-\sqrt{3} $ \q & $ 343 \q 436 \q 664 $ \cr
 &$1 $ \q & $ 115 \q 225 \q 572 \q 581 \q 757 \q 858 $ \cr
 &$-1 $ \q & $ 151 \q 252 \q 518 \q 527 \q 775 \q 885 $ \cr
 \hline \\[-0.9em]
 \end{tabular} \end{center}
 \end{small}
 \caption{First-class octet non-zero elements of the 
          $r_1$, $r_2$, $r_3$ and $s_1$, $s_2$ tensors.}
 \end{table}

 \begin{table}[!htb]
 \begin{small}
 \begin{center} 
 \begin{tabular}{crl}
 tensor & value\ \ & \ \ position \\ \hline \\[-0.9em]
 $t_1$
 &$1 $ \q & $ 115 \q 225 \q 335 \q 445 \q 665 \q 775 \q 885 $ \cr
 &$-1 $ \q & $ 518 \q 527 \q 536 \q 544 \q 563 \q 572 \q 581 $ \cr
 \hline \\[-0.9em]
 &$\sqrt{3} $ \q & $ 115 \q 225 \q 775 \q 885 $ \cr
 &$-\sqrt{3} $ \q & $ 518 \q 527 \q 572 \q 581 $ \cr
 $t_2$
 &$\sqrt{2} $ \q & $ 123 \q 216 \q 783 \q 876 $ \cr
 &$-\sqrt{2} $ \q & $ 317 \q 371 \q 628 \q 682 $ \cr
 &$1 $ \q & $ 224 \q 418 \q 481 \q 774 $ \cr
 &$-1 $ \q & $ 114 \q 427 \q 472 \q 884 $ \cr
 \hline \\[-0.9em]
 &$\sqrt{6} $ \q & $ 123 \q 216 \q 317 \q 628 $ \cr
 &$-\sqrt{6} $ \q & $ 371 \q 682 \q 783 \q 876 $ \cr
 $u_1$
 &$\sqrt{3} $ \q & $ 224 \q 427 \q 481 \q 884 $ \cr
 &$-\sqrt{3} $ \q & $ 114 \q 418 \q 472 \q 774 $ \cr
 &$1 $ \q & $ 572 \q 581 \q 775 \q 885 $ \cr
 &$-1 $ \q & $ 115 \q 225 \q 518 \q 527 $ \cr
 \hline \\[-0.9em]
 \end{tabular} \end{center}
 \end{small}
 \caption{Second-class octet non-zero elements of the $t_1$, $t_2$ 
          and $u_1$ tensors.}
 \end{table}

 \begin{table}[!htb]
 \begin{small}
 \begin{center} 
 \begin{tabular}{crl}
 tensor & value\ \ & \ \ position \\ \hline \\[-0.9em]
 &$-18 $ \q & $ 555 $ \cr
 &$14 $ \q & $ 335 \q 445 \q 536 \q 544 \q 563 \q 665 $ \cr
 &$-5 \sqrt{6} $ \q & $ 132 \q 261 \q 738 \q 867 $ \cr
 $q_1$
 &$9 $ \q & $ 151 \q 252 \q 757 \q 858 $ \cr
 &$5 \sqrt{3} $ \q & $ 141 \q 848 $ \cr
 &$-5 \sqrt{3} $ \q & $ 242 \q 747 $ \cr
 &$-6 $ \q & $ 115 \q 225 \q 353 \q 454 \q 518 \q 527 $ \cr
 & \q & $ 572 \q 581 \q 656 \q 775 \q 885 $ \cr
 \hline \\[-0.9em]
 &$18 $ \q & $ 555 $ \cr
 &$-10 $ \q & $ 353 \q 454 \q 656 $ \cr
 &$-6 $ \q & $ 335 \q 445 \q 536 \q 544 \q 563 \q 665 $ \cr
 &$2 \sqrt{6} $ \q & $ 123 \q 216 \q 317 \q 371 \q 628 \q 682 \q 783 \q 876 $ \cr
 $q_2$
 &$2 \sqrt{3} $ \q & $ 224 \q 427 \q 472 \q 774 $ \cr
 &$-2 \sqrt{3} $ \q & $ 114 \q 418 \q 481 \q 884 $ \cr
 &$3 $ \q & $ 151 \q 252 \q 757 \q 858 $ \cr
 &$\sqrt{6} $ \q & $ 132 \q 261 \q 738 \q 867 $ \cr
 &$\sqrt{3} $ \q & $ 242 \q 747 $ \cr
 &$-\sqrt{3} $ \q & $ 141 \q 848 $ \cr
 \hline \\[-0.9em]
 &$4 \sqrt{2} $ \q & $ 132 \q 261 $ \cr
 &$-4 \sqrt{2} $ \q & $ 738 \q 867 $ \cr
 &$3 \sqrt{3} $ \q & $ 115 \q 225 \q 572 \q 581 $ \cr
 &$-3 \sqrt{3} $ \q & $ 518 \q 527 \q 775 \q 885 $ \cr
 $w_1$
 &$3 \sqrt{2} $ \q & $ 317 \q 628 \q 783 \q 876 $ \cr
 &$-3 \sqrt{2} $ \q & $ 123 \q 216 \q 371 \q 682 $ \cr
 &$4 $ \q & $ 242 \q 343 \q 436 \q 664 \q 848 $ \cr
 &$-4 $ \q & $ 141 \q 334 \q 463 \q 646 \q 747 $ \cr
 &$3 $ \q & $ 114 \q 427 \q 481 \q 774 $ \cr
 &$-3 $ \q & $ 224 \q 418 \q 472 \q 884 $ \cr
 \hline \\[-0.9em]
 &$3 \sqrt{3} $ \q & $ 151 \q 252 $ \cr
 &$-3 \sqrt{3} $ \q & $ 757 \q 858 $ \cr
 &$2 \sqrt{2} $ \q & $ 123 \q 216 \q 371 \q 682 $ \cr
 &$-2 \sqrt{2} $ \q & $ 317 \q 628 \q 783 \q 876 $ \cr
 $w_2$
 &$2 $ \q & $ 224 \q 334 \q 418 \q 463 \q 472 \q 646 \q 884 $ \cr
 &$-2 $ \q & $ 114 \q 343 \q 427 \q 436 \q 481 \q 664 \q 774 $ \cr
 &$\sqrt{2} $ \q & $ 738 \q 867 $ \cr
 &$-\sqrt{2} $ \q & $ 132 \q 261 $ \cr
 &$1 $ \q & $ 141 \q 747 $ \cr
 &$-1 $ \q & $ 242 \q 848 $ \cr
 \hline \\[-0.9em]
 \end{tabular} \end{center}
 \end{small}
 \caption{First-class $27$-plet non-zero elements of the 
          $q_1$, $q_2$ and $w_1$, $w_2$ tensors.}
 \end{table}

 \begin{table}[!htb]
 \begin{small}
 \begin{center} 
 \begin{tabular}{crl}
 tensor & value\ \ & \ \ position \\ \hline \\[-0.9em]
 &$-9 \sqrt{3} $ \q & $ 555 $ \cr
 &$3 \sqrt{3} $ \q & $ 115 \q 151 \q 225 \q 252 \q 518 \q 527 $ \cr
 & \q & $ 572 \q 581 \q 757 \q 775 \q 858 \q 885 $ \cr
 $z$
 &$-\sqrt{3} $ \q & $ 335 \q 353 \q 445 \q 454 \q 536 \q 544 \q 563 \q 656 \q 665 $ \cr
 &$\sqrt{2} $ \q & $ 123 \q 132 \q 216 \q 261 \q 317 \q 371 $ \cr
 & \q & $ 628 \q 682 \q 738 \q 783 \q 867 \q 876 $ \cr
 &$1 $ \q & $ 224 \q 242 \q 427 \q 472 \q 747 \q 774 $ \cr
 &$-1 $ \q & $ 114 \q 141 \q 418 \q 481 \q 848 \q 884 $ \cr
 \hline \\[-0.9em]
 &$4 $ \q & $ 335 \q 445 \q 665 $ \cr
 &$-4 $ \q & $ 536 \q 544 \q 563 $ \cr
 &$3 $ \q & $ 518 \q 527 \q 572 \q 581 $ \cr
 $x_1$
 &$-3 $ \q & $ 115 \q 225 \q 775 \q 885 $ \cr
 &$\sqrt{6} $ \q & $ 123 \q 216 \q 783 \q 876 $ \cr
 &$-\sqrt{6} $ \q & $ 317 \q 371 \q 628 \q 682 $ \cr
 &$\sqrt{3} $ \q & $ 224 \q 418 \q 481 \q 774 $ \cr
 &$-\sqrt{3} $ \q & $ 114 \q 427 \q 472 \q 884 $ \cr
 \hline \\[-0.9em]
 &$3 \sqrt{3} $ \q & $ 115 \q 225 \q 518 \q 527 $ \cr
 &$-3 \sqrt{3} $ \q & $ 572 \q 581 \q 775 \q 885 $ \cr
 $y_1$
 &$\sqrt{2} $ \q & $ 123 \q 216 \q 317 \q 628 $ \cr
 &$-\sqrt{2} $ \q & $ 371 \q 682 \q 783 \q 876 $ \cr
 &$1 $ \q & $ 224 \q 427 \q 481 \q 884 $ \cr
 &$-1 $ \q & $ 114 \q 418 \q 472 \q 774 $ \cr
 \hline \\[-0.9em]
 \end{tabular} \end{center}
 \end{small}
 \caption{First-class $64$-plet and second-class $27$-plet non-zero 
          elements of the $z$ and $x_1$, $y_1$ tensors.}
 \label{zx1y1_appendix}
 \end{table}


\section{Alternative fan plots}
\label{doub-sing_rep}


\subsection{The doubly represented $-$ singly represented fan, the $P$-fan}


The traditional way of expressing the two ways of coupling octet
operators to octet hadrons are the $f$ and $d$ couplings. 
In terms of hadron structure, this choice is perhaps more natural 
for octet mesons than it is for octet baryons. Consider the 
eqs.~(\ref{fcombo}, \ref{dcombo}). In the $K^+$, with 
quark content $u \bar s$ the $f$ combination
$\langle K^+ |(\bar u \gamma u - \bar s \gamma s)| K^+\rangle$ 
is very natural (the difference between the two valence quarks), 
and the $d$ combination
$\langle K^+ |(\bar u \gamma u + \bar s \gamma s - 2 \bar d \gamma d) 
| K^+\rangle$
is also a natural-looking symmetric combination. 
For the $\Lambda$, the $d$ combination is also the natural
non-singlet operator to consider, 
$ d \propto \langle \Lambda | (2 \bar s \gamma s 
-\bar u \gamma u - \bar d \gamma d ) | \Lambda \rangle $
because the $u$ and $d$ in the $\Lambda$ have the same 
structure functions, while the $s$ structure is different 
(even before breaking $SU(3)$). 
 
But in the proton, it might be a bit more natural to choose the 
combinations $(\bar u \gamma u - \bar d \gamma d)$ and 
$( \bar u \gamma u + \bar d \gamma d - 2 \bar s \gamma s)$
instead. The first combination is the non-singlet combination
normally considered in discussions of proton structure, 
the second is almost (but not exactly) a measure of the 
total valence contribution, because the quark-line-disconnected
(sea) contribution to
$( \bar u \gamma u + \bar d \gamma d - 2 \bar s \gamma s)$
is zero at the symmetric point, and will probably stay
small if the nucleon's sea is approximately $SU(3)$ symmetric.

We can therefore construct a fan  plot for the 
doubly represented $-$ singly represented quark. 
\begin{eqnarray} 
   P_1 =  \sqrt{2} A_{\bar N \pi N} 
          &=& \left( \sqrt{2} f + \sqrt{6} d \right)  
                                - 2 \sqrt{2} ( r_3 - s_1 ) \delta m_l \,,
                                                        \nonumber  \\
   P_2 = \frac{1}{\sqrt{2}} ( A_{\bar \Sigma \pi \Sigma} 
                                + \sqrt{3} A_{\bar \Sigma \eta  \Sigma} )
         &=& \left( \sqrt{2} f + \sqrt{6} d \right)
                                                        \nonumber  \\
         & &                      + \frac{1}{\sqrt{2}}
    \left( \sqrt{3} r_1 + 6 r_3 - 2s_1 + \sqrt{3} s_2 \right) \delta m_l \,,
                                                        \nonumber  \\
   P_3 = - \frac{1}{\sqrt{2}} ( A_{\bar \Xi \pi \Xi}
                                + \sqrt{3} A_{\bar \Xi \eta  \Xi} ) 
          &=& \left( \sqrt{2} f + \sqrt{6} d \right)
                                                                   \\
         & & - \frac{1}{\sqrt{2}}
             \left( \sqrt{3} r_1 + {2} r_3 + {2} s_1 + \sqrt{3} s_2 \right) 
                   \delta m_l \,,
                                                        \nonumber  \\
   P_4 = A_{\bar \Sigma K \Xi}
          &=&  \left( \sqrt{2} f + \sqrt{6} d \right) 
                      + \sqrt{2}( r_3 - s_1) \delta m_l  \,.
                                                        \nonumber 
\end{eqnarray} 
We have based this fan plot on the doubly $-$ singly represented
structure, so several of the observables have very simple 
quark structures. 
\begin{eqnarray} 
   P_1 &=& \langle p | (\bar u \gamma u - \bar d \gamma d) | p \rangle \,,
                                                        \nonumber  \\
   P_2 &=& \langle \Sigma^+ | (\bar u \gamma u 
                            - \bar s \gamma s) | \Sigma^+ \rangle \,,
                                                        \label{simpleP} \\
   P_3 &=& 
      \langle \Xi^0 | (\bar s \gamma s - \bar u \gamma u) | \Xi^0 \rangle \,,
                                                        \nonumber  \\
   P_4 &=&   \langle \Sigma^+ | \bar u \gamma s | \Xi^0 \rangle \,.
\nonumber
\end{eqnarray} 
This $P$ fan only includes the `outer' octet baryons.
The natural plot for the $\Lambda$ structure is the $d$-fan. 
There are two linear constraints on the $P$-fan, 
\begin{eqnarray}
   \frac{1}{3} ( P_1 + P_2 + P_3 ) 
      &=& ( \sqrt{2} f + \sqrt{6} d) + O(\delta m_l^2) \,,
                                                        \nonumber  \\
   \frac{1}{3} ( P_1 + 2 P_4 )
      &=& ( \sqrt{2} f + \sqrt{6} d) + O(\delta m_l^2) \,.
\label{P_sum}
\end{eqnarray} 
A fan with just the four lines from eq.~(\ref{simpleP}), $P_1, P_2, P_3, P_4,$ 
is a four-line plot with just two independent slope parameters, $(r_3-s_1)$
and $(\sqrt{3} r_1 + 4 r_3 + \sqrt{3} s_2)$. 

The advantage of this fan plot is that some of the quantities 
are of immediate physical interest, for example in the 
weak decay case $P_1$ gives the neutron decay constant, 
while $P_4$ gives the semileptonic decays 
$\Xi^0 \to \Sigma^+ l^- \bar \nu_l $, 
$\Xi^- \to \Sigma^0 l^- \bar \nu_l $. 
The disadvantages are that there are fewer constraints
than the $d$-fan. Also, the $d$-fan and $f$-fan are 
independent -- they involve different parameters, 
and there are no constraints that mix $F_i$ and $D_i$
quantities.
A first attempt to show this fan plot for
the fraction of the baryon's momentum carried by a quark,
i.e.$\langle x \rangle$, is given in \cite{Gockeler:2011ze}.

Finally it is again often useful to note from eq.~(\ref{P_sum}) 
that for example
\begin{equation}
   X_P = \frac{1}{3} ( P_1 + P_2 + P_3 ) 
       =  (\sqrt{2} f + \sqrt{6} d) + O(\delta m_l^2) \,,
\end{equation} 
and to consider the quantities $P_i/X_P$.


 \subsection{The $V$-fan} 


The other natural non-singlet to look at in the proton is 
$\langle p | ( \bar u \gamma u + \bar d \gamma d - 2 \bar s \gamma s ) 
| p \rangle $. This is approximately the total valence distribution, 
the quark-line-disconnected (sea) contribution to
$( \bar u \gamma u + \bar d \gamma d - 2 \bar s \gamma s)$
is zero at the symmetric point, and will probably stay small if the 
nucleon's sea is approximately $SU(3)$ symmetric. 
\begin{eqnarray} 
   V_1 = \sqrt{6} A_{\bar N \eta N}
      &=& \sqrt{6} (\sqrt{3} f - d) + \sqrt{6} (r_1 - s_2 ) \delta m_l \,,
                                                        \nonumber  \\
   V_2 = \frac{3}{\sqrt{2}} A_{\bar \Sigma \pi \Sigma} 
                - \sqrt{\frac{3}{2}} A_{\bar \Sigma \eta \Sigma } \,,
      &=& \sqrt{6} (\sqrt{3} f - d)
                                                        \nonumber  \\
      & & - \frac{1}{\sqrt{2}}
 \left( \sqrt{3} r_1 + {6} r_3 + {6} s_1 -3 \sqrt{3} s_2 \right) \delta m_l \,,
                                                        \nonumber  \\
   V_3 = \frac{3}{\sqrt{2}} A_{\bar \Xi \pi \Xi}  
                     - \sqrt{\frac{3}{2}} A_{\bar \Xi \eta \Xi } 
      &=& \sqrt{6} (\sqrt{3} f - d) \,,
                                                                   \\
      & & - \frac{1}{\sqrt{2}}
 \left( \sqrt{3} r_1 - {6} r_3 - {6} s_1 + \sqrt{3} s_2 \right) \delta m_l \,,
                                                        \nonumber  \\
   V_4 = \sqrt{2} ( A_{\bar N \pi N} + 2 A_{\bar \Xi \pi \Xi } )  
      &=& \sqrt{6} (\sqrt{3} f - d)  + 2 \sqrt{2} ( r_3 + 3 s_1 ) \delta m_l \,,
                                                        \nonumber  \\
   V_5 = ( A_{ \bar \Sigma K \Xi } - 2 A_{\bar N K \Sigma} ) 
    &=& \sqrt{6} (\sqrt{3} f - d) - \sqrt{2} ( r_3 + 3 s_1 ) \delta m_l \,.
                                                        \nonumber 
\end{eqnarray} 
We have the two constraints 
\begin{eqnarray} 
   \frac{1}{3} ( V_1 + V_2 + V_3 ) 
      &=& \sqrt{6} (\sqrt{3} f - d) + O(\delta m_l^2) \,,
                                                        \nonumber  \\
  \frac{1}{3} ( V_4 + 2 V_5 )
      &=& \sqrt{6} (\sqrt{3} f - d) + O(\delta m_l^2) \,,
\end{eqnarray} 
and can again construct an $X_V$ from either combination, for example set
\begin{eqnarray}
   X_V = {1 \over 3} ( V_1 + V_2 + V_3 ) \,,
\end{eqnarray}
and again consider ratios such as $V_i/X_V$.


\section{LO flavour diagonal matrix elements}
\label{LO_flav_diag_ME}


To leading order we have for the representative octet baryons 
$p$, $\Sigma^+$, $\Lambda^0$ and $\Xi^0$
\begin{eqnarray}
   \lefteqn{\langle p|\bar{u}\gamma u|p \rangle}
   & &                                                    \nonumber \\
   &=& {1\over\sqrt{3}}\left(a_0 + \sqrt{6}f + \sqrt{2}d\right)
                 + {1\over\sqrt{3}}\left(
                     3a_1 + {1\over\sqrt{2}}r_1 - \sqrt{6}r_3
                     + \sqrt{6}s_1 - {1\over\sqrt{2}}s_2 \right) \delta m_l \,,
                                                          \nonumber \\
   \lefteqn{\langle p|\bar{d}\gamma d|p \rangle}
   & &                                      \label{delta_q_expan_p} \\
   &=& {1\over\sqrt{3}}\left(a_0 -2\sqrt{2}d\right)
                 + {1\over\sqrt{3}}\left(
                     3a_1 + {1\over\sqrt{2}}r_1 + \sqrt{6}r_3
                     - \sqrt{6}s_1 - {1\over\sqrt{2}}s_2 \right) \delta m_l \,,
                                                          \nonumber \\
   \lefteqn{\langle p|\bar{s}\gamma s|p \rangle}
   & &                                                    \nonumber \\
   &=& {1\over\sqrt{3}}\left(a_0 -\sqrt{6}f + \sqrt{2}d\right)
                 + {1\over\sqrt{3}}\left(
                     3a_1 - \sqrt{2}r_1 + \sqrt{2}s_2 \right) \delta m_l \,,
                                                          \nonumber
\end{eqnarray}   
\begin{eqnarray}
   \lefteqn{\langle \Sigma^+|\bar{u}\gamma u|\Sigma^+ \rangle}
     & &                                                  \nonumber \\
     &=& {1\over\sqrt{3}}\left(a_0 + \sqrt{6}f + \sqrt{2}d\right)
                 + {1\over\sqrt{3}}\left(
                    -3a_2 + {1\over\sqrt{2}}r_1 + \sqrt{6}r_3
                     - \sqrt{6}s_1 + {3\over\sqrt{2}}s_2 \right) \delta m_l \,,
                                                          \nonumber \\
   \lefteqn{\langle \Sigma^+|\bar{d}\gamma d|\Sigma^+ \rangle}
     & &                                \label{delta_q_expan_Sigma} \\
     &=& {1\over\sqrt{3}}\left(a_0 - \sqrt{6}f + \sqrt{2}d\right)
                 + {1\over\sqrt{3}}\left(
                    -3a_2 + {1\over\sqrt{2}}r_1 + \sqrt{6}r_3
                     + \sqrt{6}s_1 - {3\over\sqrt{2}}s_2 \right) \delta m_l \,,
                                                          \nonumber \\
   \lefteqn{\langle \Sigma^+|\bar{s}\gamma s|\Sigma^+ \rangle}
     & &                                                  \nonumber \\
     &=& {1\over\sqrt{3}}\left(a_0 -2\sqrt{2}d\right)
                 + {1\over\sqrt{3}}\left(
                    -3a_2 - \sqrt{2}r_1 - 2\sqrt{6}r_3 \right) \delta m_l \,,
                                                          \nonumber         
\end{eqnarray}   
\begin{eqnarray}
   \langle \Lambda|\bar{u}\gamma u|\Lambda \rangle
      &=& \langle \Lambda|\bar{d}\gamma d|\Lambda \rangle \,,
                                                          \nonumber \\
      &=& {1\over\sqrt{3}}\left(a_0 - \sqrt{2}d\right)
                 + {1\over\sqrt{3}}\left(
             3a_2 + {1\over\sqrt{2}}r_1 + \sqrt{2}r_2 \right) \delta m_l \,,
                                                          \nonumber \\
   \langle \Lambda|\bar{s}\gamma s|\Lambda \rangle
      &=& {1\over\sqrt{3}}\left(a_0 + 2\sqrt{2}d\right)
                 + {1\over\sqrt{3}}\left(
                     3a_2 - \sqrt{2}r_1 - 2\sqrt{2}r_2 \right) \delta m_l \,,
\label{delta_q_expan_Lam}                 
\end{eqnarray}   
and
\begin{eqnarray}
   \lefteqn{\langle \Xi^0|\bar{u}\gamma u|\Xi^0 \rangle}
     & &                                                  \nonumber \\
     &=& {1\over\sqrt{3}}\left(a_0 - 2\sqrt{2}d\right)
                 + {1\over\sqrt{3}}\left(
                    -3(a_1-a_2) + {1\over\sqrt{2}}r_1 + \sqrt{6}r_3
                     + \sqrt{6}s_1 + {1\over\sqrt{2}}s_2 \right) \delta m_l \,,
                                                          \nonumber \\
   \lefteqn{\langle \Xi^0|\bar{d}\gamma d|\Xi^0 \rangle}
     & &                                                  \nonumber \\
     &=& {1\over\sqrt{3}}\left(a_0 - \sqrt{6}f + \sqrt{2}d\right)
                                           \label{delta_q_expan_Xi} \\
     & & + {1\over\sqrt{3}}\left(
                    -3(a_1-a_2) + {1\over\sqrt{2}}r_1 - \sqrt{6}r_3
                     - \sqrt{6}s_1 + {1\over\sqrt{2}}s_2 \right) \delta m_l \,,
                                                          \nonumber \\
   \lefteqn{\langle \Xi^0|\bar{s}\gamma s|\Xi^0 \rangle}
     & &                                                  \nonumber \\
     &=& {1\over\sqrt{3}}\left(a_0 +\sqrt{6}f+\sqrt{2}d\right)
                 + {1\over\sqrt{3}}\left(
                 -3(a_1-a_2) - \sqrt{2}r_1 - \sqrt{2}s_2 \right) \delta m_l \,.
                                                          \nonumber
\end{eqnarray}


\section{LO disconnected flavour diagonal matrix elements}
\label{LO_flav_diag_ME_dis}


From eqs.~(\ref{dis_zero_text}, \ref{dis_zero_r2})
we have $f^{\ind{dis}}$, $d^{\ind{dis}}$, $r_2^{\ind{dis}}$, $r_3^{\ind{dis}}$,
$s_1^{\ind{dis}}$ and $s_2^{\ind{dis}}$ all vanishing at LO and
only $r_1^{\ind{dis}}$ contributing. Thus we have
\begin{eqnarray}
   \langle N | \bar{u}\gamma u | N \rangle^{\ind{dis}}
      &=& \langle N | \bar{d}\gamma d | N \rangle^{\ind{dis}}
                                                          \nonumber \\
      &=& {1\over \sqrt{3}}a_0^{\ind{dis}}
          + \left( \sqrt{3}a_1^{\ind{dis}} + {1\over\sqrt{6}}r_1^{\ind{dis}}
                         \right) \delta m_l \,,
                                         \label{delta_q_expan_dis_p} \\
   \langle N | \bar{s}\gamma s | N \rangle^{\ind{dis}}
      &=& {1\over \sqrt{3}}a_0^{\ind{dis}}
          + \left( \sqrt{3}a_1^{\ind{dis}} - \sqrt{2\over 3}r_1^{\ind{dis}}
                         \right)\delta m_l \,,
                                                         \nonumber
\end{eqnarray}
(for $n$, $p$),
\begin{eqnarray}
   \langle \Sigma | \bar{u}\gamma u | \Sigma \rangle^{\ind{dis}}
     &=& \langle \Sigma | \bar{d}\gamma d | \Sigma \rangle^{\ind{dis}}
                                                          \nonumber \\
     &=& {1\over\sqrt{3}}a_0^{\ind{dis}}
         + \left(-\sqrt{3}a_2^{\ind{dis}} + {1\over\sqrt{6}}r_1^{\ind{dis}} 
                         \right) \delta m_l \,,
                                    \label{delta_q_expan_dis_Sigma} \\
   \langle \Sigma | \bar{s}\gamma s | \Sigma \rangle^{\ind{dis}}
     &=& {1\over\sqrt{3}}a_0^{\ind{dis}} 
             + \left( -\sqrt{3}a_2^{\ind{dis}} - \sqrt{2\over 3}r_1^{\ind{dis}}
                          \right) \delta m_l \,,
                                                          \nonumber
\end{eqnarray}   
(for $\Sigma^+$, $\Sigma^0$, $\Sigma^-$),
\begin{eqnarray}
   \langle \Lambda | \bar{u}\gamma u | \Lambda \rangle^{\ind{dis}}
      &=& \langle \Lambda | \bar{d}\gamma d | \Lambda \rangle^{\ind{dis}}
                                                          \nonumber \\
      &=& {1\over\sqrt{3}}a_0^{\ind{dis}}
          + \left(
              \sqrt{3}a_2^{\ind{dis}} + {1\over\sqrt{6}}r_1^{\ind{dis}}
                             \right) \delta m_l \,,
                                      \label{delta_q_expan_dis_Lam} \\
   \langle \Lambda | \bar{s}\gamma s | \Lambda \rangle^{\ind{dis}}
      &=& {1\over\sqrt{3}}a_0^{\ind{dis}}
          + \left(
              \sqrt{3}a_2^{\ind{dis}} - \sqrt{2\over 3}r_1^{\ind{dis}}
                             \right) \delta m_l \,,
                                                          \nonumber
\end{eqnarray}   
(for $\Lambda^0$) and
\begin{eqnarray}
   \langle \Xi | \bar{u}\gamma u | \Xi \rangle^{\ind{dis}}
     &=& \langle \Xi | \bar{d}\gamma d | \Xi \rangle^{\ind{dis}}
                                                          \nonumber \\
     &=& {1\over\sqrt{3}}a_0^{\ind{dis}}
           + \left(-\sqrt{3}(a_1^{\ind{dis}}-a_2^{\ind{dis}})
                         + {1\over\sqrt{6}}r_1^{\ind{dis}}
                    \right) \delta m_l \,,
                                       \label{delta_q_expan_dis_Xi} \\
   \langle \Xi | \bar{s}\gamma s | \Xi \rangle^{\ind{dis}}
     &=& {1\over\sqrt{3}}a_0^{\ind{dis}}
         + \left(-\sqrt{3}(a_1^{\ind{dis}}-a_2^{\ind{dis}})
                         - \sqrt{2\over 3}r_1^{\ind{dis}}
                   \right) \delta m_l \,,
                                                          \nonumber
\end{eqnarray}
(for $\Xi^0$, $\Xi^-$).





\end{document}